\documentclass[12pt,bibliography=totoc,DIV=20]{scrartcl}
\pdfoutput=1
\usepackage{amsmath,amssymb,color,graphicx,array}
\usepackage[symbol]{footmisc}

\usepackage{amsmath,amsfonts}
\usepackage[mathscr]{euscript}
\usepackage{longtable}
\usepackage{booktabs}
\usepackage{multirow}
\usepackage{cite}
\usepackage{graphicx, ulem, hhline}
\usepackage[table]{xcolor}
\usepackage{xspace}
\usepackage{needspace}
\usepackage{siunitx}
\usepackage{soul}
\usepackage{wasysym}              
\usepackage{hyphenat}             

\usepackage{afterpage}
\usepackage{etoolbox,xstring,xspace,calc,xifthen}

\long\def\symbolfootnote[#1]#2{\begingroup%
\def\thefootnote{\fnsymbol{footnote}}\footnote[#1]{#2}\endgroup}

\makeatletter
\def\@fnsymbol#1{\ensuremath{\ifcase#1\or%
\ast\or \dagger\or \ddagger\or \mathsection\or \parallel\or \nparallel\or%
\mathparagraph\or \cap\or \cup\or \subset\or \supset\or%
\wedge\or \vee\or <\or >\or \diamond\or \circ\or%
\vartriangle\or \triangledown\or \triangleleft\or \triangleright\or%

\else\@ctrerr\fi}}
\makeatother

\makeatletter
\newlength{\fnhskip}
\renewcommand\@makefntext[1]{
  \settowidth{\fnhskip}{\@makefnmark}
  \leftskip=\fnhskip
  \hskip-\fnhskip
  \@makefnmark#1
}
\makeatother

\renewenvironment{subequations}[1][]{
  \refstepcounter{equation}%
  \setcounter{parentequation}{\value{equation}}
  \setcounter{equation}{0}
  \def\theequation{\theparentequation\alph{equation}}%
  \let\parentlabel\label
  \ifx\\#1\\\relax\else\label{#1}\fi
  \ignorespaces
}{%
  \setcounter{equation}{\value{parentequation}}
  \ignorespacesafterend
}

\newcommand*{\nextParentEquation}[1][]{
  \refstepcounter{parentequation}
  \setcounter{equation}{0}
  \ifx\\#1\\\relax\else\parentlabel{#1}\fi
}

\usepackage{setspace}

\usepackage[numbers,sort&compress]{natbib}
\makeatletter
\def\NAT@spacechar{\,}
\makeatother
\usepackage[
colorlinks=true
,urlcolor=blue
,anchorcolor=blue
,citecolor=blue
,filecolor=blue
,linkcolor=blue
,menucolor=blue
,linktoc=all
,unicode
,backref=page
]{hyperref}
\usepackage{hypernat}

\newrobustcmd*{\tocref}[1]{\hyperref[TOC]{\color{black}{#1}}}
\newcommand{\tocsection}[2][]{\section[\boldmath #2]{\boldmath\tocref{#2#1}}}
\newcommand{\tocsubsection}[2][]{\subsection[#2]{\boldmath\tocref{#2#1}}}
\newcommand{\tocsubsubsection}[2][]{\subsubsection[#2]{\boldmath\tocref{#2#1}}}

\usepackage[all]{hypcap}
\usepackage{footnotebackref}

\usepackage[format=plain, margin=0.5cm, font=footnotesize, labelfont=bf]{caption}

\renewcommand*{\backref}[1]{}
\renewcommand*{\backrefalt}[4]{%
  \ifcase #1%
  \or [p\,#2]%
  \else [pp\,#2]%
  \fi%
}

\newif\ifbackrefshowonlyfirst
\backrefshowonlyfirsttrue
\makeatletter
\let\BR@direct@old@hyper@natlinkstart\hyper@natlinkstart
\renewcommand*{\hyper@natlinkstart}{\phantomsection\BR@direct@old@hyper@natlinkstart}
\let\BR@direct@oldBR@citex\BR@citex
\renewcommand*{\BR@citex}{\phantomsection\BR@direct@oldBR@citex}%

\long\def\hyper@page@BR@direct@ref#1#2#3{\hyperlink{#3}{#1}}

\ifx\backrefxxx\hyper@page@backref
    \let\backrefxxx\hyper@page@BR@direct@ref
    \ifbackrefshowonlyfirst
    \fi
\else
    \ifbackrefshowonlyfirst
    \fi
\fi

\patchcmd{\Hy@backout}{Doc-Start}{\@currentHref}{}{\errmessage{I can't seem to patch backref}}
\makeatother

\let\theparentequation\theequation
\patchcmd{\theparentequation}{equation}{parentequation}{}{}

\apptocmd{\thebibliography}{\scriptsize}{}{}
\let\OLDthebibliography\thebibliography
\renewcommand\thebibliography[1]{
  \OLDthebibliography{#1}
  \setlength{\parskip}{1pt}
  \setlength{\itemsep}{1pt plus 0.3ex}
}

\addtokomafont{disposition}{\rmfamily}
\BeforeTOCHead[toc]{{\pdfbookmark[1]{\contentsname}{toc}}}

\input{definitions}

\hypersetup{
  pdfauthor={Mark D. Goodsell and Sebastian Paßehr}
  ,pdftitle={All two-loop scalar self-energies and tadpoles in general renormalisable field theories}
  ,pdfsubject={}
  ,pdfkeywords={}
}

\begin{document}

\begin{titlepage}

\begin{flushright}
TTK-19-38
\end{flushright}
  
\begin{center}  
\vspace{1cm}

{\LARGE \textbf{All two-loop scalar self-energies and tadpoles in general renormalisable field theories} }

\vspace{1cm}

{\Large Mark~D.~Goodsell$^a$ and Sebastian Pa{\ss}ehr$^{a,b}$}

\vspace*{5mm}

$^{a}$\textsl{Laboratoire de Physique Th\'eorique et Hautes Energies (LPTHE),}\\
\textsl{UMR 7589, Sorbonne Universit\'e et CNRS, 4 place Jussieu, 75252 Paris Cedex 05, France.}

\medskip

$^{b}$\textsl{Institute for Theoretical Particle Physics and Cosmology,}\\
\textsl{RWTH Aachen University, 52056 Aachen, Germany.}

\end{center}
\symbolfootnote[0]{\texttt{e-mail:}}
\symbolfootnote[0]{\texttt{goodsell@lpthe.jussieu.fr}}
\symbolfootnote[0]{\texttt{passehr@physik.rwth-aachen.de}}

\abstract{We calculate the complete tadpoles and self-energies at the
  two-loop order for scalars in general renormalisable theories, a
  crucial component for calculating two-loop electroweak corrections
  to Higgs-boson masses or for any scalar beyond the Standard
  Model. We renormalise the amplitudes using mass-independent
  renormalisation schemes, based on both dimensional regularisation
  and dimensional reduction.

  The results are presented here in Feynman gauge, with expressions
  for all $121$ self-energy and $25$ tadpole diagrams given in terms
  of scalar and tensor integrals with the complete set of rules to
  reduce them to a minimal basis of scalar integrals for any physical
  kinematic configuration. In addition, we simplify the results to a
  set of only $16$ tadpole and $58$ self-energy topologies using
  relations in order to substitute the ghost and Goldstone-boson
  couplings that we derive.

  To facilitate their application, we also provide our results in
  electronic form as a new code \ourcode. We test our results by
  applying them to the Standard Model and compare with analytic
  expressions in the literature.}

\end{titlepage}

\hypersetup{linkcolor=black}
\tableofcontents\label{TOC}
\hypersetup{linkcolor=blue}

\vspace{4ex}

\tocsection[\label{SEC:INTRO}]{Introduction}

The Higgs boson mass has been measured to an accuracy of
about~$\mathcal{O}(100)$\,MeV, making it an electroweak precision
parameter. In the Standard Model~(SM), this is used to extract the
Higgs quartic coupling~$\lambda$, which, through a significant amount
of work, can now be done with high precision, where all of the
relevant running parameters of the Lagrangian can now be extracted
from calculations at full two-loop
order\,\cite{Degrassi:2012ry,Buttazzo:2013uya} and partially at the
three- and four-loop
order\,\cite{vanRitbergen:1997va,Chetyrkin:1997dh,Vermaseren:1997fq,Chetyrkin:2004mf,Mihaila:2012fm,Chetyrkin:2012rz,Mihaila:2012pz,Bednyakov:2012rb,Bednyakov:2012en,Chetyrkin:2013wya,Bednyakov:2013eba,Marquard:2015qpa,Martin:2015eia,Chetyrkin:2016ruf}. A
code for calculating~$\lambda$ in Landau
gauge, \SMH{}\,\cite{Martin:2014cxa}, and codes for extracting all
relevant SM~parameters (the gauge couplings, top and bottom Yukawa
couplings, and Higgs quartic
coupling), \mr{}\,\cite{Kniehl:2015nwa,Kniehl:2016enc}
and \SMDR{}\,\cite{Martin:2019lqd}, exist. As a result of this effort,
the uncertainty on the measurement of the top mass is now more
important than the scalar self-energies in the~SM.

However, in theories beyond the Standard Model~(BSM), it is not
possible to make full use of the Higgs-mass measurement because the
theoretical uncertainty on the mass calculation can be much larger,
owing to the new degrees of freedom. As a result, an enormous amount
of effort has gone into refining the calculation of the Higgs-boson
mass from a given set of physical or top-down inputs, in both generic
and specific theories. This has typically been driven by the need for
accurate predictions of the Higgs mass in supersymmetric models, where
the Higgs quartic coupling is predicted from the gauge couplings (and
other top-down parameters in extended models).

The early expectation was for new coloured supersymmetric particles
near the electroweak scale, and since the Minimal Supersymmetric
SM~(MSSM) had a tree-level upper bound on the Higgs mass equal to the
mass of the~$Z$~boson, a full fixed-order calculation in the~MSSM at
the one-loop
order\,\cite{Haber:1990aw,Ellis:1990nz,Okada:1990vk,Okada:1990gg,Ellis:1991zd,Brignole:1992uf,Chankowski:1991md,Dabelstein:1994hb,Pierce:1996zz,Frank:2006yh}
is vital, but the results at the two-loop order are known to
contribute several~GeV to the SM-like Higgs-boson mass. After much
work a full fixed-order two-loop result is still not available: public
codes use an effective-potential calculation (neglecting the external
momentum) in the ``gaugeless limit'' (neglecting the electroweak gauge
couplings) with results in the ``real MSSM''
(neglecting \CP~phases)\,\cite{Hempfling:1993qq,Heinemeyer:1998kz,Heinemeyer:1998jw,Heinemeyer:1998np,Zhang:1998bm,Heinemeyer:1999be,Espinosa:1999zm,Espinosa:2000df,Brignole:2001jy,Degrassi:2001yf,Degrassi:2002fi,Martin:2002wn,Brignole:2002bz,Dedes:2002dy,Martin:2002iu,Dedes:2003km,Allanach:2004rh,Heinemeyer:2004xw}
and in the ``complex MSSM''
(including \CP~violation)\,\cite{Pilaftsis:1998pe,Demir:1999hj,Pilaftsis:1999qt,Choi:2000wz,Ibrahim:2000qj,Heinemeyer:2001qd,Ibrahim:2002zk,Heinemeyer:2007aq,Hollik:2014wea,Hollik:2014bua,Passehr:2017ufr}. Some
results also exist for three-loop strong corrections in the
effective-potential limit,
see \citeres{Martin:2007pg,Kant:2010tf,Harlander:2008ju}, now
available in
the \Himalaya~package\,\cite{Harlander:2017kuc,Harlander:2018yhj}
which also includes four-loop leading logarithms of the strong
corrections.

Some results now also exist beyond the gaugeless/effective-potential
limit: a complete \emph{effective-potential} calculation was described
in \citeres{Martin:2002iu,Martin:2002wn}, where the effective
potential was computed and the derivatives taken numerically, but
application of this was hampered by the \emph{Goldstone-Boson
Catastrophe}~(GBC) which we shall discuss below. Contributions
of~$\mathcal{O}{\left(\!\left(\alpha_t + \alpha_b
+ \alpha_\tau\right)^2\right)}$\,\cite{Martin:2004kr},
$\mathcal{O}{\left(\alpha_t\, \alpha_s\right)}$\,\cite{Borowka:2014wla,Degrassi:2014pfa,Borowka:2015ura},
$\mathcal{O}{\left(\alpha\, \alpha_s\right)}$\,\cite{Degrassi:2014pfa},
and~$\mathcal{O}{\left(\!\left(\alpha_t+\alpha_b+\alpha\right)\alpha_s\right)}$\,\cite{Borowka:2018anu}
with non-vanishing external momentum were computed in fixed-order
computations.\footnote{We make use of the common notation
that~$\alpha_i \equiv g_i^2/(4\,\pi)$ where~$g_i$ is a given coupling;
$\alpha \approx 1/137$ corresponds to the electric coupling, while
$g_s, g_t, g_b$ correspond to the strong gauge, top Yukawa and bottom
Yukawa couplings.} However, there is now some urgency to fill in the
remaining discrepancy of the full electroweak corrections and include
momentum dependence (which are of the same nominal order for the
SM-like Higgs boson)---these have remained a ``holy grail'' of the
community for some time.

If new particles beyond the~SM are very heavy compared to the
electroweak scale, a fixed-order calculation breaks down due to large
logarithms, and an effective-field-theory~(EFT) approach (or a hybrid
approach,
see \citeres{Hahn:2013ria,Athron:2016fuq,Bahl:2016brp,Athron:2017fvs,Bahl:2017aev,Staub:2017jnp,Bahl:2018ykj})
should be used. Such techniques have been applied for some
time\,\cite{Sasaki:1991qu,Casas:1994us,Carena:1995bx,Carena:1995wu,Haber:1996fp,Carena:2000dp,Carena:2000yi,Carena:2001fw,Espinosa:2001mm,Draper:2013oza,Lee:2015uza},
but it has been found that the threshold corrections when matching on
to the~SM (or split supersymmetry, or Two-Higgs-Doublet Model) can be
large. Indeed, while the two-loop renormalisation group
equations~(RGEs) have been known since the~$90$s, the one-loop
threshold corrections in the~MSSM were only evaluated
recently\,\cite{Bernal:2007uv,Giudice:2011cg,Bagnaschi:2014rsa}, and
followed by an extraction of the same two-loop corrections that are
available for fixed-order
computations\,\cite{Bagnaschi:2014rsa,Vega:2015fna,Bagnaschi:2017xid,Bagnaschi:2019esc},
essentially by matching the masses of the lightest Higgs boson in the
MSSM and the SM at the EFT matching scale. Indeed, as discussed
in \citeres{Athron:2016fuq,Braathen:2018htl}, a fixed-order
calculation can always be used to extract threshold corrections in
this way, especially when the low-energy theory contains only one
scalar field such as in the~SM. Hence development of fixed-order and
EFT calculations go hand in hand.

Since the~MSSM has been the driver for much of this work, the
accompanying work on other theories has been much less developed until
recently: in the next-to-MSSM~(NMSSM), fixed-order calculations were
done at the one-loop
order\,\cite{Ellwanger:2004xm,Ellwanger:2005dv,Ellwanger:2009dp,Degrassi:2009yq,Staub:2010ty,Ender:2011qh,Porod:2011nf,Graf:2012hh,Allanach:2013kza,Domingo:2015qaa,Domingo:2017rhb,Hollik:2018yek}
with only the dominant two-loop corrections in the effective-potential
approach
of~$\mathcal{O}{\left(\!\left(\alpha_t+\alpha_b\right)\alpha_s\right)}$\,\cite{Degrassi:2009yq,Muhlleitner:2014vsa}
and~$\mathcal{O}{\left(\alpha_t^2\right)}$\,\cite{Dao:2019qaz}
explicitly available; for Dirac-gaugino models,
the~$\mathcal{O}{\left(\alpha_t\,\alpha_s\right)}$ corrections were
computed in \citere{Braathen:2016mmb}.

However, particularly given the absence of clear signs of new physics
from the Large Hadron Collider~(LHC), it is sensible to take a
model-agnostic approach where possible, and this has led to a program
of \emph{generic} calculations. In \citere{Martin:2001vx} the full
effective potential was given for \emph{general renormalisable
theories} in Landau gauge. This was then implemented in the package
\SARAH in \citere{Goodsell:2014bna} in the gaugeless limit, where
the first and second derivatives of the potential were taken
numerically. This allowed, for the first time, two-loop corrections to
be computed for \emph{any} model at the push of a button. Moreover,
in \citere{Martin:2003it}, scalar self-energies for general theories
were computed up to \emph{second order in the gauge couplings}. This
is sufficient for a gaugeless-limit computation of the Higgs mass,
however the tadpole diagrams were lacking. In \citere{Goodsell:2015ira}
these were computed, and the self-energies simplified to the effective
potential limit---the result being again implemented in the package
\SARAH. However, even in the gaugeless effective-potential limit,
this calculation was plagued by the~GBC. The solution came for the
general case in \citere{Braathen:2016cqe}, and was implemented
in \SARAH with some additional developments to the method
in \citere{Braathen:2017izn}. This now represents the state-of-the art
for any theory---supersymmetric or otherwise---other than the~SM
or~MSSM (for example, \citeres{Goodsell:2014pla,Goodsell:2016udb}
describe the only calculations including all superpotential terms in
the~NMSSM), and in fact provides the only calculation including
flavour-violating effects at the two-loop
order\,\cite{Goodsell:2015yca}, or with pure
$\ov{\mathrm{DR}}'$~renormalisation for the
complex~MSSM\,\cite{Goodsell:2016udb}. The status for threshold
corrections lags somewhat behind: generic thresholds at the one-loop
order were computed in \citeres{Braathen:2018htl,Gabelmann:2018axh},
where in particular the former reference describes the consistent
treatment of infra-red divergences and counterterm choices that can
simplify the computation, while the latter described the
implementation in \SARAH.

The purpose of this work is to finally complete the set of scalar
self-energy diagrams in the gauge coupling for general renormalisable
theories, and provide the tadpole diagrams at the same time. This
completes the set that was promised in \citere{Martin:2003it}. In this
paper we present the analytic expressions and the technical machinery
that we have used, specialising to the Feynman gauge. However, since
the final results (and therefore this paper) are rather long, we have
created a new package \ourcode where they are available in
computer-readable form. Readers wishing to skip the details and apply
the results are invited to download the code from:
\begin{gather*}
  \website
\end{gather*}

While some of the evaluation of spinor/Lorentz traces and tensor
reduction for specific models could be accomplished
using \TwoCalc{}\,\cite{Weiglein:1993hd}
and \TARCER{}\,\cite{Mertig:1998vk} (of which we have made use) we
derived some reduction rules not available there with the help of the
general relations of \citeres{Tarasov:1997kx,Davydychev:1998si}, so
that all results can be reduced to a basis of just a few one- and
two-loop scalar functions that can be numerically evaluated
in \TSIL{}\,\cite{Martin:2005qm}. Moreover, our results are
already \emph{renormalised}, and in particular we reduce the number of
classes of diagrams by making extensive use of identities relating
couplings of ghosts and Goldstone bosons to other couplings in the
theory.

Our calculation can be used for:
\begin{itemize}
\item
Corrections to charged and/or coloured scalar masses. For example, in
supersymmetric theories this would mean \EG\ squark or sgluon masses.
\item
Electroweak corrections to the Higgs-boson mass $\leftrightarrow$
extraction of Higgs/neutral scalar quartic couplings. These ought to
be supplemented by a two-loop extraction of the electroweak
expectation value and gauge couplings (which requires the two-loop
corrections to muon decay and vector-boson self-energies, which we
hope to return to in future work).
\item
EFT matching of the Higgs quartic coupling via the pole-mass matching
technique\,\cite{Athron:2016fuq}. As described at one-loop order
in \citere{Braathen:2018htl}, although a priori this would seem to
require a calculation of the $Z$-boson mass, in fact all of the
necessary information is contained in the scalar
self-energies/tadpoles, and thus the calculations here may be
sufficient.
\end{itemize}

\tocsubsection[\label{sec:introGBC}]{Treatment of tadpoles and application of our results}

The main application of our results is expected to be the evaluation
of the pole mass for scalar bosons in any given theory. For a general
theory with scalars having indices~$i,j, \ldots$ and masses~$\m_i^2$
at the tree level, this corresponds to finding the (complex) solutions
of the equations
\begin{align}
  0 &= \mathrm{Det}{\Big[\!\left(p^2 - \m^2_i\right) \delta_{ij}
                          - \Pi_{ij}{\left(p^2\right)}\Big]}\,,
  \label{EQ:DetSelf}
\end{align}
in $p^2$, where $\Pi_{ij} (p^2)$ is the self-energy. We then write
\begin{align}
  \Pi_{ij} (p^2) &= \Pi_{ij}^{(1)} + \Pi_{ij}^{(2)} + \dots
\end{align}
where the superscripts denote the order of perturbation theory.

There are three main techniques used to solve this in practice. The
first is to iteratively evaluate \refeq{EQ:DetSelf} by starting
with~\mbox{$p^2 = \m_i^2;$} this does not respect gauge invariance or
perturbation order. The second is to perturbatively expand the
momentum as~\mbox{$p^2 = \m_i^2 + p_1^2 + p_2^2 + \ldots$} and use
matrix perturbation theory to extract the pole mass at each order;
this gives a gauge invariant result which respects the order of
perturbation theory, but can be tricky to implement in the cases of
some masses being degenerate. The third method
(see \citere{Martin:2003it}) is to solve
\begin{gather}
  0 = \mathrm{Det}{\Big[\!\left(p_1^2 - \m^2_i\right) \delta_{ij}
                          - \Pi_{ij}^{(1)}{\left(\m_i^2\right)} \Big]}
\end{gather}
iteratively, then expand the one-loop self-energy, giving
\begin{gather}
  \Pi_{ij}^{(2)}{\left(p^2\right)} \rightarrow
  \Pi_{ij}^{(1)}{\left(\m_i^2\right)}
  + \left(p_1^2 - \m_i^2\right) \Pi_{ij}^{(1)\,\prime}{\left(\m_i^2\right)}
  + \Pi_{ij}^{(2)}{\left(\m_i^2\right)}\,,
\end{gather}
and insert into \refeq{EQ:DetSelf} to solve for~$p_2^2$, and so
on. This method is gauge invariant but does not respect the
perturbation order in~$p^2$, since~$p_1^2$ will contain contributions
from $\big[\Pi_{ij}^{(1)}{\left(\m_i^2\right)}\big]^2$ and higher
powers, etc. In particular, if the aim is to use the
pole-mass-matching procedure to extract threshold corrections, then
the first or third methods will lead to uncancelled higher-order
logarithms.

We therefore recommend the use of the second approach, whereby the
pole mass at the two-loop order is simply
\begin{gather}
  p^2_{{\rm pole}, i} = \m_i^2 + \Pi_{ii}^{(1)}{\left(\m_i^2\right)}
    + \Pi_{ii}^{(2)}{\left(\m_i^2\right)}
    + \Pi_{ii}^{(1)}{\left(\m_i^2\right)}\,\Pi_{ii}^{(1)\,\prime}{\left(\m_i^2\right)}
    + \sum_{j} \frac{1}{\m_i^2 - \m_j^2}\,\Pi_{ij}^{(1)}{\left(\m_i^2\right)}\,
      \Pi_{ji}^{(1)}{\left(\m_i^2\right)}\,,
\end{gather}
and where the one-loop self-energies have already been diagonalised on
the subspace of states that are degenerate at the tree level; the sum
over~$j$ includes all scalar and vector states with the same quantum
numbers.

In general, it is also necessary to include the contributions of
tadpole diagrams. This is because, for each neutral scalar with a
non-trivial expectation value, there is a non-trivial vacuum
minimisation condition, which can be used to eliminate one parameter
from the theory. Commonly, parameters of mass dimension~$2$ are
substituted in this step, such as the $\mu^2\,\lvert H\rvert^2$ term
in the Higgs potential of the~SM. If we write the neutral component of
the Higgs field as $H^0 = \frac{1}{\sqrt{2}}\left(v + h
+ \dots\right)$ and the quartic term as $\lambda\,\lvert H\rvert^4$,
then the necessary condition for the vacuum being a minimum is
\begin{gather}
  0 = \mu^2\,v + \lambda\,v^3 + \frac{\partial \Delta V}{\partial h}  
  \label{EQ:SMtad}
\end{gather}
where $\Delta V$ are the loop corrections to the effective potential;
its derivatives correspond to tadpole diagrams. If we insist that~$v$
is the correct vacuum expectation value to all loop orders, then we
can eliminate~$\mu^2$ wherever it appears in favour of~$\lambda\,v^2$
and tadpoles. Formally, then we can expand
\begin{gather}
  \mu^2 = - \lambda\, v^2
          - \frac{1}{v}\, \frac{\partial \Delta V^{(1)}}{\partial h}
          - \frac{1}{v}\, \frac{\partial \Delta V^{(2)}}{\partial h}
          - \dots
  \label{EQ:MU2EQ}
\end{gather}  
where the superscripts denote loop orders. Since the tree-level Higgs
mass is $m_h^2 = \mu^2 + 3\, \lambda\, v^2$, this means that the
tadpole contributions modify the Higgs mass at higher orders.

Utilising this method for the computation of the mass of any particle
in the theory that depends on $\mu^2$ requires the calculation of
self-energies and tadpole diagrams. In particular, the application to
computing the mass of the Goldstone bosons leads to the~GBC: in Landau
gauge the ``tree level'' Goldstone mass squared becomes of one-loop
order and is of indeterminate sign, which causes infra-red
singularities in the two-loop tadpoles and
self-energies\,\cite{Martin:2014bca,Elias-Miro:2014pca}. The initially
proposed solution was to resum Goldstone diagrams, and this was
performed for the tadpoles of the~MSSM
in \citere{Kumar:2016ltb}. However, in general this is cumbersome to
implement; and general solutions now exist for both, tadpoles and
self-energies, where we can instead use an ``on-shell'' mass for the
Goldstone bosons\,\cite{Braathen:2016cqe}, or just perturbatively
expand the generalisations of \refeq{EQ:MU2EQ} to the loop order that
we are working to in tadpoles and
self-energies\,\cite{Braathen:2017izn}, known as taking ``consistent
tadpole equations''; for example, we would need to take
\begin{gather}
  \Pi_{ij}^{(2)}{\left(p^2\right)} \rightarrow
  \left.\Pi_{ij}^{(2)}{\left(p^2\right)}\right|_{\mu^2 = - \lambda\, v^2}
  - \frac{1}{v} \left[
    \frac{\partial \Pi_{ij}^{(1)}{\left(p^2\right)}}{\partial \mu^2}\,
    \frac{\partial \Delta V^{(1)}}{\partial h}
  \right]_{\mu^2 = - \lambda\, v^2}.
\label{EQ:ConsistantTad}
\end{gather}
In this way, the infra-red singularities cancel between the two parts
on the right-hand side, and this should continue order by order in
perturbation theory.

\begin{figure}[t]
\centering
\includegraphics[width=0.2\textwidth]{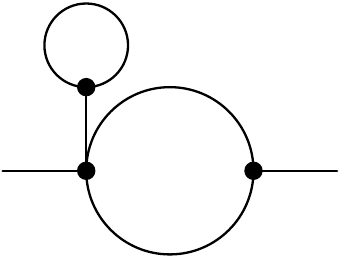}\quad
\includegraphics[width=0.2\textwidth]{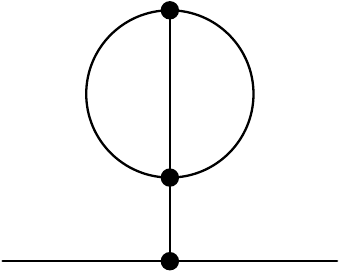} 
\caption{\label{FIG:InternalTops}
Examples of one-particle-irreducible diagrams with ``internal''
propagators that we do not include in our list of topologies. All
diagrams of these classes can be straightforwardly computed from the
one- and two-loop self-energies and tadpoles that we include here.}
\end{figure}

Another way of treating tadpoles is to work only in terms of running
parameters, so our expectation values solve the tree-level
vacuum-minimisation conditions only. That means that we must include
tadpole diagrams \emph{as part of the self energies}: these are
one-particle irreducible but contain propagators carrying no momentum
(referred to as ``internal''). This was the approach used in the
SM~calculation of \citere{Kniehl:2015nwa} and leads to a
gauge-invariant result, without needing to perform expansions of the
form of \refeq{EQ:MU2EQ}, at the expense of a proliferation of
diagrams, such as those depicted in \fig{FIG:InternalTops}.

In this paper, we shall work in the Feynman gauge. Prima facie, one
would think that our results do not suffer from infra-red
singularities, and so we sidestep the~GBC, and could avoid making an
expansion of the form of \refeq{EQ:MU2EQ}: we could (as originally
envisaged in\,\citere{Martin:2003it}) just modify~$\mu^2$ at the tree
level so that \refeq{EQ:SMtad} is satisfied and dispense with any
extra diagrams or shifts of the form of \refeq{EQ:ConsistantTad}. A
reader wishing to implement this program can use the results from the
appendix ($121$~self-energy classes and $25$~tadpoles) or the reduced
set of $89$~self-energy diagrams described
in \sect{SEC:RESULTS}. However, such an approach does not respect the
order of perturbation theory or gauge invariance.\footnote{Moreover,
theories with genuine Goldstone bosons would still suffer
the~GBC. Hence, in our integral reductions, we must still deal with
infra-red singularities. In this work, we do so by means of
dimensional regularisation: all infra-red divergent diagrams acquire
poles in the dimensional regulator~$\epsilon$. This actually provides
yet another solution to the~GBC; we shall elaborate on the connection
more clearly elsewhere.} Hence, we shall simplify the expressions
where all couplings and masses are expressed in terms of tree-level
parameters, so that the reader may use any of the other approaches. In
this way, we are also able to make extensive use of relationships
amongst the couplings in general theories, and so reduce the number of
topologies to~$16$ for tadpoles and to~$58$ for self-energies for
non-Goldstone boson scalars.

To apply the results in practice for all but the simplest of models is
a task for a computer. Even for the Standard Model it would be far too
tedious to implement by hand. Hence, we have provided our results in
electronic form as part of \ourcode. All reduction rules are included
so that the renormalised expressions can be applied for any physically
relevant configurations of masses and momenta, in the form
of \texttt{Mathematica} modules and notebooks, with code to link
from \texttt{Mathematica} to \TSIL. A user manual is provided
online. In the future we also intend to include \texttt{c++} code to
link to \TSIL for use with packages such as \SARAH, the (currently
private) results for Landau and general $R_\xi$~gauges, and extensions
to vector/mixed scalar--vector/fermion self-energies.

\tocsubsection[\label{sec:introGUIDSE}]{Guide to the paper}

The paper is organised in the following way:
\begin{itemize}
\item
  in \sect{SEC:NOTATION} we introduce our nomenclature for the fields
  and couplings that appear in renormalisable theories, and we explain
  our method of computing the two-loop diagrams and counterterms. As
  our results are valid for real fields, we also show how to apply
  them to complex scalars and Dirac fermions.
\item
  The purpose of \sect{SEC:GOLDSTONESGHOSTS} is to reduce the number
  of different diagrams by making use of relations among the couplings
  that are dictated by gauge invariance. In this way, ghosts and
  Goldstone bosons can be eliminated from the theory.
\item
  The nomenclature that is used for the loop integrals is introduced
  in \appx{SEC:REDUCTION}. There, we also describe how each integral
  can be reduced for any kinematic configuration into a basis that can
  be quickly evaluated numerically.
\item
  The full list of results for renormalised two-loop tadpoles and
  self-energies in terms of the previously defined couplings and loop
  integrals is given in \appx{SEC:DIAGLIST}.
\item
  The substitution of ghost and Goldstone couplings is applied to
  these results in \sect{SEC:RESULTS}. In the same section, we also
  compare to previously known expressions.
\item
  Our conclusions are summarised in \sect{SEC:CONCLUSIONS}.
\end{itemize}

\tocsection[\label{SEC:NOTATION}]{Notation and methods}

In this section we shall give our definitions and methods, needed to
understand the results presented in \sect{SEC:RESULTS} and the
appendix.

\tocsubsection[\label{SEC:COUPLINGS}]{Coupling definitions}

In \citere{Martin:2003it}, scalar self-energies were given using
two-component spinors and a compact notation in terms of couplings in
a general lagrangian. Such a lagrangian reads
\begin{align}
\lag &= \lag_S + \lag_{SF} + \lag_{SV} + \lag_{FV}
        + \lag_{\mathrm{gauge}} + \lag_{S \mathrm{ghost}}\,,
\end{align}
where
\begin{subequations}
\begin{align}
  \lag_S &\equiv - \frac{1}{6}\, a_{ijk}\, \Phi_i\, \Phi_j\, \Phi_k
         - \frac{1}{24}\, \lambda_{ijkl}\, \Phi_i\, \Phi_j\, \Phi_k\, \Phi_l\,,\\
  \lag_{SF} &\equiv - \frac{1}{2}\, y^{IJk}\, \psi_I\, \psi_J\, \Phi_k
         - \frac{1}{2}\, y_{IJk}\, \ov{\psi}^I\, \ov{\psi}^J\, \Phi_k\,,\\
  \lag_{FV} &\equiv g^{aJ}_I\, A_\mu^a\, \ov{\psi}^I\, \ov{\sigma}^\mu\, \psi_J\,,\\
  \lag_{SV} &\equiv \frac{1}{2}\, g^{abi}\, A^a_\mu\, A^{\mu b}\, \Phi_i
         + \frac{1}{4}\, g^{abij}\, A^a_\mu\, A^{\mu b}\, \Phi_i\, \Phi_j
         + g^{aij}\, A^a_\mu\, \Phi_i\, \partial^\mu \Phi_j\,,\\
  \lag_{\mathrm{gauge}} &\equiv g^{abc}\, A^a_\mu\, A^b_\nu\, \partial^\mu A^{\nu c}
  - \frac{1}{4}\, g^{abe}\, g^{cde}\, A^{\mu a}\, A^{\nu b}\, A_\mu^c\, A_\nu^d
  + g^{abc}\, A^a_\mu\, \omega^b\, \partial^\mu \ov{\omega}^c\,,\\
  \lag_{S \mathrm{ghost}} &\equiv \xi\, \hat{g}^{abi}\, \Phi_i\, \ov{\omega}^a\, \omega^b\, .
\end{align}
\label{EQS:GENERALLAG}\end{subequations}
The fields $\Phi_i$ with indices $\{i,j,k,l\}$ denote real scalars,
$\psi_I$ with indices $\{I,J,K,L\}$ Weyl fermions, $A_\mu^a$ with
indices $\{a,b,c,d\}$ gauge bosons, and $\omega^a, \ov{\omega}^a$
ghosts and antighosts (which carry gauge-boson
indices). The \refeqs{EQS:GENERALLAG} slightly differ from the ones
in \citere{Martin:2003it} because we work with a different metric
signature~$(+,-,-,-)$ and, since we work away from the gaugeless
limit/Landau gauge, we have couplings between scalars and ghosts.

\tocsubsubsection[\label{sec:nonghostcoups}]{Scalar, vector and fermion couplings}

In this work, due to the large numbers of diagrams and the complexity
of the expressions, we perform the generic calculations using computer
algebra, and we expect that the application of the results will be
best accomplished by implementation on computers. Hence, we use
coupling definitions that are more practical for that case; indeed,
since we use \FeynArts to generate the set of generic diagrams, we
adopt an abbreviated form of the notation of the generic model
file \texttt{Lorentz.gen}. \FeynArts works in four-spinor notation and
distinguishes particles and antiparticles; we remove this distinction
in our results by transforming all fields to real scalars, gauge
bosons and Majorana fermions. Our results are therefore given in terms
of \emph{vertices} which, in general, have more than one possible
Lorentz structure; we denote this with an index as last argument of
each coupling. Our vertices are named by their adjacent particles,
$\cpl{S}{}$ for scalars, $\cpl{F}{}$ for fermions, $\cpl{V}{}$ for
vectors, and $\cpl{U}{}$ for ghosts. The dictionary
with \refeqs{EQS:GENERALLAG} is:
\begin{subequations}
\begin{align}
  \cpl{SSS}{i,j,k,1} &= a_{ijk}\,,\\
  \cpl{SSSS}{i,j,k,l,1} &= \lambda_{ijkl}\,,\\
  \cpl{FFS}{I,J,i,1} &= y^{IJi}\,,&
  \cpl{FFS}{I,J,i,2} &= y_{IJi} = \left(y^{IJi}\right)^* ,\\
  \cpl{FFV}{I,J,a,1} &= -g_{I}^{aJ}\,,&
  \cpl{FFV}{I,J,a,2} &= -\cpl{FFV}{J,I,a,1} = \left(\cpl{FFV}{I,J,a,1}\right)^* ,\\
  \cpl{SVV}{i,a,b,1} &= -g^{abi}\,,\\
  \cpl{SSVV}{i,j,a,b,1} &= -g^{abij}\,.
\end{align}
\end{subequations}
For the terms with fermions, the Lorentz structure with index~$1$
or~$2$ refers to a left- or right-chiral projector, respectively; the
$\cpl{FFV}{}$~couplings contain a gamma matrix in addition.

The remaining couplings require more inspection. The $\cpl{SSV}{}$
vertex is given by
\begin{subequations}
\begin{align}
 \frac{\imath\,\partial^3}{\partial \Phi_i\, \partial \Phi_j\, \partial A^{\mu_a}}
 \left[g^{a'i'j'}\, A^{a',\mu}\, \Phi_{i'} \left(\imath\, p_{j'}^\mu\right) \Phi_{j'}  \right] &=
 -g^{aij}\, p_j^{\mu_a} - g^{aji}\, p_i^{\mu_a}
 = g^{aij} \left(p_i - p_j\right)^{\mu_a} \nn\\
 &= \cpl{SSV}{i,j,a,1}\, p_i^{\mu_a} + \cpl{SSV}{i,j,a,2}\, p_j^{\mu_a}\,,\\
 \rightarrow \cpl{SSV}{i,j,a,1} &= \imath\, g^{aij}\,, \qquad
 \cpl{SSV}{i,j,a,2} = -\cpl{SSV}{i,j,a,1}\,.
\end{align}
\end{subequations}
We note that we do not enforce the equality of the two terms for
the \emph{counterterm} vertex.

Next, for the pure gauge-coupling terms, we have ($\eta^{\mu\nu}$ is
the Minkowski metric)
\begin{subequations}
\begin{align}
 \begin{split}
 \frac{\imath\,\partial^3}{\partial A^{\mu_a}\, \partial A^{\mu_b}\, \partial A^{\mu_c}}
 \left[g^{abc}\, A_\mu^{a'}\, A_\nu^{b'}\, \left(\imath\, p_c^\mu\right) A^{\nu c'}\right] &= -g^{abc} \begin{aligned}[t] \,\Big[
   & \eta^{\mu_a\mu_b} \left(p_b^{\mu_c} - p_a^{\mu_c}\right)
   + \eta^{\mu_b\mu_c} \left(p_c^{\mu_a} - p_b^{\mu_a}\right)\\
   &{+}\, \eta^{\mu_c\mu_a} \left(p_a^{\mu_b} - p_c^{\mu_b}\right)
   \!\Big]\,,
   \end{aligned}
 \end{split}\\
 \rightarrow \cpl{VVV}{a,b,c,1} &= -\imath\, g^{abc}\,.
\end{align}
\end{subequations}

For the four-vector coupling, we have the generic vertex
\begin{subequations}
\begin{align}
 \frac{\imath\,\partial^4}{\partial A^{\mu_a}\, \partial A^{\mu_b}\, \partial A^{\mu_c}\, \partial A^{\mu_d}}\, \mathcal{L} &=
 -2\,\imath \left[
   g^{abe}\, g^{cde}\, \eta^{\mu_a \mu_b}\, \eta^{\mu_c \mu_d}
   + g^{ace}\, g^{bde}\, \eta^{\mu_a \mu_c}\, \eta^{\mu_b \mu_d}
   + g^{ade}\, g^{cbe}\, \eta^{\mu_a \mu_d}\, \eta^{\mu_c \mu_d}
   \right] \nn\\
 \begin{split}
 &= -\imath \begin{aligned}[t] \,\Big[
 & \cpl{VVVV}{a,b,c,d,1}\, \eta^{\mu_a \mu_b}\, \eta^{\mu_c \mu_d}
 + \cpl{VVVV}{a,b,c,d,2}\, \eta^{\mu_a \mu_c}\, \eta^{\mu_b \mu_d}\\
 &{+}\, \cpl{VVVV}{a,b,c,d,3}\, \eta^{\mu_a \mu_d}\, \eta^{\mu_c \mu_b}
 \Big]\,,
 \end{aligned}
 \end{split}\\
\intertext{and thus we have the identifications}
 \cpl{VVVV}{a,b,c,d,1} &= -2\, \cpl{VVV}{a,c,e,1}\,\cpl{VVV}{b,d,e,1}\,,\\
 \cpl{VVVV}{a,b,c,d,2} &= -2\, \cpl{VVV}{a,b,e,1}\,\cpl{VVV}{c,d,e,1}\,,\\
 \cpl{VVVV}{a,b,c,d,3} &= 2\, \cpl{VVV}{a,d,e,1}\,\cpl{VVV}{b,c,e,1}
\label{EQ:RemoveVVVV}\end{align}
\end{subequations}
with a sum on $e$.

\tocsubsubsection[\label{sec:ghostcoups}]{Ghost couplings and ghost flow}

Finally, for the ghost terms, in the \FeynArts generic model, no
particular form of the couplings is enforced: the general vertex for
the $\cpl{UUV}{}$~coupling is equal to
\begin{gather}
 -\imath\, \cpl{UUV}{a,b,c,1}\, p^{\mu_c}_a
 -\imath\, \cpl{UUV}{a,b,c,2}\, p^{\mu_c}_b\,,
\end{gather}
whereas from \refeqs{EQS:GENERALLAG} we see that in general
$R_\xi$~gauge one of these terms is always vanishing because the
vertex only contains a ghost and antighost (not ghost--ghost or
antighost--antighost) and moreover only contains a factor of the
momentum of the antighost. Hence we must preserve the distinction
between ghost and antighost in our amplitude, and we should also
include both directions of ghost flows. Then we have
\begin{subequations}
\begin{alignat}{2}
 \cpl{UUV}{-a, b, c, 1} &= \cpl{VVV}{a, b, c, 1}\,, &\qquad
 \cpl{UUV}{-a, b, c, 2} &= 0\,, \\
 \cpl{UUV}{a, -b, c, 1} &= 0\,, &\qquad
 \cpl{UUV}{a, -b, c, 2} &= -\cpl{VVV}{a, b, c, 1}\,,\\
 \cpl{SUU}{i,a, -b, 1}  &= -\xi\, \hat{g}^{bai}\,, &\qquad
 \cpl{SUU}{i,-a, b, 1}  &= -\xi\, \hat{g}^{abi}\,.
\end{alignat}
\end{subequations}
The signs here are confusing, because we expect that the ghosts are
anticommuting, and so should be the vertices. This is an artefact of
the algorithm used to construct the amplitudes, where the ordering of
the indices in the couplings is not meant to be taken literally: it is
assumed that for a given choice of particles in a vertex, either there
is only one way of combining them into a vertex, and this is the
ordering that is implied (\EG for a scalar--ghost--antighost coupling
there is only one correct choice), or the ordering does not matter.

However, there are two good reasons that the reader does not need to
worry about this issue: the first is that they can just take the above
prescriptions and plug them into our results, given in the
appendix. \emph{Importantly}, for the ghost amplitudes they should sum
over both signs of each ghost index, \EG for diagram~\lbls{025} we
have the result
\begin{subequations}
\begin{align}
\cpl{SSS}{\ind{1}, \ind{3}, \ind{6}, 1}\,
    \cpl{SUU}{\ind{2}, -\ind{7}, \ind{4}, 1}\,
    \cpl{SUU}{\ind{3}, -\ind{4}, \ind{5}, 1}\,
    \cpl{SUU}{\ind{6}, -\ind{5}, \ind{7}, 1} &\times \mathrm{loop\ function}
\intertext{that should be interpreted as}
\left(a^{\ind{1}, \ind{3}, \ind{6}}\, \hat{g}^{\ind{2}, \ind{4}, \ind{7}}\, \hat{g}^{\ind{3}, \ind{5}, \ind{4}}\, \hat{g}^{\ind{6}, \ind{7}, \ind{5}}
+ a^{\ind{1}, \ind{3}, \ind{6}}\, \hat{g}^{\ind{2}, \ind{7}, \ind{4}}\, \hat{g}^{\ind{3}, \ind{4}, \ind{5}}\, \hat{g}^{\ind{6}, \ind{5}, \ind{7}}\right) &\times \mathrm{loop\ function}.
\end{align}
\end{subequations}
However, the second reason to not worry about this is that
in \sect{SEC:GOLDSTONESGHOSTS} we demonstrate how all of the ghost
couplings can be removed from the amplitude, giving a much smaller
number of classes of diagrams to evaluate.

\Needspace{8\baselineskip}
\tocsubsection[\label{SEC:DIAGRAMS}]{Processing of diagrams}

Here we describe our approach to generating and renormalising the
diagrams.

\begin{figure}[bp!]
  \centering
  \begin{tabular}{*9{@{}>{\centering}p{.111\floatwidth}}@{}}
  &&&\includegraphics[width=.075\floatwidth]{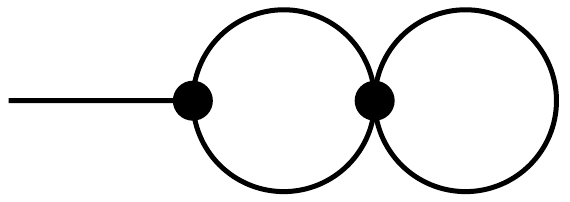}
  \hbox{\hspace{.018\floatwidth}}&
  \includegraphics[width=.075\floatwidth]{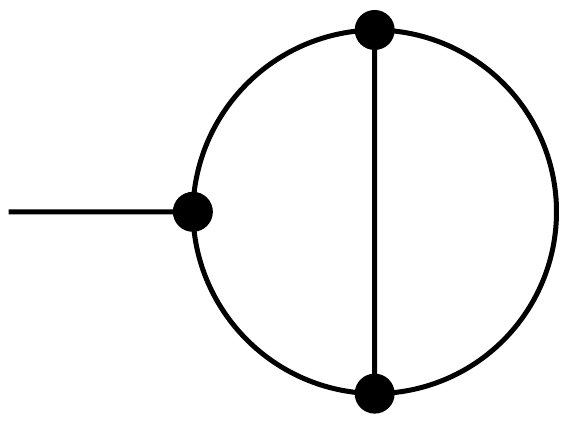}
  \hbox{\hspace{.018\floatwidth}}&
  \includegraphics[width=.075\floatwidth]{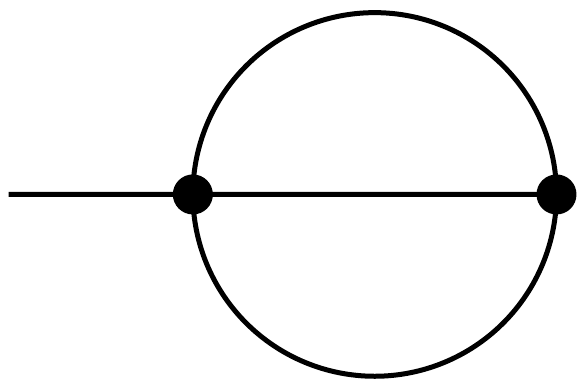}
  \hbox{\hspace{.018\floatwidth}}\tabularnewline
  &&&Top.\,$1$&Top.\,$2$&Top.\,$3$\tabularnewline\hline\\[-2ex]
  \includegraphics[width=.1\floatwidth]{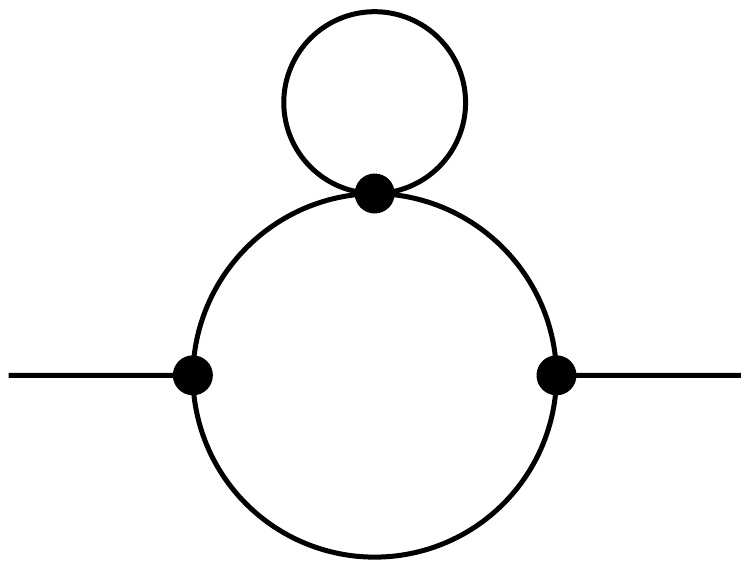}&
  \includegraphics[width=.1\floatwidth]{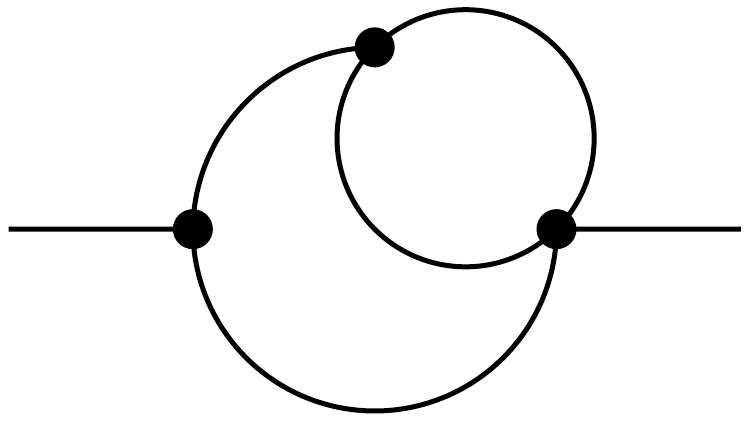}&
  \includegraphics[width=.1\floatwidth]{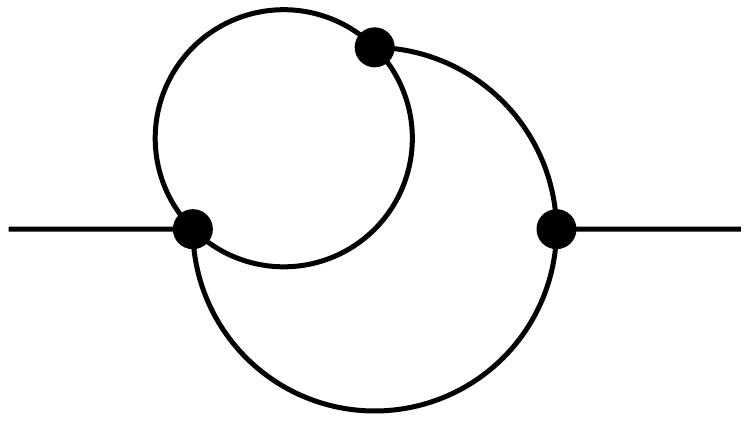}&
  \includegraphics[width=.1\floatwidth]{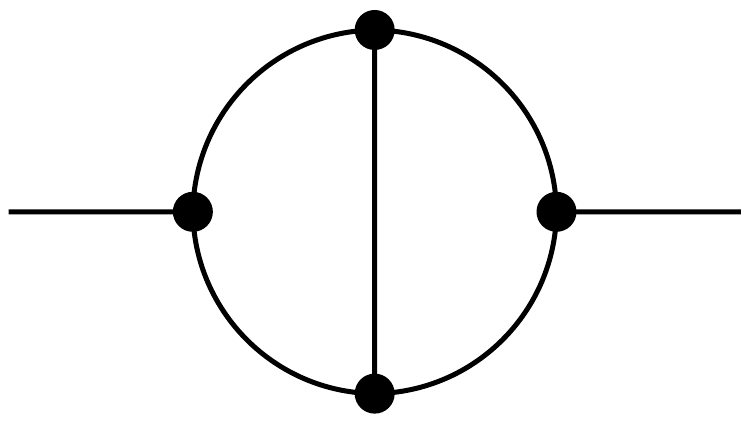}&
  \includegraphics[width=.1\floatwidth]{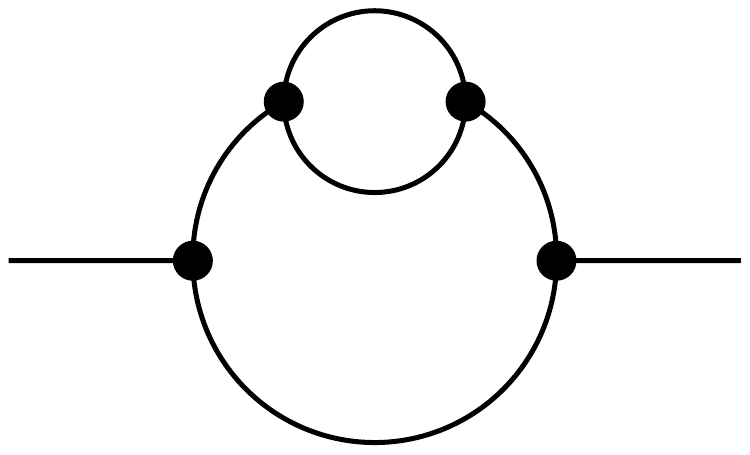}&
  \includegraphics[width=.1\floatwidth]{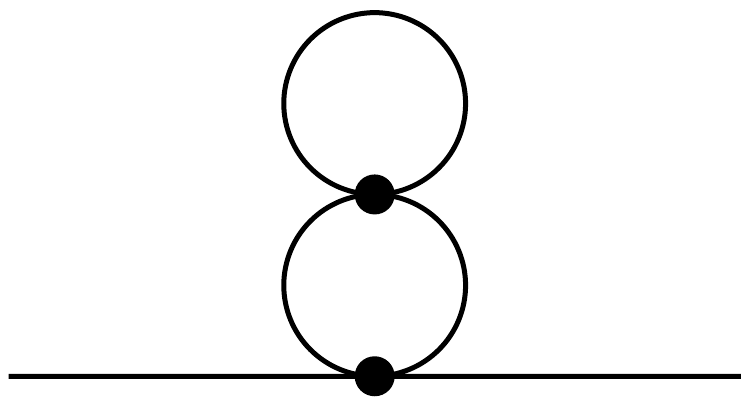}&
  \includegraphics[width=.1\floatwidth]{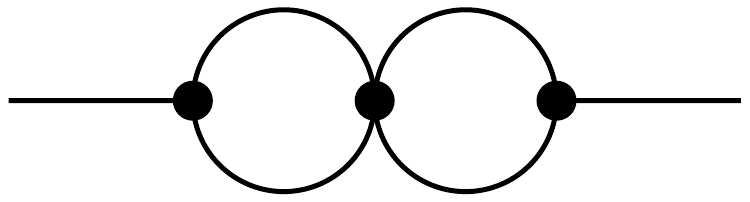}&
  \includegraphics[width=.1\floatwidth]{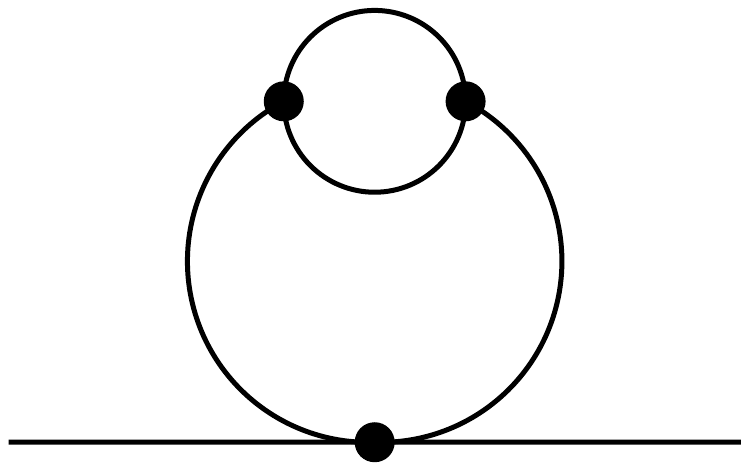}&
  \includegraphics[width=.1\floatwidth]{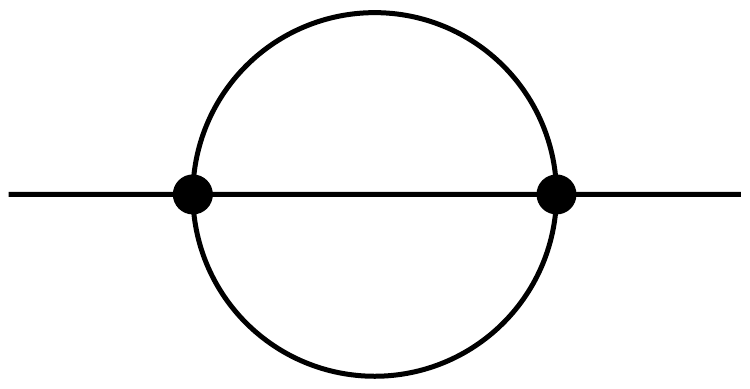}\tabularnewline
  Top.\,$1$&Top.\,$2$&Top.\,$3$&Top.\,$4$&Top.\,$5$
  &Top.\,$6$&Top.\,$7$&Top.\,$8$&Top.\,$9$
  \end{tabular}
  \caption{\label{FIG:TOPS}
  All possible topologies of one-particle irreducible Feynman diagrams
  for two-loop tadpoles (top row) and self-energies (bottom row) are
  shown.}
  \vspace{-2ex}
  \caption*{In \citere{Martin:2003it} the self-energy topologies are
  referred to as $Y$, $U$, $U$, $M$, $V$, $X$, $Z$, $W$, $S$
  respectively (topologies $2$ and $3$ are equivalent for identical
  incoming/outgoing states).}
\end{figure}

\paragraph{Feynman-diagrammatic approach:}
The two-loop Feynman diagrams are generated with the help
of \FeynArts\,\cite{Kublbeck:1990xc,Hahn:2000kx,FA-www}. The different
one-particle irreducible topologies of tadpoles and self-energies are
depicted in \fig{FIG:TOPS}. Each of these topologies is populated with
all possible combinations of fermions~$\cpl{F}{}$ (straight lines),
scalars~$\cpl{S}{}$ (dashed lines), vector~bosons~$\cpl{V}{}$ (wavy
lines), and ghosts~$\cpl{U}{}$ (dotted lines) in renormalisable
theories. The external legs are fixed to be scalars. Each particle in
the diagram is assigned a unique index
$\in \{\ind{1},\ldots,\ind{7}\}$ for full generality.

\begin{figure}[tp!]
  \centering
  \begin{tabular}{*3{@{}>{\centering}p{.3\floatwidth}}@{}}
    selected sub-loop &
    \minitab[c]{one-loop diagram with\\ counterterm vertex} &
    \minitab[c]{appropriate\\ counterterm insertion}
    \tabularnewline\hline\\[-2ex]
    \includegraphics[width=.15\floatwidth]{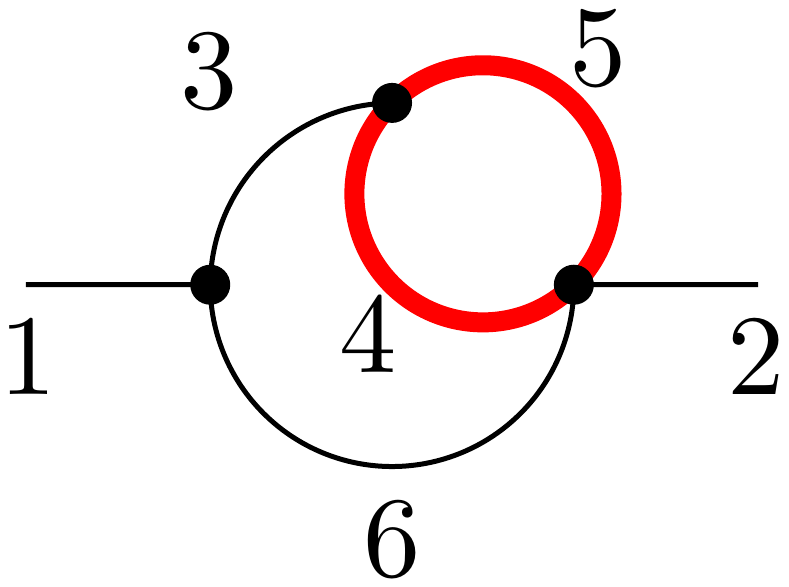} &
    \includegraphics[width=.15\floatwidth]{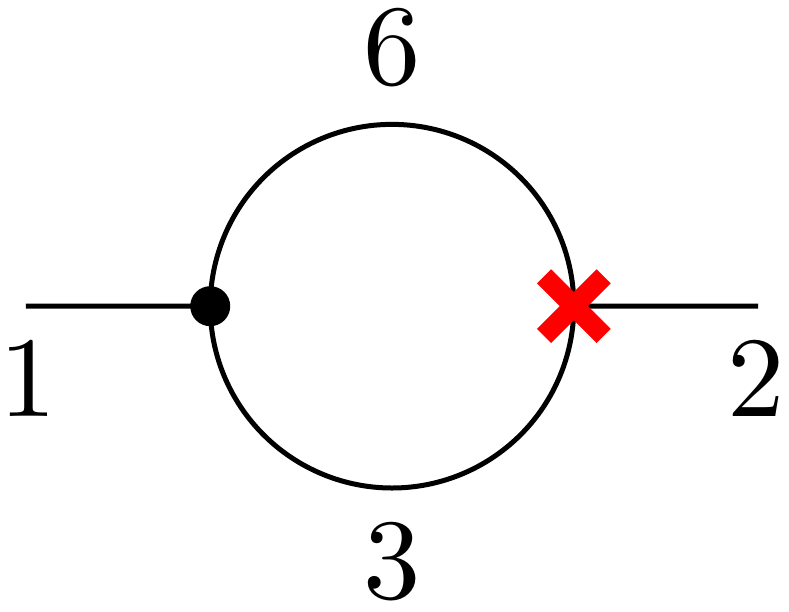} &
    \includegraphics[width=.15\floatwidth]{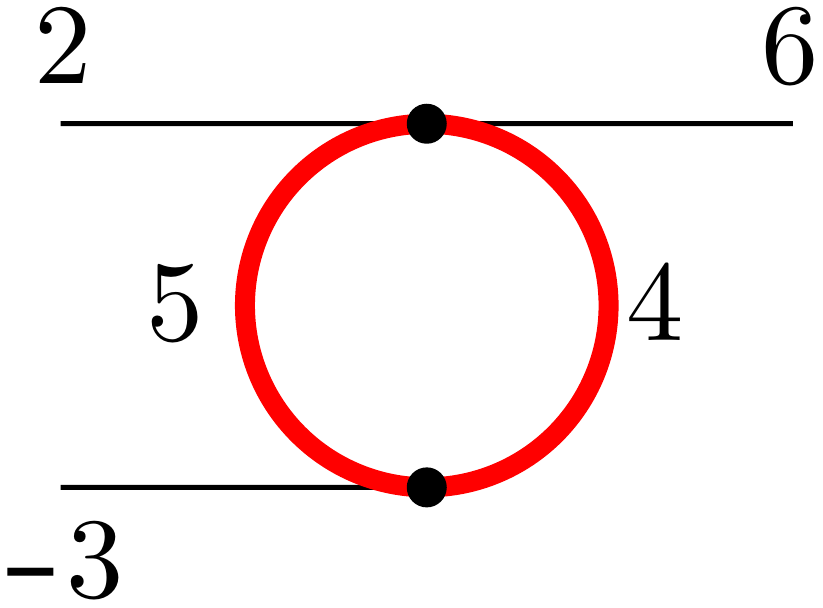}
    \tabularnewline\hline\\[-2ex]
    \includegraphics[width=.15\floatwidth]{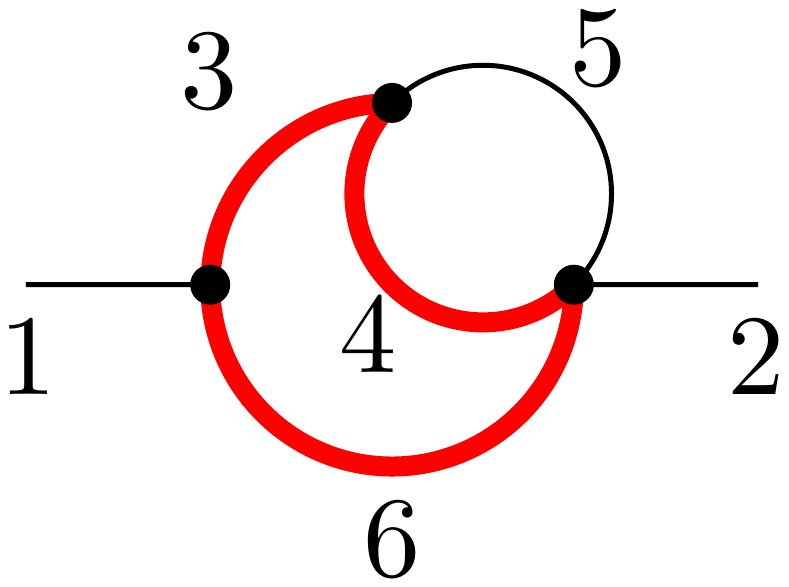} &
    \includegraphics[width=.15\floatwidth]{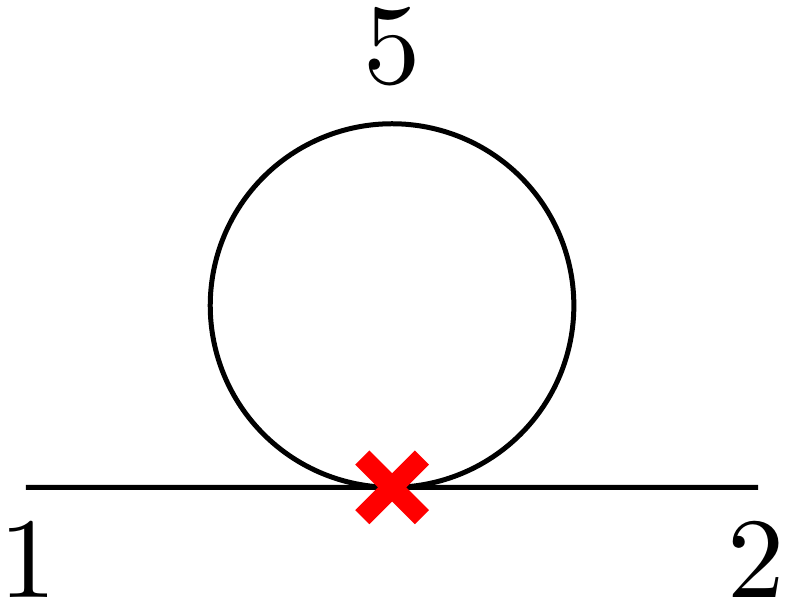} &
    \includegraphics[width=.15\floatwidth]{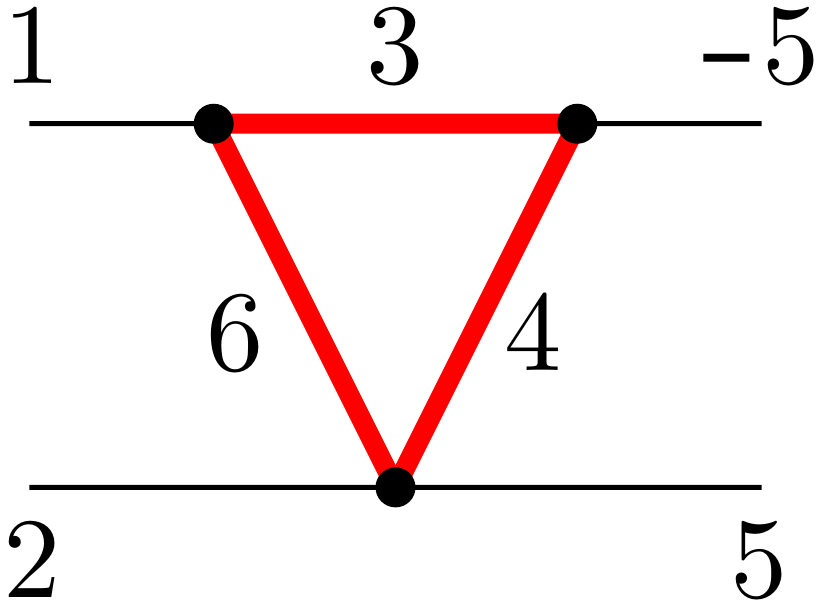}
    \tabularnewline\hline\\[-2ex]
    \includegraphics[width=.15\floatwidth]{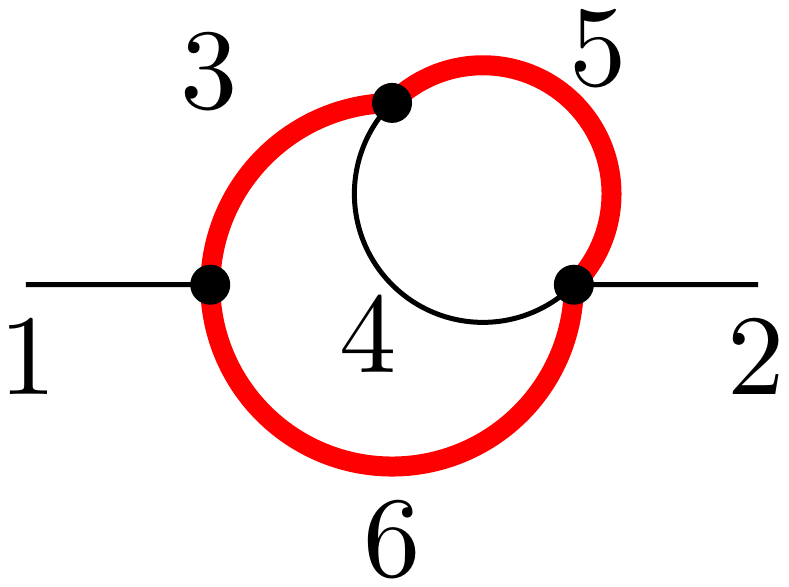} &
    \includegraphics[width=.15\floatwidth]{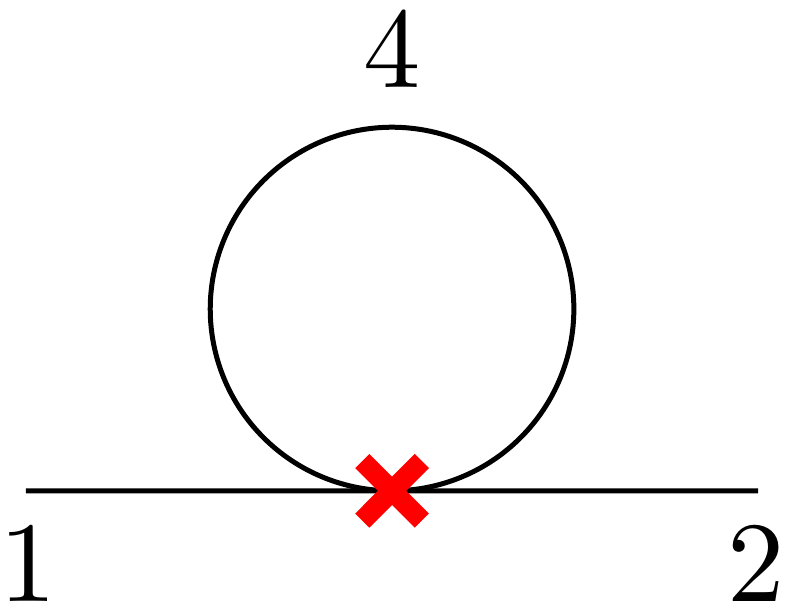} &
    \includegraphics[width=.15\floatwidth]{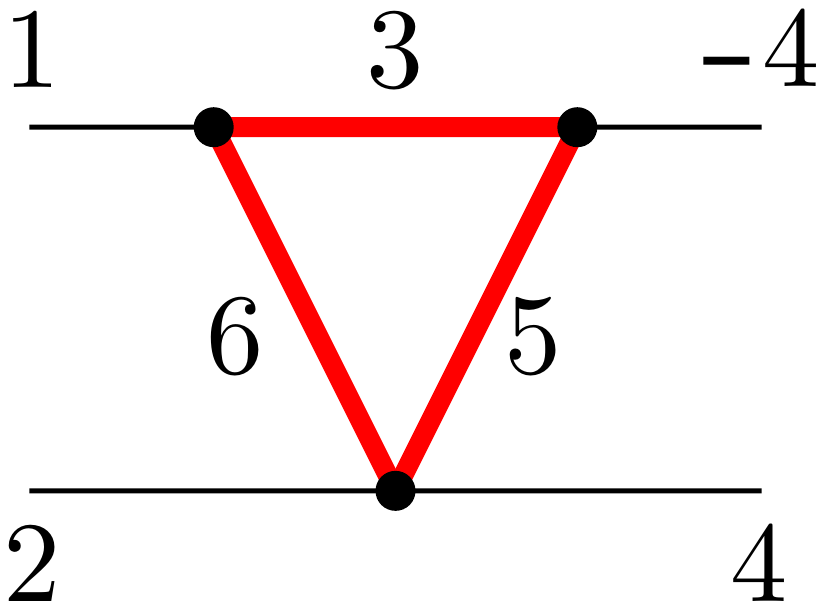}
  \end{tabular}
  \caption{\label{FIG:FOREST}
  The different sub-loops of topology~$2$ of the self-energies
  in \fig{FIG:TOPS} are marked in red at the left column. After
  shrinking the lines of these sub-loops, the remaining one-loop
  diagrams with counterterm vertex are displayed at the middle
  column. The appropriate counterterm insertion for each counterterm
  vertex is given at the right column. The consecutive numbers at the
  lines label the propagators; the same number in the diagrams of one
  row refers to the same particle at that propagator. A signed number
  indicates the antiparticle to the unsigned one (note that external
  particles are defined as incoming by default).}
  \vspace{-1.2ex}
\end{figure}

\paragraph{BPHZ method:}
In general, the integrals of the two-loop diagrams in \fig{FIG:TOPS}
are ultra-violet~(UV)~divergent. In order to regularise them, we
follow the BPHZ
prescription\,\cite{Bogoliubov:1957gp,Hepp:1966eg,Zimmermann:1969jj}.
Each two-loop diagram contains UV-divergent sub-loops of one-loop
order that can be regularised by appropriate counterterms. In general,
an additional two-loop counterterm is necessary in order to regularise
all UV~poles. While the latter can be defined as a pure polynomial in
the UV~regulator~$1/\epsilon$, the former regularise all non-local
divergences and in general give rise to UV-finite terms as well.

The realisation of the splitting of two-loop diagrams into the
so-called forest of sub-loop diagrams is carried out in an automated
way. For each sub-loop diagram, the corresponding one-loop diagram
with counterterm insertion is generated. In addition, the appropriate
counterterm topology that regularises the logarithmic divergence of
this sub-loop is determined automatically. The algorithm is based on a
graph-theoretical interpretation of Feynman diagrams: first, the
closed cycles (loops) of each diagram are identified; second, for each
cycle, the lines (internal propagators) that make up the loop are
shrunken into a point (counterterm vertex); third, the adjacencies to
the shrunken cycle in the original diagram (and the cycle itself) are
used in order to determine the required counterterm insertion. An
illustrating example of this procedure is given in \fig{FIG:FOREST}.

\paragraph{Symmetry factors:}
Analytical results in terms of amplitudes for the genuine, generic
two-loop diagrams as well as the corresponding one-loop diagrams with
counterterm vertex, and the one-loop counterterm insertions are
generated with the help
of \FeynArts, \FormCalc{}\,\cite{Hahn:1998yk,FC-www}
and \TwoCalc{}\,\cite{Weiglein:1993hd} (we also make use of \OneCalc
that is part of \TwoCalc). Note that among the~$121$ different
self-energy and~$25$ different tadpole diagrams that \FeynArts
generates, those diagrams with symmetries of internal propagators have
already been removed, whereas diagrams with symmetries with respect to
the external propagators are all kept. This processing fixes the
symmetry factors of the genuine two-loop diagrams. The symmetry
factors of the counterterm diagrams are modified accordingly in order
to match the regularised two-loop diagram.

\paragraph{Couplings:}
All occurring couplings are re-labelled by the sequence of acronyms for
all adjacent fields and carry the corresponding propagator indices as
argument. An additional numeric argument~$\ell$ at the last position
allows one to distinguish the couplings of the same fields with
different Lorentz structure, as described in \sect{SEC:COUPLINGS}
where we give the dictionary to parameters in the Lagrangian.
In \FeynArts the propagator indices may be signed in order to refer to
antiparticles.\footnote{Note that at this stage the
fields~$\cpl{F}{}, \cpl{S}{}, \cpl{V}{}, \cpl{U}{}$ are not yet
specified and may themselves be physical antiparticles. Therefore, a
signed index of an antiparticle refers to the particle.} The different
Lorentz structures of the couplings are summarised in appendix~A of
the \FeynArts manual\,\cite{FA-www}; the argument~$\ell$ refers to
the~$\ell$-th entry of the vector of couplings. The only deviation
from the default setup of the Lorentz structure applies to the
coupling~$\cpl{SSV}{\ind{a},\ind{b},\ind{c}}$ that depends on the
momenta~$k_{\ind{a}}$ and~$k_{\ind{b}}$ of the scalar fields: at the
level of the counterterm vertices not all instances are proportional
to~\mbox{$\left(k_{\ind{a}}-k_{\ind{b}}\right)$}; instead, the
dependence on~$k_{\ind{a}}$ and~$k_{\ind{b}}$ needs to be
distinguished. For this purpose, the default generic model file (and
the model file) of \FeynArts is initialised with the modifications
given in \fig{FIG:LORENTZ}; for the couplings (rather than
counterterms) we keep the original
vertex~$\cpl{SSV}{\ind{a},\ind{b},\ind{c},1}$.\footnote{The generic
model file~\texttt{Lorentz.gen} is utilised together with the model
file~\texttt{SM.mod}. The latter is required in order to load all
structures for \FeynArts. It already contains all possible
renormalisable generic couplings.}

\begin{figure}[t!]
  {
  \list{}{\leftmargin=0.5cm\rightmargin=0.5cm}\item\relax\small
\begin{verbatim}
SetOptions[InitializeModel,
 GenericModelEdit :> (M$GenericCouplings = M$GenericCouplings /. 
  AnalyticalCoupling[s1 S[j1, mom1], s2 S[j2, mom2], s3 V[j3, mom3, {li3}]] == _ :> 
  AnalyticalCoupling[s1 S[j1, mom1], s2 S[j2, mom2], s3 V[j3, mom3, {li3}]] ==
   G[-1][s1 S[j1], s2 S[j2], s3 V[j3]].
   {FourVector[mom1, li3], FourVector[mom2, li3]}
 ),
 ModelEdit :> (M$CouplingMatrices = M$CouplingMatrices /.
  (c : C[s1_. S[j1_], s2_. S[j2_], s3_. V[j3_]]) == {exp_} :> c == {exp, -exp}
 )
];
InitializeModel["SM"];
\end{verbatim}
  \endlist
  }
  \caption{\label{FIG:LORENTZ}
  The default generic model file of \FeynArts is modified in order to
  allow for different Lorentz structures of the
  coupling~$\protect\cpl{SSV}{}$.}
  \vspace{4ex}
\end{figure}

\begin{figure}[t!]
  \centering
  \begin{tabular}{@{}c@{$\;\longleftrightarrow\;$}c@{}}
  \hspace{.5cm}
  \begin{minipage}[c]{.3\floatwidth}
    \includegraphics[width=\textwidth]{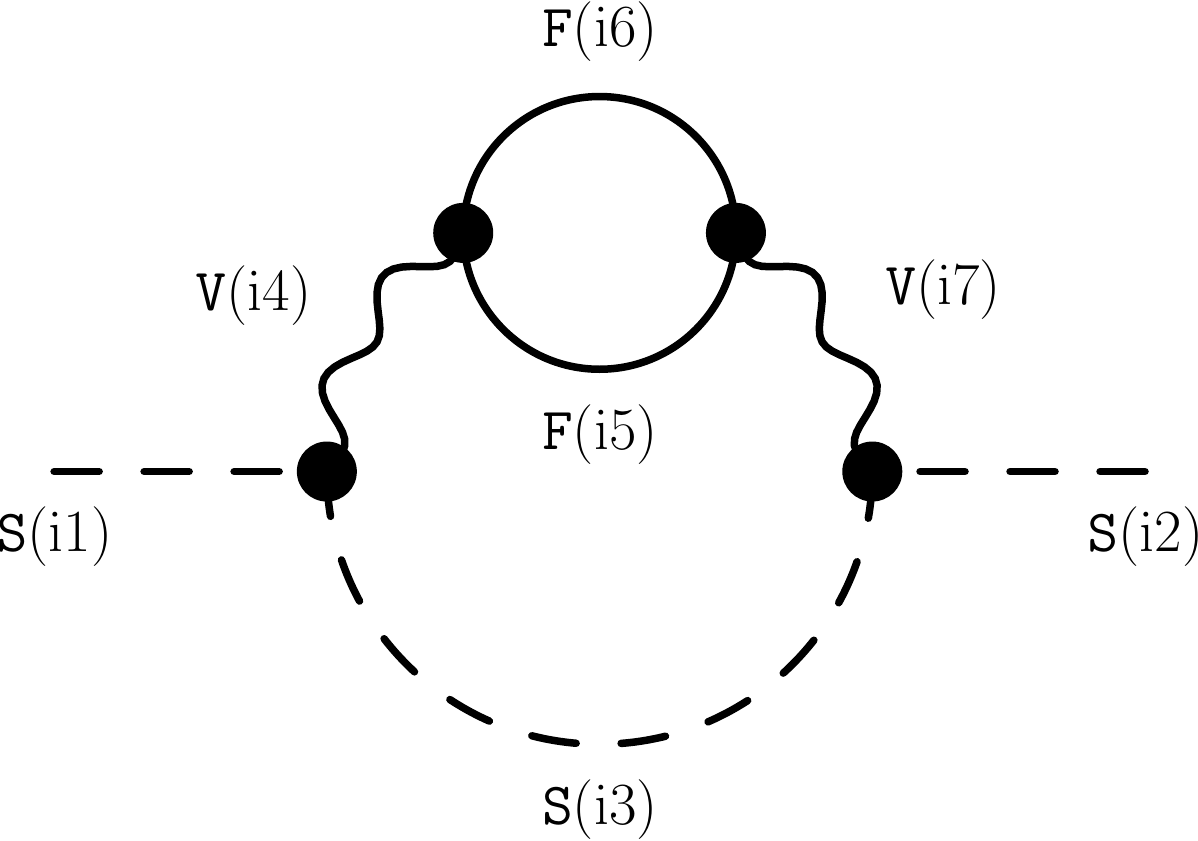}
  \end{minipage} &
\begin{minipage}[c]{.7\floatwidth}
\small
\begin{verbatim}
{edge[iv[1], v[2], S[i1]], edge[iv[3], v[4], S[i2]],
 edge[v[2], v[4], S[i3]],
 edge[v[2], v[5], V[-i4]], edge[v[4], v[6], V[-i7]],
 edge[v[5], v[6], F[-i5]], edge[v[5], v[6], F[-i6]]}
\end{verbatim}
\end{minipage}
  \end{tabular}
  \caption{\label{FIG:EDGES}
  Each Feynman diagram can be uniquely identified by the list of its
  edges when interpreted as a graph. Each~\texttt{edge} has three
  arguments: a starting vertex, an ending vertex, and a label
  (describing field type and propagator index). The
  vertices~\texttt{v[}$i$\texttt{]} are numbered
  consecutively,~$i\in\mathbb{N}$. The starting vertices of external
  fields are not connected to the other edges and
  labelled~\texttt{iv[}$i$\texttt{]}. If the starting and ending
  vertices of an edge are interchanged compared to their original
  description in \FeynArts, the propagator index receives a sign,
  indicating an antiparticle on that line.}
\end{figure}

\paragraph{Counterterm vertices:}
Since the same counterterm vertex can emerge from shrinking sub-loops
of different two-loop diagrams, it is mandatory to store information
about the two-loop topology along with the counterterm vertex in order
to determine the appropriate insertion. Our choice is to stay close to
the existing description in \FeynArts, and to assign a canonically
ordered list of edges to each Feynman diagram. The edges correspond to
the propagators (including the field type) of the diagram. The
propagator indices are stored as well since they are required for
correctly labelling the counterterm insertions; for the purpose of
sorting or identifying diagrams they are not considered. Due to the
canonical ordering of the edges, the direction of a propagator might
be reversed. In that case, the propagator index of that line receives
a sign, indicating an antiparticle. An example of this correspondence
is shown in \fig{FIG:EDGES}.

\paragraph{Counterterm insertions:}
All one-loop one-point, two-point, three-point and four-point
processes were computed at the generic level using \FormCalc, since
they can in principle appear as counterterm insertions. Each of the
Feynman diagrams that appears in these processes was evaluated
separately and stored together with its list of edges. In this way,
individual results can be looked up quickly and inserted into the
correct counterterm vertex; they may in future be provided as part
of \ourcode. In fact, to evaluate our results we use a separate
algorithm to calculate just the divergent parts of the counterterms
(which are identical in \MS or \DR at the one-loop order) on the fly,
and which works in any gauge.

\paragraph{Gauge fixing and regularisation:}
All two-loop self-energy and tadpole diagrams, the corresponding
one-loop diagrams with counterterm vertex, and the UV~divergences of
the counterterm insertions have been determined in
't\,Hooft--Feynman~gauge, Landau~gauge, and the
general~$R_\xi$~gauge. However, diagrams with a large number of
gauge~bosons can initially be expressed in a much shorter form in
't\,Hooft--Feynman~gauge. For this reason, we performed the complete
integral reduction only for the Feynman gauge
(see \appx{SEC:REDUCTION}). In addition, we computed our results using
dimensional regularisation (for \MS~renormalisation), and dimensional
reduction (for \DR~renormalisation). The extra terms in dimensional
regularisation have a coefficient~$\deltaMS$.

\paragraph{Results:}
The outcome for the combination of each genuine two-loop integral with
the corresponding counterterms for all sub-loops is given
in \appx{SEC:TADLIST} for the tadpoles, and in \appx{SEC:SELIST} for
the self-energies; first, all previously known expressions
of \citeres{Martin:2003it,Goodsell:2015ira} are given in our
nomenclature, and then all new results are stated. Note that the pure
UV~divergent two-loop counterterms are not contained in these results,
as they simply add clutter to the expressions. After carrying out the
integral reduction and extracting all UV~divergences via the relations
in \appx{SEC:REDUCTION}, polynomials in the regulator~$1/\epsilon$
will remain, which simply correspond to the genuine two-loop
counterterm of the diagram; but we may also find additional divergent
and finite parts \emph{corresponding to any infra-red divergences}: we
do not introduce infra-red counterterms. As described
in \sect{sec:introGBC} any infra-red divergences should cancel in a
full amplitude when combined with one-loop self-energy and tadpole
diagrams.

\paragraph{Basis of loop integrals:}
We have provided all of the integral reduction rules
in \appx{SEC:REDUCTION} to extract the finite part of our generic
renormalised amplitudes for \emph{any} kinematic configuration in
terms of the basis of one- and two-loop scalar functions that can be
evaluated in \TSIL. This is a basis of six two-loop scalar integrals
and two one-loop ones. The \TSIL basis are actually ``renormalised''
integrals, with specific subtractions; essentially they can be built
up from \citere{Martin:2003qz}:
\begin{subequations}
\begin{align}
  \begin{split}
  S(x,y,z) &= \lim_{\epsilon\rightarrow 0} \begin{aligned}[t] \bigg\{
    & \textbf{S}(x,y,z)
       - \frac{1}{\epsilon} \left[\textbf{A}(x) + \textbf{A}(y) + \textbf{A}(z)\right]
       - \frac{1}{2\,\epsilon^2} \left[x + y + z\right]\\
    &{-}\, \frac{1}{2\,\epsilon} \left[\frac{p^2}{2} - x - y - z\right]
       \!\bigg\}\,,
    \end{aligned}
  \end{split}\\ 
  U(x,y,z,u) &= \lim_{\epsilon\rightarrow 0} \bigg\{
    \textbf{U}(x,y,z,u)
    - \frac{1}{\epsilon}\, \textbf{B}(x,y)
    + \frac{1}{2\,\epsilon^2}
    - \frac{1}{2\,\epsilon}
    \bigg\}\,,\\
  M(x,y,z,u,v) &= \lim_{\epsilon\rightarrow 0}\,\textbf{M}(x,y,z,u,v)\,,
\end{align}
\end{subequations}
where the definitions of the boldtype integrals are given
in \appx{SEC:INTREL}, and the number of spacetime dimensions is
$d=4-2\,\epsilon$. In our results in the appendix we use the
equivalent of the boldtype integrals, and so when taking the finite
part of the diagram we must use
\begin{subequations}
\begin{align}
  \fin{\bS(x,y,z)} &= S(x,y,z) - x\, \binteps{\epsilon^1}{0,0,x}
    - y\, \binteps{\epsilon^1}{0,0,y} - z\, \binteps{\epsilon^1}{0,0,z}\,,\\
  \fin{\bU(x,y,z,u)} &= U(x,y,z,u) + \binteps{\epsilon^1}{p^2,x,y}\,,
\end{align}
\end{subequations}
where $\fin{\ldots}$ denotes the finite part as $\epsilon \rightarrow
0$, \IE\ neglecting non-zero powers of $\epsilon$, and
\begin{align}
  \binteps{\epsilon^1}{p^2,x,y} &\equiv
    \fin{\frac{1}{\epsilon}\, \mathbf{B}(x,y)}.
\end{align}
In principle, since $\mathbf{B}(x,y)$ is known analytically to all
orders in $\epsilon$, it is not difficult to evaluate
$\binteps{\epsilon}{}$ (and it is even available in \TSIL). However,
its presence is a sign of an infra-red divergence: \emph{in the
absence of infra-red divergences, all $\binteps{\epsilon}{}$ integrals
cancel in the renormalised amplitude}. We have explicitly checked that
this is the case for all the results in the appendix. So then the
reader might be curious as to why we do not give the results in terms
of the renormalised \TSIL basis, as done in \citere{Martin:2003it};
the reasons are twofold:
\begin{itemize}
\item
Diagrams with vector bosons contain many more tensor integrals. By
listing them in unreduced form we are able to give our results for
each diagram in only a few lines, whereas some diagrams could fill
pages by themselves in expanded form, even just for the most general
case where there are no vanishing or degenerate masses.
\item
The integral reduction depends crucially on whether there are
coincident or vanishing masses. The simplest cases concern scalar
integrals with repeated propagators, where for non-identical masses we
can use partial fractions to reduce them (see \refeq{EQ:PARTFRAC}).
In \citere{Martin:2003it} there were generally one or two such cases
per integral. In the implementation of the results
of \citeres{Braathen:2016cqe,Braathen:2017izn} in \SARAH, one scalar
integral contains $12$ different reductions. However, in our results,
we also have cases in the reduction where some vanishing gauge-boson
or ghost masses naively lead to poles; furthermore, there are also
special cases with the external momentum taken equal to the scalar
mass that are relevant (in particular) for charged scalar
propagators. Hence our results (for example for \lbls{086}) have up to
$47$ different special kinematic configurations! In \ourcode we
provide the different special cases for each diagram, and our
reduction rules transform them to the appropriate renormalised
integral; clearly it would be impractical (and not especially useful)
to list all of the final results here.
\end{itemize}

\tocsubsection[]{Charged scalars, Dirac fermions and \texorpdfstring{$\gamma_5$}{\unicodegamma\unicodesubfive}}

Our results are calculated using Majorana four-spinors, and
in \sect{SEC:COUPLINGS} we gave a prescription of how to translate
them to Weyl-spinor notation---but under the condition that the
spinors are also Majorana, \IE\ have diagonal (and real)
masses. In \citere{Martin:2003it} the results were presented in
Weyl-spinor notation without this condition, by allowing Dirac spinors
to have off-diagonal masses~$M_{IJ}$ where~$M_{IK}\,M_{KJ} =
m_{I}^2\,\delta_{IJ}$; such mass terms simply link the left and right
Weyl spinors that form a Dirac spinor. To translate our results to
that notation is surprisingly easy, because we pull out linear factors
in the mass for each diagram including fermions. For example, we can
write for \lbls{074}
\begin{alignat}{2}
\lbls{074} = \m_{\ind{5}}\,
    \Real{\cpl{FFS}{\ind{5}, \ind{6}, \ind{4}, 1}\,
          \cpl{FFV}{\ind{5}, \ind{6}, \ind{7}, 2}}\,
          \cpl{SSV}{\ind{1}, \ind{4}, \ind{3}, 1}\,
          \cpl{SVV}{\ind{2}, \ind{3}, \ind{7}, 1}\times
    \f{74}{}{\ind{3}, \ind{4}, \ind{5}, \ind{6}, \ind{7}} + [\ind{5} \leftrightarrow \ind{6}]\span\omit\span\omit\span\omit \nn\\
\rightarrow \Pi_{ij} &\supset&& 2\,\imath\,
  \Real{M_{I}\, y^{IJ j}\, g^{aI}_J}\,
  g^{bi k}\, g^{ab j}\,\f{74}{}{b, k, I, J, a}\nn\\
&\rightarrow\ && 2\,\imath\, \Real{M_{IK}\, y^{IJ j}\, g^{aK}_J}\,
  g^{bi k}\, g^{ab j}\,\f{74}{}{b, k, I, J, a}\,.
\end{alignat}

For practical applications, however, it is likely most useful to
retain the four-spinor notation; a translation to Dirac spinors as a
sum of two Majorana spinors, and to complex scalars as a sum of two
real scalars is straightforward. Naively, for each complex scalar or
Dirac spinor the number of amplitudes would be doubled in this way,
but actually there must be a \emph{symmetry} preventing the two
components of a complex scalar or a Dirac spinor from mixing or
splitting their masses---hence for any three-point coupling there is a
unique way of combining them into a coupling. For example, when
considering three complex scalars $\phi_1, \phi_2,\phi_3$ where the
lagrangian contains
\begin{gather}
  \mathcal{L} \supset - \frac{1}{6} \left(a\, \phi_1\, \phi_2\, \phi_3
           + a^*\, \phi_1^*\, \phi_2^*\, \phi_3^*\right)  
\end{gather}
the terms $\phi_1^*\, \phi_2\, \phi_3$, $\phi_1\, \phi_2^*\, \phi_3$,
$\phi_1\, \phi_2\, \phi_3^*$ or their complex conjugates are not
allowed because they would violate the symmetries keeping the fields
complex. Therefore, the notation with Majorana spinors and real
scalars can be very easily applied to Dirac spinors and complex
scalars: when inserting fields into our results we just choose for
each set of couplings the \emph{one} combination of fields that is
allowed in the theory. Indeed, this is the algorithm used in \SARAH
(albeit currently for the diagrams in the gaugeless limit only).

Finally, another motivation for retaining the Majorana-spinor notation
is the problem of~$\gamma_5$. For two-loop self-energies, we can use a
naive anticommuting prescription for~$\gamma_5$, but for two-loop
decays or three-loop self-energies this is known to be inconsistent
and it would be necessary to choose another definition, for example
involving the~$\epsilon^{\mu\nu\rho\kappa}$~tensor. In the
two-component formalism, spinors are automatically split by chirality,
corresponding to a naive $\gamma_5$~prescription, and it is not known
how to make this consistent for such higher-order calculations.

\tocsection[\label{SEC:GOLDSTONESGHOSTS}]{Removing Goldstone and ghost couplings}

In this section, we shall derive tree-level relations among the
Goldstone and ghost couplings of general theories, which extend those
in \citere{Goodsell:2017pdq,Arkani-Hamed:2017jhn}, and eliminate
four-point couplings involving vectors. Indeed, the first such
relation is already implicitly included in \refeq{EQS:GENERALLAG}; the
four-vector coupling is just related to the product of three-vector
couplings, as can be seen from either the requirement of unitarity of
the amplitude, or just read off from the standard kinetic term. We
shall derive all the necessary couplings using the second approach;
starting with fields in the gauge eigenstate basis of scalars~$R_i$
and gauge bosons~$V_\mu^{a} $
\begin{subequations}
\begin{align}
  F_{\mu\nu}^a &= \partial_\mu V^a_\nu - \partial_\nu V_\mu^a
               + g\, f^{abc}\, V_\mu^b\, V_\nu^c\,,\\
  - \frac{1}{4}\, F^a_{\mu \nu}\, F^{a\,\mu \nu} &\supset
               - g\, f^{cab}\, V^{a,\mu}\, V^{b,\nu} \partial_\mu V_\nu^c
               - \frac{1}{4}\, g^2\, f^{abe}\, f^{cde}\, V_\mu^a\, V_\nu^b\,
               V_\rho^c\, V_\kappa^d\, \eta^{\mu \rho}\, \eta^{\nu \kappa}\,. 
\end{align}
\end{subequations}
Once we break the gauge symmetries, we must diagonalise the vector
masses. Naively this just involves orthogonal rotations and therefore
the sum over intermediate states is not affected; however, in the
presence of kinetic mixing of $U(1)$ gauge bosons we must first make a
\emph{non-orthogonal} transformation. We must first unravel the
kinetic mixing via
\begin{align}
  V_\mu^a &= Z_{ab}\, \ov{A}_\mu^b\,, \qquad
  \sum_b Z_{ab}\, Z_{cb} \ne \delta_{ac}\,,
\end{align}
but then, since only $U(1)$ gauge bosons may mix, we must have
$f^{abc} Z_{cd} = f^{abd} $. We subsequently make an orthogonal
transformation to diagonalise the vector masses;
\begin{subequations}
\begin{alignat}{2}
 \mathrm{if}&&\ V_\mu^a &= N^{(V)}_{ab}\, A^b_\mu = Z_{ac}\, O^{(V)}_{cb}\, A^b_\mu\,,\\
 \mathrm{then}&&\ g^{abc} &= g\, f^{def}\, N^{(V)}_{da}\, N^{(V)}_{eb}\, N^{(V)}_{fc} 
\end{alignat}
\end{subequations}
and the relation between quartic and cubic gauge couplings is clear;
also that $g^{abc}$ is antisymmetric.

For four-point scalar--vector interactions, we must look at the
covariant derivative of the scalars in the gauge-eigenstate basis:
\begin{align}
  D_\mu R_i &= \partial_\mu \phi_i - \theta^a_{ij}\, R_j\, V^a_\mu\, ,
\end{align}
where $\theta^a_{ij}$ are real antisymmetric matrices. They obey the
group algebra \mbox{$[\theta^a, \theta^b] = -
f^{abc}\, \theta^c$}. This then yields
\begin{align}
  \frac{1}{2}\, D_\mu R_i\, D^\mu R_i &\supset
  V_\mu^a\, \theta^a_{ij}\, R_i\, \partial^\mu R_j
  + \frac{1}{2}\,\theta^a_{ki}\, \theta^b_{kj}\, R_i\, R_j\, V^a_\mu\, V^{b\, \mu}\,.
\end{align}
The scalars are rotated by an orthogonal transformation (there is no
kinetic mixing at tree level) so we have
\begin{align}
  g^{aij} &= \theta^b_{kl}\, N^{(V)}_{ba}\, O^{(S)}_{ki}\, O^{(S)}_{lj}\,, \qquad
  g^{abij} = g^{aki}\, g^{bkj} + g^{akj}\, g^{bki}\, .
\label{EQ:RemoveSSVV}\end{align}
The couplings $g^{aij}$ are antisymmetric on the exchange of the two
scalars. It should be noted that the assumption that the scalar rotation is orthogonal will be violated if the running parameters do not sit at the minimum of the tree-level potential, e.g. if we work with running parameters that sit at the minimum of the full loop-corrected potential. Such an approach, however, leads to many complications in the calculations and we do not recommend it. Alternatively a choice of finite counterterms (such as using an on-shell scheme) may cause the identity above and in the following to be violated.

\tocsubsection[\label{SEC:GHOSTS}]{Ghosts and gauge fixing}

To derive the Goldstone and ghost couplings, we first need the
gauge-fixing terms. Once we give expectation values to the scalars, so
$R_i = v_i + \hat{R}_i$, and defining
\begin{align}
  F_i^a &\equiv \theta^{b}_{ji}\, v_j\, Z_{ba},
\end{align}
the scalar kinetic terms contain
\begin{align}
\label{EQ:ScalarKinetic}
  \frac{1}{2}\, D_\mu R_i\, D^\mu R_i &\supset
  \ov{A}_\mu^a\,  F_i\, \partial^\mu \hat{R}_j
  + \frac{1}{2} \left(
      F^a_k\, \theta^c_{kj}\, Z_{cb} + F^b_k\, \theta^c_{kj}\, Z_{ca}
    \right) \hat{R}_i\,  \ov{A}^a_\mu\, \ov{A}^{b\, \mu}
  + \frac{1}{2}\,F_i^a\, F_i^a\, \ov{A}^a_\mu\, \ov{A}^{b\, \mu}\,.
\end{align}
We thus have the $R_\xi$ gauge-fixing terms 
\begin{align}
  G^a &= \frac{1}{\sqrt{\xi}} \left(
    \partial_\mu \ov{A}^a - \xi\, F_i^a\, \hat{R}_i
    \right), \qquad
  \mathcal{L}_{\xi} = -\frac{1}{2}\, G^a\, G^a\,,
\end{align}
defined so as to remove tree-level kinetic mixing between scalars and
vectors.

Rotating the gauge transformations in the original gauge basis
$\alpha^a$ so that $\alpha^a \equiv Z_{ab}\, \ov{\alpha}^b$, we have
\begin{subequations}
\begin{align}
  \delta \ov{A}^a_\mu &= \partial_\mu \ov{\alpha}^a
  - f^{abc}\, \ov{A}^b\, \ov{\alpha}^c
  = \left(D_\mu \ov{\alpha}\right)^a,\\
  \delta \hat{R}_i &= \delta R
  = \ov{\alpha}^b \left[
      Z_{ab}\, \theta^{a}_{ij} \left(v_j + \hat{R}_j\right)
    \right]
  = \ov{\alpha}^a \left[-F^a_i + Z_{ba}\,\theta^{b}_{ij}\,\hat{R}_j \right] .
\end{align}
\end{subequations}
This gives
\begin{subequations}
\begin{align}
  \frac{\delta G^a}{\delta \ov{\alpha}^b} &=
  \frac{1}{ \sqrt{\xi}} \left(
    \partial_\mu D^\mu  + \xi\, F_i^a\, F_i^b
    - \xi\, F_i^a\, Z_{cb}\, \theta^{c}_{ij}\, \hat{R}_j\right),\\
  \lag_{\mathrm{ghost}} &= - \ov{c}^a\, \frac{\delta G^a}{\delta \ov{\alpha}^b}\, c^b\,.
\end{align}
\end{subequations}
From this we can read off the ghost mass matrix and the ghost
couplings. For ghost--vector couplings, we have
\begin{align}
  \lag_{\mathrm{ghost}} &\supset \left(\partial_\mu \ov{\omega} \right) D^\mu \omega
  \supset - \left(\partial_\mu \ov{\omega}^a \right) f^{abc}\, \ov{A}^b_\mu\, \omega^c
  = g^{abc}\, A^c_\mu\, \omega^b \left(\partial_\mu \ov{\omega}^a \right) .
\end{align}

Also as expected,
\begin{align}
  m^2_{ab} &\equiv F_i^a\, F_i^b
\end{align}
is the mass matrix for the \emph{gauge bosons} too, and so we can
diagonalise both ghosts and vectors with the same \emph{orthogonal}
rotation $\mathcal{O}^{(V)}_{ab}$, and for each massive vector of mass
$m_a$ there will be a ghost with mass $\sqrt{\xi}\, m_a$. However, it
is important that, since the ghosts and antighosts are not identical,
we can treat them as complex fields, and we actually have the liberty
of defining them with an additional phase.

From \refeq{EQ:ScalarKinetic}, after diagonalising the scalars we read
off that, as noted in \citere{Martin:2018emo},
\begin{align}
  g^{abi} &=  \hat{g}^{abi} + \hat{g}^{bai},
\end{align}
but since $\hat{g}^{abi}$ has an antisymmetric piece we cannot simply
invert this relation. In order to write~$\hat{g}^{abi} $ in terms of
the other couplings of the theory we will have to consider the
Goldstone bosons.

\tocsubsection[\label{SEC:GOLDSTONES}]{Goldstone bosons}

The gauge-fixing terms are also expected to give mass to the Goldstone
bosons (except in Landau gauge); they contain scalar mass terms
\begin{align}
  \mathcal{L}_\xi &\supset
  - \frac{\xi}{2}\, F^a_i\, F^a_j\, \hat{R}_i\, \hat{R}_j\,.
\end{align}
To see that these concern just the Goldstone bosons, consider the
standard perturbative proof of Goldstone's theorem: we expand the
potential \emph{without gauge-fixing terms} \mbox{$V_0(R_i
+ \alpha^a\,\delta^a_i) = V_0(R_i)$} where \mbox{$\delta^a_i
= \theta^b_{ij}\, R_j\, Z_{ba},$} and differentiate the relation once:
\begin{align}
  \alpha^a\, \theta^a_{ij}\, R_j\, \frac{\partial V}{\partial R_i} &=
  0\,, \qquad
  \frac{\partial{\left(\alpha^a\, \delta^a_i\right)}}{\partial R_j}\,
  \frac{\partial V_0}{\partial R_i}
  + \alpha^a\, \delta^a_i\, \frac{\partial^2 V_0}{\partial R_i\, \partial R_j} =
  0\,.
  \label{EQ:MasterV0}
\end{align}
If we work at the minimum of the potential, this becomes
\begin{align}
  0 &= \alpha^a\, \delta^a_i \left. \frac{\partial^2 V}{\partial R_i\, \partial R_j}\right|_{R_k = v_k} 
  = -\ov{\alpha}^a\,F^a_i\, \frac{\partial^2 V_0}{\partial \hat{R}_i\, \partial \hat{R}_j}\,.
\end{align}
This is true for any $\ov{\alpha}^a$; the $F^a_i$ are null
eigenvectors of the mass matrix until we add the gauge-fixing
terms. Adding the gauge fixing terms we have
\begin{align}
  \mathcal{M}^2_{ij} &=
  \frac{\partial^2 V_0}{\partial \hat{R}_i\, \partial \hat{R}_j}
  + \xi\, F^a_i\, F^a_j\,. 
\label{EQ:ScalarMass}\end{align}
Now let us use the singular value decomposition of $F^a_i$ and define,
suggestively:
\begin{align}
  F^a_i &\equiv \mathcal{O}^{(V)}_{ab}\, (F_D)^b_j\, \mathcal{O}^{(S)}_{ij}
  \label{EQ:DefFD}
\end{align}
where $\mathcal{O}^{(V)}_{ab},\mathcal{O}^{(S)}_{ji} $ are orthogonal
and $F_D $ is a diagonal---but not in general square---matrix. Since
\begin{align}
  0 &= \mathcal{O}^{(V)}_{ba}\,F^b_i\, \frac{\partial^2 V_0}{\partial \hat{R}_i\, \partial \hat{R}_j}
  = (F_D)^b_k\, \mathcal{O}^{(S)}_{ik}\, \frac{\partial^2 V_0}{\partial \hat{R}_i\, \partial \hat{R}_j} 
\end{align}
clearly $\mathcal{O}^{(S)}_{ji}$ is arbitrary when acting on
Goldstone-boson indices; it can be chosen to simultaneously
diagonalise both matrices in \refeq{EQ:ScalarMass}, splitting the
scalars into would-be Goldstone bosons and a remaining set. Then, in
the diagonal basis~$\Phi_i$, $F_D$~becomes a \emph{projector onto
Goldstone bosons:}
\begin{align}
  (F_D)_{j}^a &= \left\{
  \begin{array}{@{}rr@{\;}l@{}}
    0\,, & a &> N_G \text{ or } j > N_G \\
    m_a\, \delta_{aj}\,, & a, j &\le N_G
  \end{array}
  \right. ,
\end{align}
where $N_G$ is the number of Goldstone bosons/massive vectors.

Armed with this, we can write 
\begin{align}
  \hat{g}^{abi} &= g^{aij}\, (F_D)^b_j
  = \frac{1}{2}\, g^{abi} - \frac{1}{2}\, g^{abc}\, (F_D)^c_i\,,
\label{EQ:SUUidentity}\end{align}
\IE\ we can exchange the scalar--ghost coupling for scalar--vector
couplings. However, because the coupling has a different form
depending on whether the scalar is a Goldstone boson or not, this
introduces some complications in calculating amplitudes: we should
either sum separately over Goldstone indices and the remaining scalars
(which is necessary for finding a gauge-invariant result), or just use
these pieces to remove the ghost couplings. In this work we take the
latter approach.

Another point is in order: there is actually some ambiguity in the
above definitions because we have the freedom to introduce
signs/phases of the Goldstone bosons and ghosts. In all calculations
such signs should drop out. This is in particular notable because
implementations of gauge-fixing in the literature are defined via the
standard procedure but, in order to verify the above relation (and,
indeed, those that we shall introduce below), it is necessary to
introduce such signs/phases---we checked our identities against
the \FeynArts model file of the~SM, for example, where we need to
introduce a sign for the $Z$-boson Goldstone and a factor of~$\imath$
for the $W$-boson Goldstone.

\tocsubsection[\label{SEC:ELIMINATION}]{Eliminating all Goldstone-boson couplings}

Now that we have given explicit forms for the ghost couplings in terms
of scalar and vector couplings, and found that the form depends on
whether the scalar is a Goldstone boson or not, we can also consider
relating the couplings of \emph{Goldstone bosons} to other couplings
in the theory. Partial results for these can be found in
\citere{Goodsell:2017pdq,Arkani-Hamed:2017jhn}. The general strategy
will be to use our projector~$F_D$ and invert the relation
in \refeq{EQ:DefFD}, and use the identity
\begin{align}
  m_a^2\, \delta_{ab} &= \OV_{ca}\, \OV_{db}\, F_i^c\, F_i^d
  = - \OV_{ca}\, \OV_{db} \left(v^T\, \theta^e\, \theta^f\, v\right)
  Z_{ec}\, Z_{fd}\,.
\end{align}
Throughout we shall distinguish would-be Goldstone bosons from
ordinary scalars $\cpl{S}{}$ by using the letter $\cpl{G}{}$ to
represent them, and we use $G_a, G_b, G_c, \ldots $ for their indices
(instead of $i,j,k,\ldots$): the subscript $a,b,c,\ldots$ of course
indicates that they correspond to the gauge boson of that index.

\tocsubsubsection{Scalar--vector couplings}

\paragraph{GGV}

Inserting our projector into $g^{aij}$ we have: 
\begin{align}
\begin{split}
  g^{a G_b G_c} &= \frac{1}{m_b\, m_c}\, g^{aij}\, (F_D )^b_i\, (F_D)^c_j
  = \frac{1}{2\,m_b\, m_c}\, g^{aij}
    \left[(F_D )^b_i\, (F_D)^c_j - (F_D )^c_i\, (F_D)^b_j\right]\\
  &= \frac{1}{2\,m_b\, m_c}\, Z_{a' a''}\, Z_{b' b''}\, Z_{b' b''}\,
    \mathcal{O}^{(V)}_{a''a}\, \mathcal{O}^{(V)}_{b''b}\,\mathcal{O}^{(V)}_{c''c}\,
    v^T \left(\theta^{b'}\, \theta^{a'}\, \theta^{c'}
    - \theta^{c'}\, \theta^{a'}\, \theta^{b'}\right) v\,.
\end{split}
\end{align}
Next we use
\begin{align}
  v^T \left(\theta^b\, \theta^a\, \theta^c \right) v &=
  - f^{acd}\, v^T\, \theta^b\, \theta^d\, v
  - f^{bcd}\, v^T\, \theta^d\, \theta^a\, v
  - f^{bad}\, v^T\, \theta^c\, \theta^d\, v
  + v^T \left(\theta^c\, \theta^a\, \theta^b \right) v\,,
\end{align}
and therefore the coupling of two Goldstone bosons to a gauge boson is:
\begin{align}
  g^{a G_b G_c} &= \frac{1}{2\,m_b\, m_c}\, g^{abc} \left(m^2_a - m_b^2 - m_c^2\right) .
\end{align}
This is the expression found in \citere{Arkani-Hamed:2017jhn} by
requiring that high-energy scattering amplitudes of theories with
massive gauge bosons have the correct behaviour.

\paragraph{SGV}

For a general scalar, coupled to a Goldstone boson and a vector, we
can derive
\begin{align}
  g^{a i G_b} &= \frac{1}{m_b}\, g^{aij}\, (F_D)^b_j
  = \frac{1}{m_b}\, \hat{g}^{abi}
  = \frac{1}{2\,m_b} \left[ g^{abi} - (F_D)^c_i\, g^{abc}\right] .
\end{align}

\paragraph{GVV} 

Consistent with the previous two expressions, we can derive
\begin{align}
  g^{ab G_c} &= -\frac{1}{m_c}\, g^{abc} \left(m_a^2 - m_b^2\right) .
\end{align}

\tocsubsubsection{Scalar--Goldstone couplings}

The pure-scalar interactions involving Goldstone bosons can also be
related to couplings involving vectors, thus allowing them to be
eliminated. To derive the required relations, we continue to apply
derivatives to \refeq{EQ:MasterV0}:
\begin{align}
\begin{split}
 0 &= \frac{\partial \delta^a_i}{\partial R_j}\,
      \frac{\partial^2 V_0}{\partial R_i\, \partial R_k}
      + \frac{\partial \delta^a_i}{\partial R_k}\,
        \frac{\partial^2 V_0}{\partial R_i\, \partial R_j}
      + \delta^a_i\, \frac{\partial^3 V_0}{\partial R_i\, \partial R_j\, \partial R_k}\\
   &= \frac{\partial \delta^a_i}{\partial R_j}\,
      \frac{\partial^3 V_0}{\partial R_i\, \partial R_k\, \partial R_l}
      + \frac{\partial \delta^a_i}{\partial R_k}\,
        \frac{\partial^3 V_0}{\partial R_i\, \partial R_j\, \partial R_l}
      + \frac{\partial \delta^a_i}{\partial R_l}\,
        \frac{\partial^3 V_0}{\partial R_i\, \partial R_j\, \partial R_k}
      + \delta^a_i\, \frac{\partial^4 V_0}{\partial R_i\, \partial R_j\, \partial R_k\, \partial R_l}\,.
\end{split}
\label{EQ:V0DERIVS}\end{align}

\paragraph{SSS}

From the first line of \refeq{EQ:V0DERIVS} we get the $\cpl{GSS}{}$
coupling
\begin{align}
  a_{G_a j k} &= \frac{1}{m_a}\, g^{ajk} \left[m_{0,j}^2 - m_{0,k}^2\right] , 
\label{EQ:GSS}\end{align}
where we have denoted by~$m_{0,i}^2$ the eigenvalues of~$(\partial^2
V_0)/(\partial R_i\, \partial R_j)|$. These are equal to the scalar
masses in the full potential, except that they \emph{do not include
contributions from gauge fixing:} would-be Goldstone bosons
have \mbox{$m_{0, G_a}^2 = 0$}.

It is also useful to have the expression for an $\cpl{GGS}{}$ coupling:
\begin{align}
  a_{G_a G_b k} &= \frac{m_{0,k}^2 }{2\, m_a\, m_b}\, g^{abk}\, ,
\end{align}
where we note (as shown in \citere{Braathen:2016cqe}) that a
triple-Goldstone coupling vanishes.

To avoid using $m_{0,i}^2$ we can write 
\begin{align}
  m_{0,i}^2 &= m_i^2 - \xi\, m_a\, (F_D)_i^a 
  \label{EQ:M0trick}
\end{align}
and therefore
\begin{align}
  a_{G_a jk} &= \frac{1}{m_a}\, g^{ajk} \left[m_j^2 - m_k^2\right]
    + (F_D)^b_j\, \frac{m_b}{m_a}\, g^{a k G_b}
    + (F_D)^b_k\, \frac{m_b}{m_a}\, g^{a j G_b} \nn\\
  &= \frac{1}{m_a}\, g^{ajk} \left[m_j^2 - m_k^2\right]
  + (F_D)^b_j\, \frac{1}{2\,m_a}\, g^{a b k}
  + (F_D)^b_k\, \frac{1}{2\, m_a}\, g^{a b j}\,.
  \label{EQ:GSSalt}
\end{align}

\paragraph{SSSS}

From the second line of \refeq{EQ:V0DERIVS} we retrieve the
$\cpl{GSSS}{}$ coupling
\begin{align}
  \lambda_{G_a jkl} &= \frac{1}{m_a} \left[
    g^{aij}\, a_{ikl} + g^{aik}\, a_{ijl} + g^{ail}\, a_{ijk}
  \right] .
\label{EQ:GSSS}\end{align}
Here there is a sum over all scalars~$i$ including Goldstone bosons,
and indeed $j,k,l$ can be Goldstone fields. To eliminate couplings
with more Goldstone bosons we then just need to insert the formulae
that we have given above into this equation; for example (as we shall
later require) a four-point coupling of two Goldstone bosons and two
scalars \emph{that are not Goldstone bosons} $\hat{k}, \hat{l}$ is
\begin{align}
  \lambda_{G_a G_b \hat{k} \hat{l}} = \frac{1}{2\,m_a\, m_b}\, \bigg[
  & g^{abi}\, a_{i \hat{k}\hat{l}} - \frac{1}{2}\, g^{a c \hat{k}}\, g^{bc \hat{l}}
    - \frac{1}{2}\, g^{a c \hat{l}} g^{bc \hat{k}}\nn\\
  &{+}\, g^{ai\hat{k}}\, g^{bi \hat{l}}
    \left(2\,m_i^2 - m_{\hat{l}}^2 - m_{\hat{k}}^2\right)
    + g^{ai\hat{l}}\, g^{bi \hat{k}}
    \left(2\, m_i^2 - m_{\hat{k}}^2 - m_{\hat{l}}^2\right) \!\bigg]\,.
\end{align}
Also particularly interesting (but not needed for this work) is the
four-Goldstone coupling, which can be written as (summing over all
non-Goldstone scalars~$i$)
\begin{align}
  \lambda_{G_a G_b G_c G_d} &= \frac{m_{0,i}^2}{4\, m_a\, m_b\, m_c\, m_d} \left[
    g^{abi}\, g^{cd i} + g^{aci}\, g^{b d i} + g^{adi}\, g^{bc i}
  \right].
\end{align}

\tocsubsubsection{Fermion couplings}

To complete the removal of Goldstone boson couplings, we would require
the couplings of fermions to Goldstone bosons. We will not actually
use these in this work, and they were given
in \citere{Goodsell:2017pdq}; we give them again here in our notation
for future reference:
\begin{align}
 y^{IJ G_a} &= \imath\, \frac{1}{m_a } \left(m_J - m_I\right) g^{aI}_{J}\,.
\end{align}

\tocsection[\label{SEC:RESULTS}]{Results}

In this section we shall describe our results. The renormalised
expressions for all of the basic classes of diagrams are given
in \appx{SEC:TADLIST} for the tadpoles and \appx{SEC:SELIST} for the
self-energies, since they are rather long; initially, there are $25$
tadpole and $121$ self-energy classes which is a much larger set than
in the gaugeless limit. Therefore we shall describe how we can reduce
this set to $89$ or $92$ for generic self-energies (depending on
whether we choose to exchange the ghost--ghost--vector coupling for a
triple-gauge coupling) or $58$ for non-Goldstone scalars; and just
$16$ for the tadpoles.

\tocsubsection[\label{SEC:TADPOLES}]{Tadpole diagrams}

\tocsubsubsection[]{Unreduced diagrams}

To present our results in a readable way, and to make the connection
with the diagrams in \citere{Goodsell:2015ira}, we shall denote
tadpole topologies $1$, $3$ and $2$ with all scalar propagators as
$T_{SS}$, $T_{SSS}$, $T_{SSSS}$; the subscripts are modified for
different fields accordingly. The total tadpole can be written as
\begin{align}
  \frac{\partial V^{(2)}}{\partial \Phi_{\ind{1}}} &= - T_{\ind{1}}^{(2)}
  = - \sum_{n=0}^3 T^{(2,n)}\,,
\end{align}
where the superscript $(2,n)$ indicates $\mathcal{O} (2n)$ in the
gauge couplings. Then the unreduced set of tadpole topologies is:
\begin{subequations}
\begin{flushright}
\begin{tabular}{r@{\ }*6{c@{}>{$}c<{$}@{}}c>{\raggedleft\arraybackslash}p{2.5cm}@{}}
\cmidrule{1-14}\morecmidrules\cmidrule{1-14}
$T^{(2,0)} =$& \minitab[c]{$T_{SS}$\\\lblt{01}}
&+& \minitab[c]{$T_{FFFS}$\\\lblt{05}} &+& \minitab[c]{$T_{SSFF}$\\\lblt{06}}
&+& \minitab[c]{$T_{SSSS}$\\\lblt{07}} &+& \minitab[c]{$T_{SSS}$\\\lblt{24}}\,,
&&&&& \ \refstepcounter{equation}(\theequation)\\
\cmidrule{1-14}\morecmidrules\cmidrule{1-14}
$T^{(2,1)} =$ & \minitab{$T_{SV}$\\\lblt{02}}
&+& \minitab[c]{$T_{FFFV}$\\\lblt{10}} &+& \minitab[c]{$T_{SVFF}$\\\lblt{11}}
&+& \minitab[c]{$T_{SVSS}$\\\lblt{12}} &+& \minitab[c]{$T_{SSSV}$\\\lblt{13}}\,,
&&&&& \ \refstepcounter{equation}(\theequation)\\
\cmidrule{1-14}\morecmidrules\cmidrule{1-14}
$T^{(2,2)} =$ & \minitab[c]{$T_{VS}$\\\lblt{03}}
&+& \minitab[c]{$T_{VV}$\\\lblt{04}} &+& \minitab[c]{$T_{SSUU}$\\\lblt{08}}
&+& \minitab[c]{$T_{SVUU}$\\\lblt{14}} &+& \minitab[c]{$T_{UUUV}$\\\lblt{15}}
&+& \minitab[c]{$T_{VVFF}$\\\lblt{16}} &+& \minitab[c]{$T_{VVSS}$\\\lblt{17}}\\
\cmidrule{2-14}
$+$ & \minitab[c]{$T_{SVSV}$\\\lblt{18}} &+& \minitab[c]{$T_{SSVV}$\\\lblt{19}}
&+& \minitab[c]{$T_{VVUU}$\\\lblt{20}} &+& \minitab[c]{$T_{SVVV}$\\\lblt{22}}
&+& \minitab[c]{$T_{VVVV}$\\\lblt{23}} &+& \minitab[c]{$T_{SVV}$\\\lblt{25}}\,,
&&& \ \refstepcounter{equation}(\theequation)\\
\cmidrule{1-14}\morecmidrules\cmidrule{1-14}
$T^{(2,3)} =$ & \minitab[c]{$T_{UUSU}$\\\lblt{09}}
&+& \minitab[c]{$T_{VVSV}$\\\lblt{21}}\smash{\,.}
&&&&&&&&&&& \ \refstepcounter{equation}(\theequation)\\
\cmidrule{1-14}\morecmidrules\cmidrule{1-14}
\end{tabular}
\end{flushright}
\end{subequations}
The integrals $\lblt{N}$ can be found in \appx{SEC:TADLIST}. Of these,
the expressions in $T^{(2,0)}$ are equivalent to equations (2.32),
(2.33), (2.34), (2.36) and (2.37) in \citere{Goodsell:2015ira}, since
they are independent of the gauge fixing (and the results in the
gaugeless limit were given there).

\tocsubsubsection[]{Combined diagrams}

The diagrams with fermions are irreducible, but we can exchange all
four-point couplings including vectors, and all ghost couplings, for
three-point couplings involving vectors and scalars using the
identities in \sect{SEC:GOLDSTONESGHOSTS}. This means that we can
reduce the number of topologies by combining the loop functions
together. To do this, we need some notation: we write each integral in
the appendix as a sum over the different combinations of Lorentz
structures multiplied by a loop function. Suppose we have an integral
with $n$ propagators and $m$ couplings with $p$ indices in total; then
let us write the couplings generically as
\begin{gather}
  \cpl{C}{\ind{1},\ind{2},\dots, \text{L}_1}, \dots,
  \cpl{C}{\dots,\ind{(p)}, \text{L}_m },
\end{gather}
where $\{\text{L}_1,\dots,\mathrm{L}_m\}$ denote the Lorentz structure
of the couplings. The diagrams can be written as
\begin{align}
  \lblt{N} &= \sum_{\{\mathrm{L}\}}
    \tI{N}{\mathrm{l}_1, \cdots, \mathrm{l}_m}{\ind{3}, \dots,\ind{p}}
    \,\prod_{j=1}^m \cpl{C}{\dots,\mathrm{L}_j}\,.
\end{align}
In the cases where there is only one function, we omit the superscript
with the Lorentz indices. For example, we can write
\begin{align}
  \lblt{04}  &= \cpl{SVV}{\ind{1}, \ind{2}, \ind{3}, 1}\,
                \cpl{VVVV}{\ind{2}, \ind{3}, \ind{4}, \ind{4}, 1}\,
                \tI{4}{1,1}{\ind{2}, \ind{3}, \ind{4}}\nn\\
             &\quad+ \cpl{SVV}{\ind{1}, \ind{2}, \ind{3}, 1}\,
                     \cpl{VVVV}{\ind{2}, \ind{3}, \ind{4}, \ind{4}, 2}\,
                     \tI{4}{1,2}{\ind{2}, \ind{3}, \ind{4}} \nn\\
             &\quad+ \cpl{SVV}{\ind{1}, \ind{2}, \ind{3}, 1}\,
                     \cpl{VVVV}{\ind{2}, \ind{3}, \ind{4}, \ind{4}, 3}\,
                     \tI{4}{1,3}{\ind{2}, \ind{3}, \ind{4}}\,,
\end{align}
but
\begin{align}
  \lblt{01}  &= \cpl{SSS}{\ind{1}, \ind{2}, \ind{3}, 1}\,
                \cpl{SSSS}{\ind{2}, \ind{3}, \ind{4}, \ind{4}, 1}\,
                \tI{1}{}{\ind{2}, \ind{3}, \ind{4}}\,.
\end{align}
For fermions, there are several Lorentz structures and the loop
functions differ, while for amplitudes without fermions there are at
most three, and the loop functions are typically equal or differ very
little (in this example, $\tI{4}{1,3}{} = \tI{4}{1,2}{}$).

Armed with this notation, we can then combine the amplitudes. For all
the combinations, we will also be able to reduce each class to just a
single Lorentz structure. It is therefore straightforward to convert
our expressions in \FeynArts-based coupling notation to the notation
of \refeqs{EQS:GENERALLAG} using the identities
in \sect{SEC:COUPLINGS}. The result for the tadpoles is:
\begin{subequations}
\begin{align}
  \ov{T}^{(2,0)} &= T^{(2,0)}\,, \\
  \ov{T}^{(2,1)} &= T_{FFFV} + T_{SVFF} + T_{SVSS} +\ov{T}_{SSSV}\,, \\
  \ov{T}^{(2,2)} &= T_{VVFF} + \ov{T}_{VVSS} + \ov{T}_{SVSV} + \ov{T}_{SSVV}+ \ov{T}_{SVVV} + \ov{T}_{VVVV}\,, \\
  \ov{T}^{(2,3)} &= \ov{T}_{VVSV}\,.
\end{align}
\end{subequations}
We see that the original set of $25$ topologies is reduced to just
$16$. A more detailed description of the subtleties in the reduction
involving scalar--ghost couplings is given in the next section. Here
we simply present the results for the reduced expressions in turn.

The simpler combinations are
\begin{subequations}
\begin{align}
  \ov{T}_{SSSV} &= \cpl{SSS}{\ind{1}, \ind{2}, \ind{5}, 1}\,
                  \cpl{SSV}{\ind{2}, \ind{3}, \ind{4}, 1}\,
                  \cpl{SSV}{\ind{3}, \ind{5}, \ind{4}, 1}
                  \left[ \tI{13}{}{\ind{2}, \ind{3},\ind{4}, \ind{5}}
                    - 2\, \tI{2}{}{\ind{2}, \ind{5},\ind{4}} \right], \\
\begin{split}
  \ov{T}_{VVSS} &= \cpl{SSV}{\ind{3}, \ind{4}, \ind{2}, 1}\,
                  \cpl{SSV}{\ind{3}, \ind{4}, \ind{5}, 1}\,
                  \cpl{SVV}{\ind{1}, \ind{2}, \ind{5}, 1}\,\times \\
                  &\quad\, \left[
                    \tI{17}{}{\ind{2}, \ind{3},\ind{4}, \ind{5}}
                    + \tI{3}{}{\ind{2}, \ind{5},\ind{3}}
                    + \tI{3}{}{\ind{2}, \ind{5},\ind{4}} \right],
                  \end{split}\\
  \ov{T}_{SVSV} &= \cpl{SSV}{\ind{1}, \ind{2}, \ind{5}, 1}\,
                  \cpl{SSV}{\ind{2}, \ind{3}, \ind{4}, 1}\,
                  \cpl{SVV}{\ind{3}, \ind{4}, \ind{5}, 1}
                  \left[ \tI{18}{}{\ind{2}, \ind{3},\ind{4}, \ind{5}}
                    - 2\, \tI{25}{}{\ind{3}, \ind{4},\ind{5}} \right], \\
  \ov{T}_{SSVV} &= \cpl{SSS}{\ind{1}, \ind{2}, \ind{5}, 1}\,
                  \cpl{SVV}{\ind{2}, \ind{3}, \ind{4}, 1}\,
                  \cpl{SVV}{\ind{5}, \ind{3}, \ind{4}, 1}
                  \left[ \tI{19}{}{\ind{2}, \ind{3},\ind{4}, \ind{5}}
                    + \frac{1}{2}\, \tI{8}{}{\ind{2}, \ind{3},\ind{4},\ind{5}}
                  \right], \\
\begin{split}
  \ov{T}_{SVVV} &= \cpl{SSV}{\ind{1}, \ind{2}, \ind{5}, 1}\,
                  \cpl{SVV}{\ind{2}, \ind{3}, \ind{4}, 1}\,
                  \cpl{VVV}{\ind{3}, \ind{4}, \ind{5}, 1}\,\times\\
               &\quad \left[ \tI{22}{}{\ind{2}, \ind{3},\ind{4}, \ind{5}}
                    + \frac{1}{4}\, \tI{14}{}{\ind{2}, \ind{3},\ind{4}, \ind{5}}
                    - \frac{1}{4}\, \tI{14}{}{\ind{2}, \ind{4},\ind{3}, \ind{5}}
                    \right], 
                  \end{split} \\
 \ov{T}_{VVSV} &= \cpl{SVV}{\ind{1}, \ind{2}, \ind{5}, 1}\,
                  \cpl{SVV}{\ind{3}, \ind{2}, \ind{4}, 1}\,
                  \cpl{SVV}{\ind{3}, \ind{4}, \ind{5}, 1}
                  \left[ \tI{21}{}{\ind{2}, \ind{3},\ind{4}, \ind{5}}
                    - \frac{1}{4}\, \tI{9}{}{\ind{2}, \ind{3},\ind{4},\ind{5}}
                  \right],
\end{align}
while the more complicated combination is
\begin{align}
\begin{split}
  \ov{T}_{VVVV} &= \cpl{SVV}{\ind{1}, \ind{2}, \ind{5}, 1}\,
                  \cpl{VVV}{\ind{2}, \ind{3}, \ind{4}, 1}\,
                  \cpl{VVV}{\ind{3}, \ind{4}, \ind{5}, 1}\,\times\\
               &\quad \begin{aligned}[t] \,\bigg[
                 & \tI{23}{}{\ind{2}, \ind{3},\ind{4}, \ind{5}}
                   - 2\, \tI{20}{}{\ind{2}, \ind{3},\ind{4}, \ind{5}}\\
                 & {-}\, \tI{4}{1,1}{\ind{2},\ind{5},\ind{4}}
                   - \tI{4}{1,1}{\ind{2},\ind{5},\ind{3}}
                   + \tI{4}{1,3}{\ind{2},\ind{5},\ind{4}}
                   + \tI{4}{1,3}{\ind{2},\ind{5},\ind{3}}\\
                 & {-}\, \frac{1}{4}\,\m_{\ind{1}}^2\,
                     \tI{8}{}{\ind{2}, \ind{3},\ind{4}, \ind{5}}
                   - \frac{1}{8} \left(\m_{\ind{3}}^2 + \m_{\ind{4}}^2\right)
                     \tI{9}{}{\ind{2}, \ind{3},\ind{4}, \ind{5}}\\
                 & {+}\, \frac{1}{8}\, \tI{14}{}{\ind{2}, \ind{3},\ind{4}, \ind{5}}
                   + \frac{1}{8}\, \tI{14}{}{\ind{2}, \ind{4},\ind{3}, \ind{5}}
                   + \frac{1}{2}\, \tI{15}{}{\ind{2}, \ind{3},\ind{4}, \ind{5}}
                   + \frac{1}{2}\, \tI{15}{}{\ind{2}, \ind{4},\ind{3}, \ind{5}}
              \bigg].
            \end{aligned}
\end{split}
\end{align}
\end{subequations}


\tocsubsection[\label{SEC:SELFENERGIES}]{Self-energies}

\tocsubsubsection[]{Unreduced diagrams}

To make the connection with the diagrams up to second order in the
gauge coupling in \citere{Martin:2003it}
and \citere{Goodsell:2015ira}, we write the total self-energy as
\begin{align}
  \Pi_{\ind{1},\ind{2}}^{(2)} &\equiv
    - \sum_{n=0}^4 \frac{1}{2} \left(\Pi^{(2,n)}_{\ind{1},\ind{2}}
      + \Pi^{(2,n)}_{\ind{2},\ind{1}}\right)
\end{align}
where the superscript $(2,n)$ indicates $\mathcal{O} (2n)$ in the
gauge couplings. The minus sign reflects the difference in our
conventions compared to \FeynArts. The symmetrisation on the external
legs is to account for the fact that some diagrams are not explicitly
symmetric (for example, we eliminate~$\lbls{078}$ since it is
identical to~$\lbls{077}$ with $\ind{1} \leftrightarrow \ind{2}$,
$\ind{4} \leftrightarrow \ind{7}$) and it is far faster to symmetrise
the total final result rather than evaluate twice as many diagrams.

The expressions for $\Pi^{(2,0)}$ and $\Pi^{(2,1)}$ are known, and we
can write them as
\begin{subequations}
\begin{align}
  \Pi^{(2,0)}_{\ind{1},\ind{2}} &\equiv \Pi^{S} + \Pi^{SF}\,,\\
  \Pi^{(2,1)}_{\ind{1},\ind{2}} &\equiv
    \Pi^{(I) S V,g^2} + \Pi^{(R) S V,g^2} + \Pi^{S F V,g^2}\,.
\end{align}
\end{subequations}
All of these simple terms except $\Pi^{(R) S V,g^2} $ are irreducible.
\newpage
The diagrams without vectors are
\begin{subequations}
\begin{flushright}\begin{tabular}{r@{\ }*7{c@{}>{$}c<{$}@{}}c>{\raggedleft\arraybackslash}p{1cm}@{}}
\cmidrule{1-16}\morecmidrules\cmidrule{1-16}
$\Pi^{S} =$ & \minitab[c]{$W_{SSSS}$\\\lbls{111}}
&+& \minitab[c]{$X_{SSS}$\\\lbls{101}} &+& \minitab[c]{$Y_{SSSS}$\\\lbls{001}}
&+& \minitab[c]{$Z_{SSSS}$\\\lbls{105}} &+& \minitab[c]{$S_{SSS}$\\\lbls{120}}
&+& \minitab[c]{$U_{SSSS}$\\\;$2$\,\lbls{009}\;}
&+& \minitab[c]{$V_{SSSSS}$\\\lbls{059}}
&+& \minitab[c]{$M_{SSSSS}$\\\lbls{024}}\,,
& \hspace{-1em}\ \refstepcounter{equation}(\theequation)\\
\cmidrule{1-16}\morecmidrules\cmidrule{1-16}
$\Pi^{(I)SF} =$ & \minitab[c]{$W_{SSFF}$\\\lbls{110}}
&+& \minitab[c]{$M_{FSFSF}$\\$2$\,\lbls{022}}
&+& \minitab[c]{$M_{FFFFS}$\\\lbls{021}} &+& \minitab[c]{$V_{FFFFS}$\\\lbls{057}}
&+& \minitab[c]{$V_{SSSFF}$\\\lbls{058}}\smash{\,,}
&&&&&&& \hspace{-1em}\ \refstepcounter{equation}(\theequation)\\
\cmidrule{1-16}\morecmidrules\cmidrule{1-16}
\end{tabular}\end{flushright}
\end{subequations}
%
%
while the diagrams with only one vector propagator are
\begin{subequations}
\begin{flushright}\begin{tabular}{r@{\ }*5{c@{}>{$}c<{$}@{}}c>{\raggedleft\arraybackslash}p{2.3cm}@{}}
\cmidrule{1-12}\morecmidrules\cmidrule{1-12}
$\Pi^{(I)SV,g^2} =$ & \minitab[c]{$Y_{VSSS}$\\\lbls{003}}
&+& \minitab[c]{$U_{SVSS}$\\$2$\,\lbls{010}}
&+& \minitab[c]{$M_{SVSSS}$\\$2$\,\lbls{031}}
&+& \minitab[c]{$M_{SSSSV}$\\\lbls{033}} &+& \minitab[c]{$V_{VSSSS}$\\\lbls{066}}
&+& \minitab[c]{$V_{SSVSS}$\\$2$\,\lbls{067}}\,,
& \ \refstepcounter{equation}(\theequation)\\
\cmidrule{1-12}\morecmidrules\cmidrule{1-12}
$\Pi^{(R) SV,g^2} =$ & \minitab[c]{$Y_{SSSV}$\\\lbls{002}}
&+& \minitab[c]{$V_{VSSSS}$\\\lbls{069}} &+& \minitab[c]{$X_{SSV}$\\\lbls{102}}
&+& \minitab[c]{$W_{SSSV}$\\\lbls{113}}\smash{\,,}
&&&&& \ \refstepcounter{equation}(\theequation)\\
\cmidrule{1-12}\morecmidrules\cmidrule{1-12}
$\Pi^{SFV,g^2} =$ & \minitab[c]{$M_{FFFFV}$\\\lbls{028}}
&+& \minitab[c]{$M_{FVFSF}$\\$2$\,\lbls{029}}
&+& \minitab[c]{$V_{FFFFV}$\\\lbls{062}} &+& \minitab[c]{$V_{VSSFF}$\\\lbls{063}}
&+& \minitab[c]{$V_{SSVFF}$\\$2$\,\lbls{064}}\smash{\,.}
&&& \ \refstepcounter{equation}(\theequation)\\
\cmidrule{1-12}\morecmidrules\cmidrule{1-12}
\end{tabular}\end{flushright}
\end{subequations}
In \citere{Martin:2003it}, the expressions are given in unreduced
form; those in $\Pi^{(R) SV,g^2}$ can be simplified by removing the
four-point scalar--vector interaction, as we shall do in the next
section.



Next we can write
\begin{align}
  \Pi_{\ind{1},\ind{2}}^{(2,2)} &= \Pi^{(I),g^4} + \Pi^{(R) SV,g^4} + \Pi^{(R) FV,g^4}
                   + \Pi^{\ddag,g^4} + \Pi^{\S,g^4}\,. 
\end{align}
The letters $I, R$ represent irreducible and reducible classes
respectively; the final two contain special classes. The expressions
are
\begin{subequations}
\begin{flushright}
\begin{tabular}{r@{\ }*6{c@{}>{$}c<{$}@{}}c>{\raggedleft\arraybackslash}p{1.2cm}@{}}
\cmidrule{1-14}\morecmidrules\cmidrule{1-14}
$\Pi^{(I),g^4} =$& \minitab[c]{$M_{FVFVF}$\\$2$\,\lbls{037}}
&+& \minitab[c]{$M_{VVSSS}$\\\lbls{039}}
&+& \minitab[c]{$V_{VSVFF}$\\$2$\,\lbls{074}}
&+& \minitab[c]{$V_{VSVSS}$\\$2$\,\lbls{077}}
&+& \minitab[c]{$V_{SSVSV}$\\$2$\,\lbls{081}}
&+& \minitab[c]{$W_{SSVV}$\\\lbls{116}}\smash{\,,}
&&& \ \refstepcounter{equation}(\theequation)\\
\cmidrule{1-14}\morecmidrules\cmidrule{1-14}
$\Pi^{(R) SV,g^4} =$ & \minitab[c]{$U_{SSVV}$\\$2$\,\lbls{011}}
&+& \minitab[c]{$U_{SVVV}$\\$2$\,\lbls{013}}
&+& \minitab[c]{$M_{SVSVS}$\\$2$\,\lbls{040}}
&+& \minitab[c]{$M_{VSSVS}$\\\lbls{041}}
&+& \minitab[c]{$M_{SVSSV}$\\$2$\,\lbls{043}}
&+& \minitab[c]{$M_{VVSSV}$\\\lbls{049}} &+& \minitab[c]{$V_{SSSVV}$\\\lbls{083}}\\
\cmidrule{2-14}
$+$& \minitab[c]{$V_{SSVVV}$\\$2$\,\lbls{093}}
&+& \minitab[c]{$Z_{SSVV}$\\$2$\,\lbls{106}}
&+& \minitab[c]{$Y_{SVVS}$\\\lbls{004}} &+& \minitab[c]{$Y_{VSSV}$\\\lbls{005}}
&+& \minitab[c]{$Y_{SVVV}$\\\lbls{006}}
&+& \minitab[c]{$U_{VSSV}$\\$2$\,\lbls{012}}
&+& \minitab[c]{$V_{SSSUU}$\\\lbls{060}} \\
\cmidrule{2-14}
$+$& \minitab[c]{$V_{SSVUU}$\\$2$\,\lbls{071}}
&+& \minitab[c]{$V_{SVVSS}$\\\lbls{079}} &+& \minitab[c]{$V_{VSSSV}$\\\lbls{080}}
&+& \minitab[c]{$X_{VVS}$\\\lbls{103}} &+& \minitab[c]{$X_{VVV}$\\\lbls{104}}
&+& \minitab[c]{$W_{VVSS}$\\\lbls{115}} &+& \minitab[c]{$S_{SVV}$\\\lbls{121}}\,,
& \ \refstepcounter{equation}(\theequation) \\
\cmidrule{1-14}\morecmidrules\cmidrule{1-14}
$\Pi^{(R) FV,g^4} =$ & \minitab[c]{$V_{SVVFF}$\\\lbls{076}}
&+& \minitab[c]{$W_{VVFF}$\\\lbls{114}}\smash{\,,}
&&&&&&&&&&& \ \refstepcounter{equation}(\theequation)\\
\cmidrule{1-14}\morecmidrules\cmidrule{1-14}
$\Pi^{\ddag,g^4} =$ & \minitab[c]{$V_{SVVVV}$\\\lbls{099}}
&+& \minitab[c]{$V_{SVVUU}$\\\lbls{086}} &+& \minitab[c]{$W_{VVVV}$\\\lbls{119}}
&+& \minitab[c]{$W_{VVUU}$\\\lbls{117}}\smash{\,,}
&&&&&&& \ \refstepcounter{equation}(\theequation)\\
\cmidrule{1-14}\morecmidrules\cmidrule{1-14}
$\Pi^{\S,g^4} =$ & \minitab[c]{$Z_{VSVS}$\\\lbls{107}}
&+& \minitab[c]{$W_{SSUU}$\\\lbls{112}}\smash{\,.}
&&&&&&&&&&& \ \refstepcounter{equation}(\theequation)\\
\cmidrule{1-14}\morecmidrules\cmidrule{1-14}
\end{tabular}
\end{flushright}
\end{subequations}

For the third-order terms, we have
\begin{align}
  \Pi_{\ind{1},\ind{2}}^{(2,3)} &= \Pi^{(I),g^6} + \Pi^{(R) SV,g^6} + \Pi^{\ddag,g^6}\,.
\end{align}
\newpage
The expressions read
\begin{subequations}
\begin{flushright}\begin{tabular}{r@{\ }*6{c@{}>{$}c<{$}@{}}c>{\raggedleft\arraybackslash}p{1.3cm}@{}}
\cmidrule{1-14}\morecmidrules\cmidrule{1-14}
$\Pi^{(I),g^6} =$ & \minitab[c]{$M_{VSSVV}$\\\lbls{051}}
&+& \minitab[c]{$V_{VVVFF}$\\\lbls{087}}\smash{\,,}
&&&&&&&&&&& \ \refstepcounter{equation}(\theequation)\\
\cmidrule{1-14}\morecmidrules\cmidrule{1-14}
$\Pi^{(R) SV,g^6} =$& \minitab[c]{$U_{SSVV}$\\$2$\,\lbls{011}}
&+& \minitab[c]{$U_{SVVV}$\\$2$\,\lbls{013}}
&+& \minitab[c]{$M_{SVSVS}$\\$2$\,\lbls{040}}
&+& \minitab[c]{$M_{VSSVS}$\\\lbls{041}}
&+& \minitab[c]{$M_{SVSSV}$\\$2$\,\lbls{043}}
&+& \minitab[c]{$M_{VVSSV}$\\\lbls{049}} &+& \minitab[c]{$V_{SSSVV}$\\\lbls{083}}\\
\cmidrule{2-14}
$+$& \minitab[c]{$V_{SSVVV}$\\$2$\,\lbls{093}}
&+& \minitab[c]{$Z_{SSVV}$\\$2$\,\lbls{106}}
&+& \minitab[c]{$Y_{SVVS}$\\\lbls{004}} &+& \minitab[c]{$Y_{VSSV}$\\\lbls{005}}
&+& \minitab[c]{$Y_{SVVV}$\\\lbls{006}}
&+& \minitab[c]{$U_{VSSV}$\\$2$\,\lbls{012}}
&+& \minitab[c]{$V_{SSSUU}$\\\lbls{060}}\\
\cmidrule{2-14}
$+$& \minitab[c]{$V_{SSVUU}$\\$2$\,\lbls{071}}
&+& \minitab[c]{$V_{SVVSS}$\\\lbls{079}} &+& \minitab[c]{$V_{VSSSV}$\\\lbls{080}}
&+& \minitab[c]{$X_{VVS}$\\\lbls{103}} &+& \minitab[c]{$X_{VVV}$\\\lbls{104}}
&+& \minitab[c]{$W_{VVSS}$\\\lbls{115}} &+& \minitab[c]{$S_{SVV}$\\\lbls{121}}\,,
& \ \refstepcounter{equation}(\theequation)\\
\cmidrule{1-14}\morecmidrules\cmidrule{1-14}
$\Pi^{\ddag,g^6} =$ & \minitab[c]{$V_{VVVUU}$\\\lbls{100}}
&+& \minitab[c]{$V_{VVVVV}$\\\lbls{095}}\smash{\,.}
&&&&&&&&&&& \ \refstepcounter{equation}(\theequation)\\
\cmidrule{1-14}\morecmidrules\cmidrule{1-14}
\end{tabular}\end{flushright}
\end{subequations}

Finally, the fourth-order terms are given by
\begin{flushright}\begin{tabular}{r@{\ }*3{c@{}>{$}c<{$}@{}}c>{\raggedleft\arraybackslash}p{4.3cm}@{}}
\cmidrule{1-8}\morecmidrules\cmidrule{1-8}
$\Pi_{\ind{1},\ind{2}}^{(2,4)} =$ & \minitab[c]{$M_{UUUUS}$\\\lbls{027}}
&+& \minitab[c]{$M_{VVVVS}$\\\lbls{053}} &+& \minitab[c]{$V_{UUUSU}$\\\lbls{061}}
&+& \minitab[c]{$V_{VVVSV}$\\\lbls{096}}\,.
& \ \refstepcounter{equation}(\theequation)\\
\cmidrule{1-8}\morecmidrules\cmidrule{1-8}
\end{tabular}\end{flushright}

Together, these describe $92$ classes. However, we can immediately
reduce this to $89$ by removing the ghost--ghost--vector couplings
inside $\Pi^{\ddag,g^4}$ and $\Pi^{\ddag,g^6}$, as we describe in the
next sections.

\tocsubsubsection[]{Combination of classes}

As in the case of the tadpole diagrams, we shall exchange four-point
couplings involving vectors, and all ghost couplings, for three-point
couplings involving scalars and vectors. We therefore use the same
notation as in the tadpole case to represent our loop functions and
their couplings; the self-energy diagrams can be written as
\begin{align}
\lbls{N} &= \sum_{\{\mathrm{L}\}}
            \f{N}{\mathrm{l}_1, \cdots, \mathrm{l}_m}{\ind{3},\dots,\ind{(n+2)}}\,
            \prod_{j=1}^m \cpl{C}{\dots,\mathrm{L}_j}\,.
\end{align}
In the cases where there is only one function, we omit the superscript
with the Lorentz indices. For example, we can write
\begin{align}
  \lbls{104} &= \cpl{SSVV}{\ind{1}, \ind{2}, \ind{3}, \ind{4}, 1}\,
                \cpl{VVVV} {\ind{3}, \ind{4}, \ind{5}, \ind{5}, 1}\,
                \f{104}{1,1}{\ind{3}, \ind{4}, \ind{5}} \nn\\
             &\quad+ \cpl{SSVV}{\ind{1}, \ind{2}, \ind{3}, \ind{4}, 1}\,
                     \cpl{VVVV} {\ind{3}, \ind{4}, \ind{5}, \ind{5}, 2}\,
                     \f{104}{1,2}{\ind{3}, \ind{4}, \ind{5}} \nn\\
             &\quad+ \cpl{SSVV}{\ind{1}, \ind{2}, \ind{3}, \ind{4}, 1}\,
                     \cpl{VVVV}{\ind{3}, \ind{4}, \ind{5}, \ind{5}, 3}\,
                     \f{104}{1,3}{\ind{3}, \ind{4}, \ind{5}}\,,
\end{align}
but
\begin{align}
\lbls{001} &= \cpl{SSS}{\ind{1}, \ind{3}, \ind{4}, 1}\,
    \cpl{SSS}{\ind{2}, \ind{3}, \ind{5}, 1}\,
    \cpl{SSSS}{\ind{4}, \ind{5}, \ind{6}, \ind{6}, 1}\,
    \f{1}{}{\ind{3}, \ind{4}, \ind{5},\ind{6}}\,.
\end{align}


The first trivial application of this is to remove the diagrams with
ghosts that do not couple to scalars. In the above notation we can
write
\begin{align}
  \Pi^{\ddag,g^6} &= \cpl{SVV}{\ind{1}, \ind{3}, \ind{4}, 1}\,
                  \cpl{SVV}{\ind{2}, \ind{3}, \ind{7}, 1}\,
                  \cpl{VVV}{\ind{4}, \ind{5}, \ind{6}, 1}\,
                  \cpl{VVV}{\ind{5}, \ind{6}, \ind{7}, 1}\,\times \nn\\
     &\quad \left[
        \f{100}{1,1,1,1}{\ind{3}, \ind{4}, \ind{5}, \ind{6}, \ind{7}}
       - 4\, \f{95}{1,1,1,1}{\ind{3}, \ind{4}, \ind{5}, \ind{6}, \ind{7}}
       \right] ,
\end{align}
where we note that~\mbox{$\f{95}{1,1,1,1}{} = \f{95}{1,1,2,2}{}$}, and
$\Pi^{\ddag,g^4} $ consists of two classes corresponding to
diagrams~\lbls{099} and~\lbls{119}:
\begin{align}
  \Pi^{\ddag,g^4} &= \cpl{SSV}{\ind{1}, \ind{3}, \ind{4}, 1}\,
    \cpl{SSV}{\ind{2}, \ind{3}, \ind{7}, 1}\,
    \cpl{VVV}{\ind{4}, \ind{5}, \ind{6}, 1}\,
    \cpl{VVV}{\ind{5}, \ind{6}, \ind{7}, 1}\,\times \nn\\
            &\quad\left[
            \f{99}{1,1,1,1}{\ind{3}, \ind{4}, \ind{5}, \ind{6}, \ind{7}}
            - 4\, \f{86}{1,1,1,1}{\ind{3}, \ind{4}, \ind{5}, \ind{6}, \ind{7}}
            \right] \nn\\
   &\quad+ \cpl{SSVV}{\ind{1}, \ind{2}, \ind{5}, \ind{6}, 1}\,
    \cpl{VVV}{\ind{3}, \ind{4}, \ind{5}, 1}\,
    \cpl{VVV}{\ind{3}, \ind{4}, \ind{6}, 1}\,\times\nn\\
   &\qquad\! \left[
     \f{119}{1,1,1}{\ind{3}, \ind{4}, \ind{5}, \ind{6}}
     - 4\, \f{117}{1,1,1}{\ind{3}, \ind{4}, \ind{5}, \ind{6}}
   \right].
\end{align}
This reduces the number of classes of diagrams to evaluate to $89$,
which will speed up evaluation (since the loop functions are all of
the same class, they require no substantial extra time to evaluate,
whereas performing loops over the couplings and evaluating the loop
functions each time is slow). In fact, by using \refeq{EQ:RemoveSSVV}
we can combine these two classes of diagrams into one graph of
topology~$V_{SVVVV} $; but we shall apply this systematically in the
next section.

\tocsubsubsection[]{Ghostbusting}

In the previous section we eliminated diagrams containing only
ghost--ghost--vector couplings and no ghost--ghost--scalar
couplings. In this section we shall remove all ghost couplings from
the amplitude, and also apply \refeq{EQ:RemoveSSVV}
and \refeq{EQ:RemoveVVVV} in order to obtain only $58$ classes to
evaluate in total (compared to the $28$ for up to $\mathcal{O}(g^2)$
terms) at the expense of requiring that the external scalars
are \emph{not would-be Goldstone bosons}.

The key equation that we shall apply is \refeq{EQ:SUUidentity} which
shows how we can relate $\cpl{SUU}{}$ couplings to scalar and vector
couplings. However, the form of those couplings has an extra
contribution for would-be Goldstone bosons:
\begin{align}
  \cpl{SUU}{\ind{1},-\ind{2},\ind{3},1} &=
  - \frac{\xi}{2}\, \cpl{SVV}{\ind{1},\ind{2},\ind{3},1}
  - \frac{\imath}{2}\, \sum_{\ind{4}} \xi\, (F_D)^{\ind{4}}_{\ind{1}}\,
    \cpl{VVV}{\ind{4},\ind{2},\ind{3},1}\,.
  \label{EQ:SUUid}
\end{align}
We therefore have to make some distinction between would-be Goldstone
bosons and the other scalars in the summation. One approach to dealing
with this would be to introduce the Goldstone bosons as a new class of
fields (separate from other scalars) from the start, and remove them
via the identities in \sect{SEC:GOLDSTONESGHOSTS} afterwards. This
would be necessary to obtain an explicitly gauge-invariant result, and
should lead to a faster evaluation of the final result, but comes at
the expense of complicating the expressions. We leave this to future
work.

Instead, we deal with the above problem by noting that the first term
on the right-hand side of \refeq{EQ:SUUid} is universal for all
scalars whether they are Goldstone bosons or not, so we can split any
diagram containing an $\cpl{SUU}{}$ vertex into two (or more) and
explicitly sum over the second part, since it effectively becomes a
vector propagator. As an example, consider $\lbls{027}$:
\begin{align}
\lbls{027}&= \cpl{SUU}{\ind{1}, -\ind{3}, \ind{6}, 1}\,
    \cpl{SUU}{\ind{2}, -\ind{7}, \ind{4}, 1}\,
    \cpl{SUU}{\ind{5}, -\ind{4}, \ind{3}, 1}\,
    \cpl{SUU}{\ind{5}, -\ind{6}, \ind{7}, 1}\,
    \f{27}{}{\ind{3},\ind{4},\ind{5},\ind{6},\ind{7}} \nn\\
&= \cpl{SVV}{\ind{1}, \ind{3}, \ind{6}, 1}\,
    \cpl{SVV}{\ind{2}, \ind{4}, \ind{7}, 1}\,
    \cpl{SVV}{\ind{5}, \ind{3}, \ind{4}, 1}\,
    \cpl{SVV}{\ind{5}, \ind{6}, \ind{7}, 1}\,
    \frac{1}{8}\, \f{27}{}{\ind{3},\ind{4},\ind{5},\ind{6},\ind{7}} \nn\\
&+ \cpl{SVV}{\ind{1}, \ind{3}, \ind{6}, 1}\,
    \cpl{SVV}{\ind{2}, \ind{4}, \ind{7}, 1}\,
    \cpl{VVV}{\ind{3}, \ind{4}, \ind{5}, 1}\,
    \cpl{VVV}{\ind{5}, \ind{6}, \ind{7}, 1}\,
  \frac{1}{8}\, \m_{i5}^2\, \f{27}{}{\ind{3},\ind{4},\ind{5},\ind{6},\ind{7}}\,.
\end{align}
Two combinations here have dropped out:
\begin{subequations}
\begin{align}
\frac{\imath}{8}\, \m_{i5}\, \f{27}{}{\ind{3},\ind{4},\ind{5},\ind{6},\ind{7}}  \Big[
  & \cpl{SVV}{\ind{1}, \ind{3}, \ind{6}, 1}\,
    \cpl{SVV}{\ind{2}, \ind{4}, \ind{7}, 1}\,
    \cpl{SVV}{\ind{5}, \ind{3}, \ind{4}, 1}\,
    \cpl{VVV}{\ind{5}, \ind{6}, \ind{7}, 1} \nn\\
  &{+}\,\cpl{SVV}{\ind{1}, \ind{3}, \ind{6}, 1}\,
    \cpl{SVV}{\ind{2}, \ind{4}, \ind{7}, 1}\,
    \cpl{VVV}{\ind{5}, \ind{3}, \ind{4}, 1}\,
    \cpl{SVV}{\ind{5}, \ind{6}, \ind{7}, 1} \Big]\! = 0\,,\\
\frac{\imath}{8}\, \m_{i5}\, \f{27}{}{\ind{4},\ind{3},\ind{5},\ind{7},\ind{6}}  \Big[
  & \cpl{SVV}{\ind{2}, \ind{4}, \ind{7}, 1}\,
    \cpl{SVV}{\ind{1}, \ind{3}, \ind{6}, 1}\,
    \cpl{SVV}{\ind{5}, \ind{4}, \ind{3}, 1}\,
    \cpl{VVV}{\ind{5}, \ind{7}, \ind{6}, 1} \nn\\
  &{+}\,\cpl{SVV}{\ind{2}, \ind{4}, \ind{7}, 1}\,
    \cpl{SVV}{\ind{1}, \ind{3}, \ind{6}, 1}\,
    \cpl{VVV}{\ind{5}, \ind{4}, \ind{3}, 1}\,
    \cpl{SVV}{\ind{5}, \ind{7}, \ind{6}, 1} \Big]\! = 0\,.
\end{align}
\end{subequations}
Schematically we then have
\begin{center}
\begin{tabular}{@{}*4{c@{\ }}c}
\begin{minipage}{0.3\textwidth}
\includegraphics[width=\textwidth]{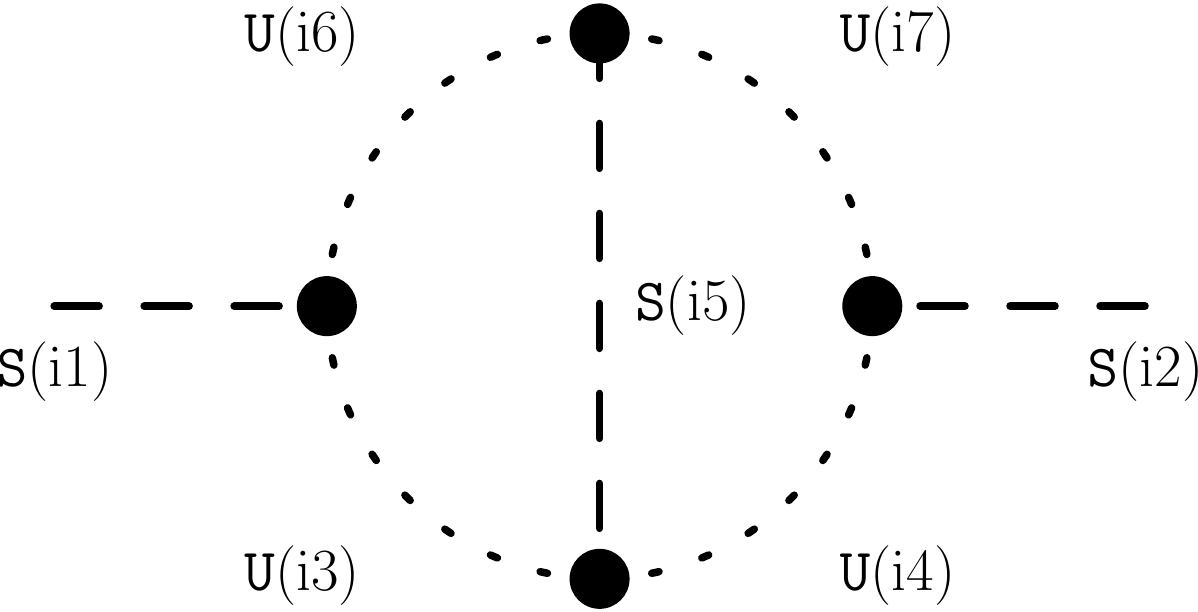}
\end{minipage}
&\begin{minipage}{0.03\textwidth}
$\longrightarrow$
\end{minipage}
&\begin{minipage}{0.3\textwidth}
\includegraphics[width=\textwidth]{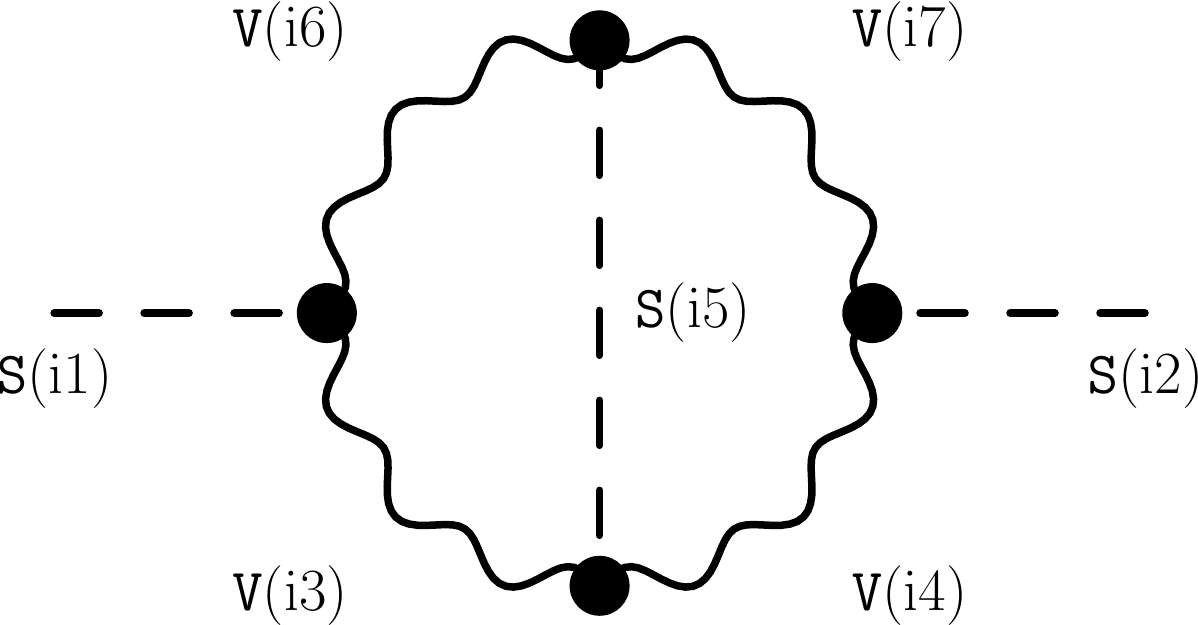}
\end{minipage}
&\begin{minipage}{0.02\textwidth}
$+$
\end{minipage}
&\begin{minipage}{0.3\textwidth}
\includegraphics[width=\textwidth]{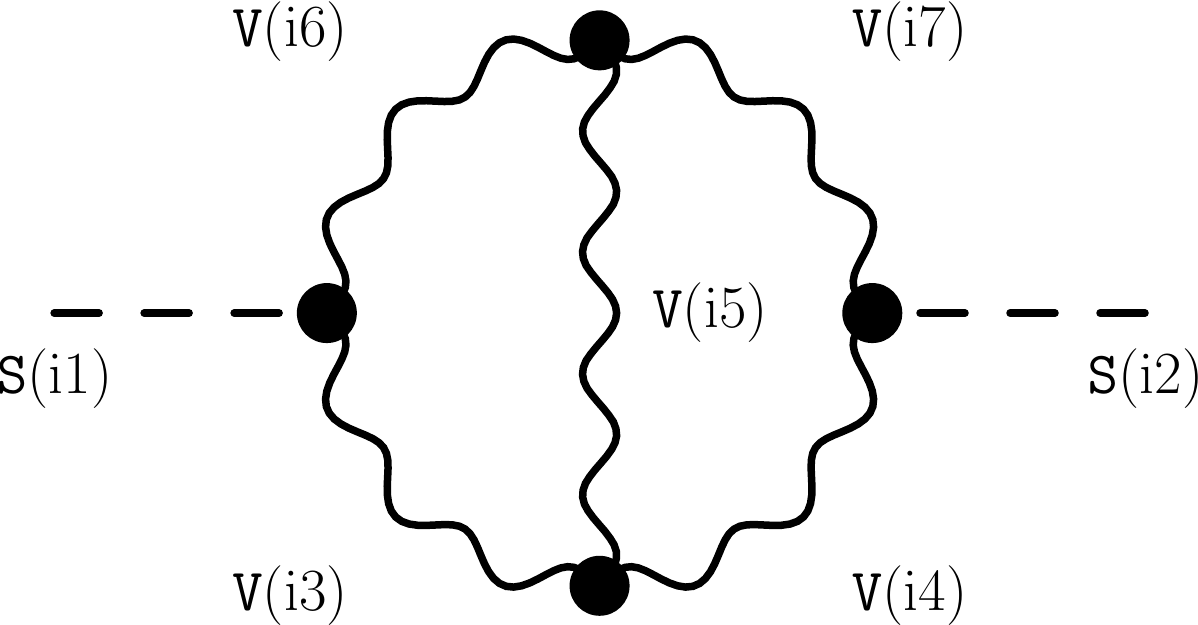}
\end{minipage} \\
$M_{UUUUS}$ & &$M_{VVVVS}$ & & $M_{VVVVV}$ 
\end{tabular}
\end{center}
and we see that the diagrams that drop out would not have fit with the
above picture (note that almost exactly the same pattern reproduces
for diagram \lbls{061}). In the first diagram on the left, the sum
over scalar propagators is indeed a sum over all scalars in the
theory. We restrict to the case that the \emph{external} states are
not Goldstone bosons here, because otherwise the couplings for the
diagrams on the right hand side of the above relations would include
gauge bosons as external legs, and we would therefore need to combine
those amplitudes with mixed scalar--vector and vector--vector
amplitudes.

There are also complications when the ``Goldstone'' leg attaches to a
triple or quartic scalar vertex. The reason for this can be traced
back to \refeq{EQ:GSS}: masses appear in the $\cpl{GSS}{}$ coupling
relation that are different for Goldstone bosons than for other
scalars (in any gauge except Landau gauge), and this then feeds into
the relation for the $\cpl{GSSS}{}$ coupling. There would also
potentially be a similar problem for the $\cpl{SGV}{}$ vertex, but in
all our examples this is avoided because we assert that we have no
external Goldstone bosons. There is only one diagram with a quartic
scalar coupling and ghosts---\lbls{112}---that we will discuss in more
detail below, but diagrams \lbls{025}, \lbls{060} and \lbls{071}
contain triple scalar couplings, that we treat by using
\refeq{EQ:GSSalt} and again splitting the diagram into sums over all
scalars and sums over just Goldstone indices, explicitly trading the
Goldstone bosons for vectors.



\tocsubsubsection[]{Final simplified results for self-energies}

The final result for our set of $58$ self-energy topologies can be
expressed as:
\begin{subequations}
\begin{align}
\Pi^{(2,0)}_{\ind{1},\ind{2}} &= \Pi^{S} + \Pi^{SF}\,,\\
\Pi^{(2,1)}_{\ind{1},\ind{2}} &= \Pi^{(I) S V,g^2} + \Pi^{(I) S F V,g^2} + \ov{\Pi}^{S V,g^2}\,,\\
\Pi^{(2,2)}_{\ind{1},\ind{2}} &= \Pi^{(I),g^4} + \ov{\Pi}^{FV, g^4} + \ov{\Pi}^{SUV, g^4} + \Pi^{\S, g^4}\,,\\
\Pi^{(2,3)}_{\ind{1},\ind{2}} &= \Pi^{(I),g^6} + \ov{\Pi}^{g^6}\,,\\
\Pi^{(2,4)}_{\ind{1},\ind{2}} &= \ov{\Pi}^{g^8}\,.
\end{align}
\end{subequations}
The pieces without bars are unchanged from above, the ones with bars
are explained in the following.

\tocsubsubsection[]{Reduced diagrams of \texorpdfstring{$\mathcal{O}(g^2)$}{\unicodescriptO(\unicodeitalicg\unicodesuptwo)}}

The combined diagrams start at $\mathcal{O}(g^2)$:
\begin{align}
 \ov{\Pi}^{S V,g^2} &= \cpl{SSS}{\ind{1}, \ind{3}, \ind{4}, 1}\,
                     \cpl{SSS}{\ind{2}, \ind{3}, \ind{7}, 1}\,
                     \cpl{SSV}{\ind{4}, \ind{5}, \ind{6}, 1}\,
                     \cpl{SSV}{\ind{5}, \ind{7}, \ind{6}, 1}\,\times\nn\\*
  &\quad\, \Big\{ \f{69}{}{\ind{3}, \ind{4}, \ind{5}, \ind{6}, \ind{7}}
                - \f{2}{}{\ind{3}, \ind{4}, \ind{7}, \ind{6}}
                - \f{2}{}{\ind{3}, \ind{7}, \ind{4}, \ind{6}}  \!\Big\}\nn\\*
  &\quad+ \cpl{SSSS}{\ind{1}, \ind{2}, \ind{5}, \ind{6}, 1}\,
          \cpl{SSV}{\ind{3}, \ind{5}, \ind{4}, 1}\,
          \cpl{SSV}{\ind{3}, \ind{6}, \ind{4}, 1}\,\times \nn\\*
  &\qquad \Big\{ \f{113}{}{\ind{3}, \ind{4}, \ind{5}, \ind{6}}
             + 2\,\f{102}{}{\ind{5}, \ind{6}, \ind{4}} \!\Big\}\,.
\end{align}

\tocsubsubsection[]{Reduced diagrams of \texorpdfstring{$\mathcal{O}(g^4)$}{\unicodescriptO(\unicodeitalicg\unicodesupfour)}}

The only new diagrams with fermions enter at order $\mathcal{O}(g^4)$
and are given by:
\begin{align}
\ov{\Pi}^{FV, g^4} &= 2\,\Real{\cpl{FFV}{\ind{5}, \ind{6}, \ind{4}, 1}\,
                             \cpl{FFV}{\ind{5}, \ind{6}, \ind{7}, 1}\,
                             \cpl{SSV}{\ind{1}, \ind{3}, \ind{4}, 1}\,
                             \cpl{SSV}{\ind{2}, \ind{3}, \ind{7}, 1}}\,\times
                             \nn\\*
 &\quad\, \Big\{ \f{76}{1,1,1,1}{\ind{3}, \ind{4}, \ind{5}, \ind{6}, \ind{7}}
            + 2\,\f{114}{1,1,1}{\ind{5}, \ind{6}, \ind{4}, \ind{7}}
          \!\Big\} \nn\\*
 &\quad + 2\,\Real{\cpl{FFV}{\ind{5}, \ind{6}, \ind{4}, 1}\,
                   \cpl{FFV}{\ind{5}, \ind{6}, \ind{7}, 2}\,
                   \cpl{SSV}{\ind{1}, \ind{3}, \ind{4}, 1}\,
                   \cpl{SSV}{\ind{2}, \ind{3}, \ind{7}, 1}}\,\times \nn\\*
 &\qquad \Big\{ \f{76}{1,2,1,1}{\ind{3}, \ind{4}, \ind{5}, \ind{6}, \ind{7}}
           + 2\,\f{114}{1,2,1}{\ind{5}, \ind{6}, \ind{4}, \ind{7}} \!\Big\}\,.
\end{align}
The other combined diagrams at order $\mathcal{O}(g^4)$ are:
\begin{align}
\ov{\Pi}^{SUV, g^4} &=
  \ov{M}_{SVSSV} + \ov{M}_{SVSVS} + \ov{M}_{VSSVS} + \ov{M}_{VVSSV}\nn\\*
  &\quad+ \ov{V}_{SSSVV} + \ov{V}_{SVVSS} + \ov{V}_{SVVVV} + \ov{V}_{VSSSV} \,.
\end{align}
These new topologies are:
{\allowdisplaybreaks
\begin{subequations}
\begin{align}
\ov{M}_{SVSSV} &= \cpl{SSS}{\ind{1}, \ind{3}, \ind{6}, 1}\,
                 \cpl{SSV}{\ind{2}, \ind{4}, \ind{7}, 1}\,
                 \cpl{SSV}{\ind{3}, \ind{4}, \ind{5}, 1}\,
                 \cpl{SVV}{\ind{6}, \ind{5}, \ind{7}, 1}\,\times\nn\\*
  &\quad\, \Big\{ 2\,\f{43}{}{\ind{3}, \ind{4}, \ind{5}, \ind{6}, \ind{7}}
                + 4\,\f{11}{}{\ind{6}, \ind{5}, \ind{7}, \ind{3}} \!\Big\}\,,\\
\ov{M}_{SVSVS} &= \cpl{SSS}{\ind{1}, \ind{3}, \ind{6}, 1}\,
                 \cpl{SSV}{\ind{3}, \ind{5}, \ind{4}, 1}\,
                 \cpl{SSV}{\ind{5}, \ind{6}, \ind{7}, 1}\,
                 \cpl{SVV}{\ind{2}, \ind{4}, \ind{7}, 1}\,\times\nn\\*
  &\quad\, \Big\{ 2\,\f{40}{}{\ind{3}, \ind{4}, \ind{5}, \ind{6}, \ind{7}}
                - 4\,\f{106}{}{\ind{3}, \ind{6}, \ind{4}, \ind{7}} \!\Big\}\,,\\
\ov{M}_{VSSVS} &= \cpl{SSV}{\ind{1}, \ind{3}, \ind{6}, 1}\,
                 \cpl{SSV}{\ind{2}, \ind{7}, \ind{4}, 1}\,
                 \cpl{SSV}{\ind{3}, \ind{5}, \ind{4}, 1}\,
                 \cpl{SSV}{\ind{5}, \ind{7}, \ind{6}, 1}\,\times\nn\\*
  &\quad\, \Big\{ \f{41}{}{\ind{3}, \ind{4}, \ind{5}, \ind{6}, \ind{7}}
                + \f{12}{}{\ind{3}, \ind{5}, \ind{4}, \ind{6}}
                + \f{12}{}{\ind{7}, \ind{5}, \ind{6}, \ind{4}}
             - 2\,\f{121}{}{\ind{5}, \ind{4}, \ind{6}} \!\Big\}\,,\\
\ov{M}_{VVSSV} &= \cpl{SSV}{\ind{1}, \ind{3}, \ind{6}, 1}\,
                \cpl{SSV}{\ind{2}, \ind{4}, \ind{7}, 1}\,
                \cpl{SSV}{\ind{3}, \ind{4}, \ind{5}, 1}\,
                \cpl{VVV}{\ind{5}, \ind{6}, \ind{7}, 1}\,\times\nn\\*
  &\quad\, \Big\{ \f{49}{}{\ind{3}, \ind{4}, \ind{5}, \ind{6}, \ind{7}}
            - 2\, \f{13}{}{\ind{6}, \ind{5}, \ind{7}, \ind{3}}
            - 2\, \f{13}{}{\ind{7}, \ind{5}, \ind{6}, \ind{4}} \!\Big\}\,,\\
\ov{V}_{SSSVV} &= \cpl{SSS}{\ind{1}, \ind{3}, \ind{4}, 1}\,
                \cpl{SSS}{\ind{2}, \ind{3}, \ind{7}, 1}\,
                \cpl{SVV}{\ind{4}, \ind{5}, \ind{6}, 1}\,
                \cpl{SVV}{\ind{7}, \ind{5}, \ind{6}, 1}\,\times\nn\\*
  &\quad\, \bigg\{ \f{83}{}{\ind{3}, \ind{4}, \ind{5}, \ind{6}, \ind{7}}
   + \frac{1}{2}\, \f{60}{}{\ind{3}, \ind{4}, \ind{5}, \ind{6}, \ind{7}}
   \!\bigg\}\,,\\
\ov{V}_{SVVSS} &= \cpl{SSV}{\ind{1}, \ind{3}, \ind{4}, 1}\,
                \cpl{SSV}{\ind{2}, \ind{3}, \ind{7}, 1}\,
                \cpl{SSV}{\ind{5}, \ind{6}, \ind{4}, 1}\,
                \cpl{SSV}{\ind{5}, \ind{6}, \ind{7}, 1}\,\times\nn\\*
 &\quad\, \begin{aligned}[b] \bigg\{
   & \f{79}{}{\ind{3}, \ind{4}, \ind{5}, \ind{6}, \ind{7}}
     + 2\, \f{115}{}{\ind{5}, \ind{6}, \ind{4}, \ind{7}}
     + 2\, \f{103}{}{\ind{4}, \ind{7}, \ind{5}}
     + 2\, \f{103}{}{\ind{4}, \ind{7}, \ind{6}}\\
   &{+}\, \frac{1}{2} \Big[ \f{4}{}{\ind{3}, \ind{4}, \ind{7}, \ind{5}}
        + \f{4}{}{\ind{3}, \ind{4}, \ind{7}, \ind{6}}
        + \f{4}{}{\ind{3}, \ind{7}, \ind{4}, \ind{5}}
        + \f{4}{}{\ind{3}, \ind{7}, \ind{4}, \ind{6}} \Big] \!\bigg\}\,,
  \end{aligned}\\
\ov{V}_{SVVVV} &= \cpl{SSV}{\ind{1}, \ind{3}, \ind{4}, 1}\,
                \cpl{SSV}{\ind{2}, \ind{3}, \ind{7}, 1}\,
                \cpl{VVV}{\ind{4}, \ind{5}, \ind{6}, 1}\,
                \cpl{VVV}{\ind{5}, \ind{6}, \ind{7}, 1}\,\times\nn\\*
  &\quad\,\begin{aligned}[b] \bigg\{
    & \f{99}{}{\ind{3}, \ind{4}, \ind{5}, \ind{6}, \ind{7}}
      + 2\,\f{119}{}{\ind{5}, \ind{6}, \ind{4}, \ind{7}}
      - \frac{1}{2} \left(\m_{\ind{1}}^2 - \m_{\ind{3}}^2\right)
        \left(\m_{\ind{2}}^2 - \m_{\ind{3}}^2\right)
        \f{60}{}{\ind{3}, \ind{4}, \ind{5}, \ind{6}, \ind{7}}\hspace{-2em} \\
    &{+}\, \frac{1}{4} \left(\m_{\ind{1}}^2 + \m_{\ind{2}}^2 - 2\, \m_{\ind{3}}^2\right)  \!\Big[
    \f{71}{}{\ind{3}, \ind{4}, \ind{5}, \ind{6}, \ind{7}}
      + \f{71}{}{\ind{3}, \ind{4}, \ind{6}, \ind{5}, \ind{7}}\Big] &
    \\
    &{-}\, \f{6}{1,1,1}{\ind{3}, \ind{4}, \ind{7}, \ind{5}}
      - \f{6}{1,1,1}{\ind{3}, \ind{4}, \ind{7}, \ind{6}}
      + \f{6}{1,1,3}{\ind{3}, \ind{4}, \ind{7}, \ind{5}}
      + \f{6}{1,1,3}{\ind{3}, \ind{4}, \ind{7}, \ind{6}}\\
    &{-}\, 2\,\f{86}{}{\ind{3}, \ind{4}, \ind{5}, \ind{6}, \ind{7}}
      - 4\,\f{117}{}{\ind{3}, \ind{4}, \ind{5}, \ind{6}, \ind{7}}\\
    &{-}\, 2\,\f{104}{1,1}{\ind{4}, \ind{7}, \ind{5}}
      - 2\,\f{104}{1,1}{\ind{4}, \ind{7}, \ind{6}}
      + 2\,\f{104}{1,3}{\ind{4}, \ind{7}, \ind{5}}
      + 2\,\f{104}{1,3}{\ind{4}, \ind{7}, \ind{6}} \!\bigg\}\,,
    \end{aligned}\\
\ov{V}_{VSSSV} &= \cpl{SSV}{\ind{1}, \ind{4}, \ind{3}, 1}\,
                \cpl{SSV}{\ind{2}, \ind{7}, \ind{3}, 1}\,
                \cpl{SSV}{\ind{4}, \ind{5}, \ind{6}, 1}\,
                \cpl{SSV}{\ind{5}, \ind{7}, \ind{6}, 1}\,\times\nn\\*
 &\quad\, \begin{aligned}[b] \Big\{
   & \f{80}{}{\ind{3}, \ind{4}, \ind{5}, \ind{6}, \ind{7}}
     - \f{5}{}{\ind{3}, \ind{4}, \ind{7}, \ind{6}}
     - \f{5}{}{\ind{3}, \ind{7}, \ind{4}, \ind{6}} \\
   &{-}\, 2\, \f{121}{}{\ind{5}, \ind{3}, \ind{6}}
     + \f{12}{}{\ind{4}, \ind{5}, \ind{6},\ind{3}}
     + \f{12}{}{\ind{7}, \ind{5}, \ind{6},\ind{3}} \!\Big\}\,.
   \end{aligned}
\end{align}
\end{subequations}
}%

\tocsubsubsection[]{Reduced diagrams of \texorpdfstring{$\mathcal{O}(g^6)$}{\unicodescriptO(\unicodeitalicg\unicodesupsix)}}

The $10$ combined diagrams involving scalar and vector couplings
of~$\mathcal{O}(g^6)$ are given by
\begin{align}
\ov{\Pi}^{g^6} &=
  \ov{M}_{SVSVV} + \ov{M}_{VVSVS} + \ov{M}_{VVSVV} + \ov{M}_{VVVVV} \nn\\
   &\quad + \ov{V}_{SVVSV} + \ov{V}_{VSSVV} + \ov{V}_{VSVSV} + \ov{V}_{VSVVV} + \ov{V}_{VVVSS} + \ov{V}_{VVVVV}\,.
\end{align}
The topologies of type ``$M$'' are:
{\allowdisplaybreaks
\begin{subequations}
\begin{align}
\ov{M}_{SVSVV} &= \cpl{SSS}{\ind{1}, \ind{3}, \ind{6}, 1}\,
    \cpl{SVV}{\ind{2}, \ind{4}, \ind{7}, 1}\,
    \cpl{SVV}{\ind{3}, \ind{4}, \ind{5}, 1}\,
    \cpl{SVV}{\ind{6}, \ind{5}, \ind{7}, 1}\,\times\nn\\*
  &\quad \left\{ 2\, \f{50}{}{\ind{3}, \ind{4}, \ind{5}, \ind{6}, \ind{7}}
  - \frac{1}{2}\, \f{25}{}{\ind{3}, \ind{4}, \ind{5}, \ind{6}, \ind{7}}
  \!\right\},\\
\ov{M}_{VVSVS} &= \cpl{SSV}{\ind{1}, \ind{3}, \ind{6}, 1}\,
    \cpl{SSV}{\ind{3}, \ind{5}, \ind{4}, 1}\,
    \cpl{SVV}{\ind{2}, \ind{4}, \ind{7}, 1}\,
    \cpl{SVV}{\ind{5}, \ind{6}, \ind{7}, 1}\,\times \nn\\*
  &\quad\, \Big\{ 2\, \f{47}{}{\ind{3}, \ind{4}, \ind{5}, \ind{6}, \ind{7}}
  - 2\, \f{14}{}{\ind{7}, \ind{5}, \ind{6}, \ind{4}} \!\Big\},\\
\ov{M}_{VVSVV} &= \cpl{SSV}{\ind{1}, \ind{3}, \ind{6}, 1}\,
    \cpl{SVV}{\ind{2}, \ind{4}, \ind{7}, 1}\,
    \cpl{SVV}{\ind{3}, \ind{4}, \ind{5}, 1}\,
    \cpl{VVV}{\ind{5}, \ind{6}, \ind{7}, 1}\,\times \nn\\*
  &\quad \left\{ 2\, \f{54}{}{\ind{3}, \ind{4}, \ind{5}, \ind{6}, \ind{7}}
  - \frac{1}{2}\f{34}{1,1,1,1}{\ind{3}, \ind{4}, \ind{5}, \ind{6}, \ind{7}}
  + \frac{1}{2}\f{34}{1,1,1,2}{\ind{3}, \ind{4}, \ind{5}, \ind{6}, \ind{7}} \!\right\},\\
\ov{M}_{VVVVV} &= \cpl{SVV}{\ind{1}, \ind{3}, \ind{6}, 1}\,
    \cpl{SVV}{\ind{2}, \ind{4}, \ind{7}, 1}\,
    \cpl{VVV}{\ind{3}, \ind{4}, \ind{5}, 1}\,
    \cpl{VVV}{\ind{5}, \ind{6}, \ind{7}, 1}\,\times \nn\\*
  &\quad\, \begin{aligned}[b] \bigg\{
    & \f{56}{}{\ind{3}, \ind{4}, \ind{5}, \ind{6}, \ind{7}}
      + \frac{1}{8} \left(\m_{\ind{1}}^2 + \m_{\ind{2}}^2\right)
        \f{25}{}{\ind{3}, \ind{4}, \ind{5}, \ind{6}, \ind{7}}\\
    &{-}\, 2\, \f{109}{1,1,1}{\ind{3}, \ind{6}, \ind{4}, \ind{7}}
      + 2\, \f{109}{1,1,3}{\ind{3}, \ind{6}, \ind{4}, \ind{7}}
      + \frac{1}{8}\, \m_{\ind{5}}^2\, \f{27}{}{\ind{3}, \ind{4}, \ind{5}, \ind{6}, \ind{7}}\\
    &{-}\, \frac{1}{16} \begin{aligned}[t] \Big[
      & \f{34}{1,1,1,1}{\ind{3}, \ind{4}, \ind{5}, \ind{6}, \ind{7}}
       + \f{34}{1,1,1,2}{\ind{3}, \ind{4}, \ind{5}, \ind{6}, \ind{7}}
       + \f{36}{}{\ind{3}, \ind{4}, \ind{5}, \ind{6}, \ind{7}}\\
      &{+}\, \f{34}{1,1,1,1}{\ind{4}, \ind{3}, \ind{5}, \ind{7}, \ind{6}}
       + \f{34}{1,1,1,2}{\ind{4}, \ind{3}, \ind{5}, \ind{7}, \ind{6}}
       + \f{36}{}{\ind{4}, \ind{3}, \ind{5}, \ind{7}, \ind{6}}\\
      &{+}\, \f{34}{1,1,1,1}{\ind{6}, \ind{7}, \ind{5}, \ind{3}, \ind{4}}
       + \f{34}{1,1,1,2}{\ind{6}, \ind{7}, \ind{5}, \ind{3}, \ind{4}}
       + \f{36}{}{\ind{6}, \ind{7}, \ind{5}, \ind{3}, \ind{4}}\\
      &{+}\, \f{34}{1,1,1,1}{\ind{7}, \ind{6}, \ind{5}, \ind{4}, \ind{3}}
       + \f{34}{1,1,1,2}{\ind{7}, \ind{6}, \ind{5}, \ind{4}, \ind{3}}
       + \f{36}{}{\ind{7}, \ind{6}, \ind{5}, \ind{4}, \ind{3}}
       \Big]
     \end{aligned}\\
    &{+}\, \frac{1}{4} \Big[
      \f{45}{}{\ind{3}, \ind{4}, \ind{5}, \ind{6}, \ind{7}}
      + \f{45}{}{\ind{4}, \ind{3}, \ind{5}, \ind{7}, \ind{6}}
      + \f{45}{}{\ind{6}, \ind{7}, \ind{5}, \ind{3}, \ind{4}}
      + \f{45}{}{\ind{7}, \ind{6}, \ind{5}, \ind{4}, \ind{3}} \Big]
    \!\bigg\}\,.
   \end{aligned}
 \end{align}
\end{subequations}
}%
The last term in the above appears fearsome, but actually its finite
part simplifies dramatically:
\begin{align}
\fin{\ov{M}_{VVVVV}} &= \cpl{SVV}{\ind{1}, \ind{3}, \ind{6}, 1}\,
                      \cpl{SVV}{\ind{2}, \ind{4}, \ind{7}, 1}\,
                      \cpl{VVV}{\ind{3}, \ind{4}, \ind{5}, 1}\,
                      \cpl{VVV}{\ind{5}, \ind{6}, \ind{7}, 1}\,\times \nn\\
 &\quad \begin{aligned}[b] \,\bigg\{
   & \deltaMS \left[5 - \frac{9}{4}\, B(\m_{\ind{3}}^2,\m_{\ind{6}}^2)
     - \frac{9}{4}\, B(\m_{\ind{4}}^2,\m_{\ind{7}}^2)
     + \frac{15}{8}\, B(\m_{\ind{3}}^2,\m_{\ind{6}}^2)\,
       B(\m_{\ind{4}}^2,\m_{\ind{7}}^2)\right]\\
   &{+}\, \frac{1}{16} \left[
 34\, \m_{\ind{3}}^2 + 34\, \m_{\ind{4}}^2 + 34\, \m_{\ind{6}}^2  + 34\, \m_{\ind{7}}^2
 + 65\, \m_{\ind{5}}^2 - 66\, p^2 - \m_{\ind{1}}^2 - \m_{\ind{2}}^2
     \right]\times\\
   &\quad M(\m_{\ind{3}}^2,\m_{\ind{7}}^2,\m_{\ind{6}}^2,\m_{\ind{4}}^2,\m_{\ind{5}}^2)\\
   &{-}\,\frac{17}{8}\,U(\m_{\ind{3}}^2,\m_{\ind{6}}^2,\m_{\ind{5}}^2,\m_{\ind{7}}^2)
     - \frac{17}{8}\,U(\m_{\ind{4}}^2,\m_{\ind{7}}^2,\m_{\ind{5}}^2,\m_{\ind{6}}^2)\\
   &{-}\,\frac{17}{8}\,U(\m_{\ind{6}}^2,\m_{\ind{3}}^2,\m_{\ind{4}}^2,\m_{\ind{5}}^2)
     - \frac{17}{8}\,U(\m_{\ind{7}}^2,\m_{\ind{4}}^2,\m_{\ind{3}}^2,\m_{\ind{5}}^2)
  \!\bigg\}\,.
  \end{aligned}
\end{align}
This expression is untroubled if any masses are coincident or
vanishing.

The topologies of type ``$V$'' are:
\begin{subequations}\begin{align}
\ov{V}_{SVVSV} &= \cpl{SSV}{\ind{1}, \ind{3}, \ind{4}, 1}\,
    \cpl{SSV}{\ind{2}, \ind{3}, \ind{7}, 1}\,
    \cpl{SVV}{\ind{5}, \ind{4}, \ind{6}, 1}\,
    \cpl{SVV}{\ind{5}, \ind{6}, \ind{7}, 1}\,\times\nn\\*
  &\quad\, \Big\{ \f{91}{}{\ind{3}, \ind{4}, \ind{5}, \ind{6}, \ind{7}}
            + 2\, \f{118}{}{\ind{5}, \ind{6}, \ind{4}, \ind{7}} \!\Big\}\,,\\
\ov{V}_{VSSVV} &= \cpl{SSV}{\ind{1}, \ind{4}, \ind{3}, 1}\,
    \cpl{SSV}{\ind{2}, \ind{7}, \ind{3}, 1}\,
    \cpl{SVV}{\ind{4}, \ind{5}, \ind{6}, 1}\,
    \cpl{SVV}{\ind{7}, \ind{5}, \ind{6}, 1}\,\times\nn\\*
  &\quad \left\{ \f{92}{}{\ind{3}, \ind{4}, \ind{5}, \ind{6}, \ind{7}}
  + \frac{1}{2}\, \f{70}{}{\ind{3}, \ind{4}, \ind{5}, \ind{6}, \ind{7}}
  \!\right\},\\
\ov{V}_{VSVSV} &= \cpl{SSV}{\ind{1}, \ind{4}, \ind{3}, 1}\,
    \cpl{SSV}{\ind{4}, \ind{5}, \ind{6}, 1}\,
    \cpl{SVV}{\ind{2}, \ind{3}, \ind{7}, 1}\,
    \cpl{SVV}{\ind{5}, \ind{6}, \ind{7}, 1}\,\times\nn\\*
  &\quad\, \Big\{ 2\,\f{89}{}{\ind{3}, \ind{4}, \ind{5}, \ind{6}, \ind{7}}
  - 2\, \f{14}{}{\ind{7}, \ind{5}, \ind{6}, \ind{3}} \!\Big\}\,,\\
\ov{V}_{VSVVV} &= \cpl{SSV}{\ind{1}, \ind{4}, \ind{3}, 1}\,
    \cpl{SVV}{\ind{2}, \ind{3}, \ind{7}, 1}\,
    \cpl{SVV}{\ind{4}, \ind{5}, \ind{6}, 1}\,
    \cpl{VVV}{\ind{5}, \ind{6}, \ind{7}, 1}\,\times\nn\\*
  &\quad \left\{ 2\,\f{97}{}{\ind{3}, \ind{4}, \ind{5}, \ind{6}, \ind{7}}
      + \frac{1}{2}\, \f{84}{}{\ind{3}, \ind{4}, \ind{5}, \ind{6}, \ind{7}}
      - \frac{1}{2}\, \f{84}{}{\ind{3}, \ind{4}, \ind{6}, \ind{5}, \ind{7}}
    \!\right\}.\\
\ov{V}_{VVVSS} &= \cpl{SSV}{\ind{5}, \ind{6}, \ind{4}, 1}\,
    \cpl{SSV}{\ind{5}, \ind{6}, \ind{7}, 1}\,
    \cpl{SVV}{\ind{1}, \ind{3}, \ind{4}, 1}\,
    \cpl{SVV}{\ind{2}, \ind{3}, \ind{7}, 1}\,\times\nn\\*
  &\quad\, \Big\{ \f{88}{}{\ind{3}, \ind{4}, \ind{5}, \ind{6}, \ind{7}}
                 + \f{7}{}{\ind{3}, \ind{4}, \ind{7}, \ind{5}}
                 + \f{7}{}{\ind{3}, \ind{4}, \ind{7}, \ind{6}} \!\Big\}\,,\\
V_{VVVVV} &= \cpl{SVV}{\ind{1}, \ind{3}, \ind{4}, 1}\,
    \cpl{SVV}{\ind{2}, \ind{3}, \ind{7}, 1}\,
    \cpl{VVV}{\ind{4}, \ind{5}, \ind{6}, 1}\,
    \cpl{VVV}{\ind{5}, \ind{6}, \ind{7}, 1}\,\times\nn\\*
  &\quad\, \begin{aligned}[b] \bigg\{
    & \f{100}{}{\ind{3}, \ind{4}, \ind{5},\ind{6}, \ind{7}}
      - 2\, \f{95}{}{\ind{3}, \ind{4}, \ind{5},\ind{6}, \ind{7}}
      - \frac{1}{8}\, \f{70}{}{\ind{3}, \ind{4}, \ind{5},\ind{6}, \ind{7}}\\
    &{-}\, \f{8}{1,1,1}{\ind{3}, \ind{4}, \ind{7}, \ind{5}}
      - \f{8}{1,1,1}{\ind{3}, \ind{4}, \ind{7}, \ind{6}}
      + \f{8}{1,1,3}{\ind{3}, \ind{4}, \ind{7}, \ind{5}}
      + \f{8}{1,1,3}{\ind{3}, \ind{4}, \ind{7}, \ind{6}}\\
    &{-}\, \frac{\m_{\ind{3}}^2}{8} \left(
      1 - \frac{\m_{\ind{1}}^2}{\m_{\ind{4}}^2} - \frac{\m_{\ind{2}}^2}{\m_{\ind{7}}^2}
      \right) \f{60}{}{\ind{3}, \ind{4}, \ind{5},\ind{6}, \ind{7}}
      + \frac{1}{16} \left(\m_{\ind{5}}^2 + \m_{\ind{6}}^2\right)
        \f{61}{}{\ind{3}, \ind{4}, \ind{5},\ind{6}, \ind{7}}\\
    &{-}\, \frac{1}{8} \Big[
        \f{71}{}{\ind{3}, \ind{4}, \ind{5},\ind{6}, \ind{7}}
        + \f{71}{}{\ind{3}, \ind{4}, \ind{6},\ind{5}, \ind{7}}\Big]\\
    &{+}\, \frac{1}{4} \Big[
        \f{84}{}{\ind{3}, \ind{4}, \ind{5},\ind{6}, \ind{7}}
        + \f{84}{}{\ind{3}, \ind{4}, \ind{6},\ind{5}, \ind{7}}
        - \f{73}{}{\ind{3}, \ind{4}, \ind{5},\ind{6}, \ind{7}}
        - \f{73}{}{\ind{3}, \ind{4}, \ind{6},\ind{5}, \ind{7}}
    \Big]\!\bigg\}\,.
  \end{aligned}
\end{align}
\end{subequations}
For the last expression, the factors of $\m_{\ind{4}}^2$ and
$\m_{\ind{7}}^2$ in the denominators may at first appear to lead to
divergences if the corresponding gauge bosons are massless. However,
it is straightforward to show that, in that case and for
$\ind{1}, \ind{2}$ \emph{not} would-be Goldstone bosons, the
$\cpl{SVV}{\ind{1}, \ind{3}, \ind{4}, 1}$,
$\cpl{SVV}{\ind{2}, \ind{3}, \ind{7}, 1}$ vertices also vanish. This
expression simplifies to a relatively short form similarly to
$\ov{M}_{VVVVV}$, except that the reduction is different for
$\m_{\ind{4}}^2 = \m_{\ind{7}}^2$ as compared to the non-degenerate
case.

\tocsubsubsection[]{Reducible diagrams containing scalars and vectors of \texorpdfstring{$\mathcal{O}(g^8)$}{\unicodescriptO(\unicodeitalicg\unicodesupeight)}}

The nominally highest-order diagrams in the gauge coupling consist of
just two topologies, reduced from the original four:
\begin{align}
  \ov{\Pi}^{g^8} &= \ov{M}_{VVVVS} + \ov{V}_{VVVSV}\,.
\end{align}

The expressions for these are:
\begin{subequations}
\begin{align}
\ov{M}_{VVVVS} &= \cpl{SVV}{\ind{1}, \ind{3}, \ind{6}, 1}\,
    \cpl{SVV}{\ind{2}, \ind{4}, \ind{7}, 1}\,
    \cpl{SVV}{\ind{5}, \ind{3}, \ind{4}, 1}\,
    \cpl{SVV}{\ind{5}, \ind{6}, \ind{7}, 1}\,\times\nn\\
  &\quad\, \bigg\{ \f{53}{}{\ind{3}, \ind{4}, \ind{5},\ind{6}, \ind{7}}
   + \frac{1}{8}\, \f{27}{}{\ind{3}, \ind{4}, \ind{5},\ind{6}, \ind{7}}
   \!\bigg\}\,,\\
\ov{V}_{VVVSV} &= \cpl{SVV}{\ind{1}, \ind{3}, \ind{4}, 1}\,
    \cpl{SVV}{\ind{2}, \ind{3}, \ind{7}, 1}\,
    \cpl{SVV}{\ind{5}, \ind{4}, \ind{6}, 1}\,
    \cpl{SVV}{\ind{5}, \ind{6}, \ind{7}, 1}\,\times\nn\\
  &\quad\, \bigg\{ \f{96}{}{\ind{3}, \ind{4}, \ind{5},\ind{6}, \ind{7}}
   + \frac{1}{8}\, \f{61}{}{\ind{3}, \ind{4}, \ind{5},\ind{6}, \ind{7}}
   \!\bigg\}\,.
\end{align}
\end{subequations}

\tocsubsubsection[]{Special topologies: \texorpdfstring{$\Pi^{\S, g^4}$}{\unicodePi\unicodecircumflex\S,\unicodeitalicg\unicodesupfour}}

Recall that $ \Pi^{\S, g^4} = \lbls{107} + \lbls{112}$. These two
diagrams contain reducible couplings; the first has a four-point
scalar--vector coupling, and the second contains ghosts. However,
when we remove the four-point vector/scalar interactions and the
ghosts we find topologies that are either \emph{not 1PI} or
contain \emph{internal} propagators. The first of these is \lbls{107}:
\begin{align}
\lbls{107} &= \cpl{SSV}{\ind{1}, \ind{3}, \ind{4}, 1}\,
    \cpl{SSV}{\ind{2}, \ind{5}, \ind{6}, 1}\,
    \cpl{SSVV}{\ind{3}, \ind{5}, \ind{4}, \ind{6}, 1}\,
    \f{107}{}{\ind{3}, \ind{4}, \ind{5},\ind{6}} \nn\\
&= -\cpl{SSV}{\ind{1}, \ind{3}, \ind{6}, 1}\,
    \cpl{SSV}{\ind{2}, \ind{7}, \ind{4}, 1}\,
    \cpl{SSV}{\ind{3}, \ind{5}, \ind{4}, 1}\,
    \cpl{SSV}{\ind{5}, \ind{7}, \ind{6}, 1}\,
    \f{107}{}{\ind{3}, \ind{6}, \ind{7},\ind{4}} \nn\\
&\quad+ \cpl{SSV}{\ind{1}, \ind{3}, \ind{4}, 1}\,
    \cpl{SSV}{\ind{2}, \ind{5}, \ind{6}, 1}\,
    \cpl{SSV}{\ind{3}, \ind{7}, \ind{4}, 1}\,
    \cpl{SSV}{\ind{5}, \ind{7}, \ind{6}, 1}\,
    \f{107}{}{\ind{3}, \ind{4}, \ind{5},\ind{6}}\,.
\end{align}
The first diagram contributes to topology $M_{VSSVS}$ or \lbls{041},
while the second is a \emph{non-1PI diagram}; schematically:
\begin{center}
\begin{tabular}{@{}*4{c@{\ }}c}
\begin{minipage}{0.28\textwidth}
\includegraphics[width=\textwidth]{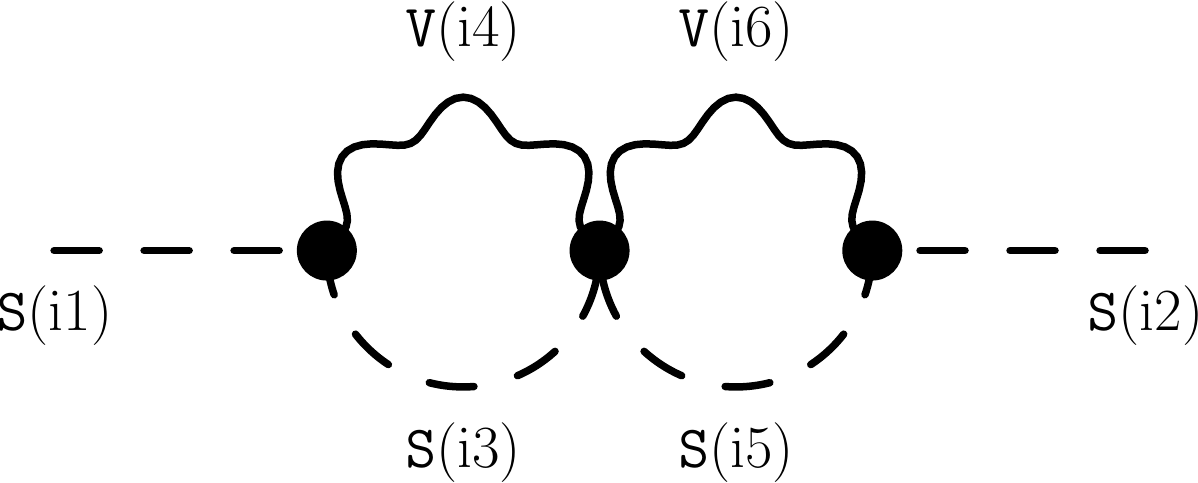}
\end{minipage}
&\begin{minipage}{0.03\textwidth}
$\longrightarrow$
\end{minipage}
&\begin{minipage}{0.28\textwidth}
\includegraphics[width=\textwidth]{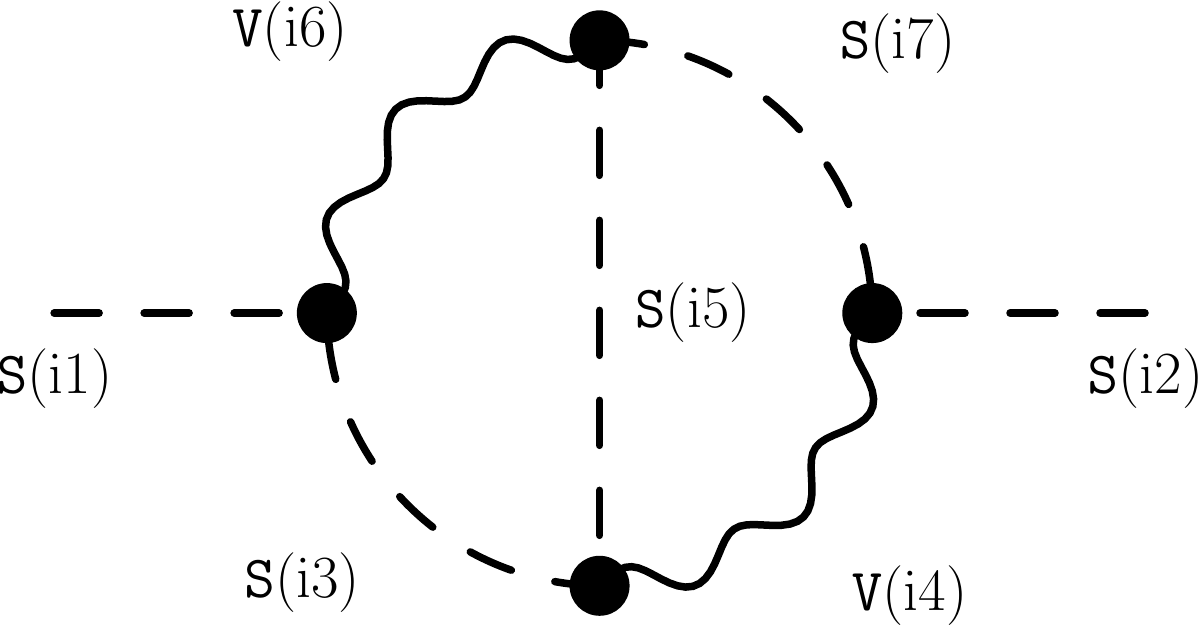}
\end{minipage}
&\begin{minipage}{0.02\textwidth}
$+$
\end{minipage}
&\begin{minipage}{0.32\textwidth}
\includegraphics[width=\textwidth]{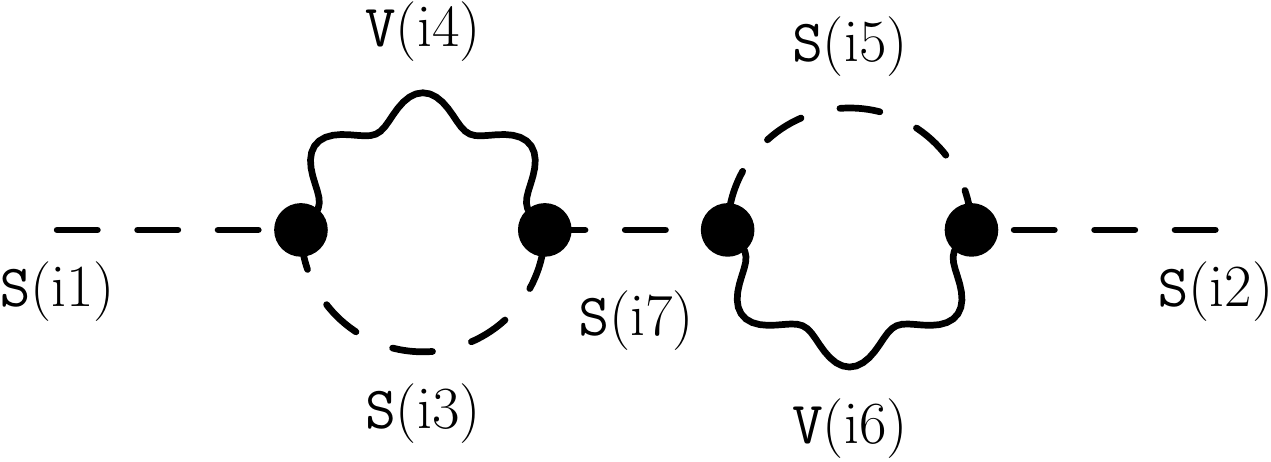}
\end{minipage} \\
$Z_{VSVS}$ & & $M_{VSSVS}$ & 
\end{tabular}
\end{center}

\bigskip

The second interesting topology is \lbls{112}:
\begin{align}
\lbls{112} &= \cpl{SSSS}{\ind{1}, \ind{2}, \ind{5}, \ind{6}, 1}\,
    \cpl{SUU}{\ind{5}, -\ind{4}, \ind{3}, 1}\,
    \cpl{SUU}{\ind{6}, -\ind{3}, \ind{4}, 1}\,
    \f{112}{}{\ind{3}, \ind{4}, \ind{5},\ind{6}} \nn\\
&= \frac{1}{2} \left(\m_{\ind{1}}^2 + \m_{\ind{2}}^2 - 2 \m_{\ind{3}}^2\right)
    \cpl{SSV}{\ind{1}, \ind{3}, \ind{4}, 1}\,
    \cpl{SSV}{\ind{2}, \ind{3}, \ind{7}, 1}\,
    \cpl{VVV}{\ind{4}, \ind{5}, \ind{6}, 1}\,
    \cpl{VVV}{\ind{5}, \ind{6}, \ind{7}, 1}\,
    \f{112}{}{\ind{5}, \ind{6}, \ind{4},\ind{7}} \nn\\
&\quad - \frac{1}{8}\,\cpl{SVV}{\ind{1}, \ind{3}, \ind{4}, 1}\,
    \cpl{SVV}{\ind{2}, \ind{3}, \ind{7}, 1}\,
    \cpl{VVV}{\ind{4}, \ind{5}, \ind{6}, 1}\,
    \cpl{VVV}{\ind{5}, \ind{6}, \ind{7}, 1}\,
    \f{112}{}{\ind{5}, \ind{6}, \ind{4},\ind{7}} \nn\\
&\quad + \frac{1}{2}\,\cpl{SSSS}{\ind{1}, \ind{2}, \ind{5}, \ind{6}, 1}\,
    \cpl{SVV}{\ind{5}, \ind{3}, \ind{4}, 1}\,
    \cpl{SVV}{\ind{6}, \ind{3}, \ind{4}, 1}\,
    \f{112}{}{\ind{5}, \ind{6}, \ind{4},\ind{7}} \nn\\
&\quad - \frac{1}{4}\, \cpl{SSS}{\ind{1}, \ind{2}, \ind{7}, 1}\,
    \cpl{SVV}{\ind{7}, \ind{5}, \ind{6}, 1}\,
    \cpl{VVV}{\ind{3}, \ind{4}, \ind{5}, 1}\,
    \cpl{VVV}{\ind{3}, \ind{4}, \ind{6}, 1}\,
    \f{112}{}{\ind{3}, \ind{4}, \ind{5},\ind{6}}\,.
\label{EQ:Reduce112}
\end{align}
Here we have used \refeq{EQ:M0trick} and substituted \refeq{EQ:GSSalt}
into \refeq{EQ:GSSS}. The topologies on the right-hand side correspond
to \lbls{099} ($V_{SVVVV}$), \lbls{100} ($V_{VVVVV}$), \lbls{116}
($W_{SSVV}$) and an extra diagram, schematically:
\begin{center}
\begin{tabular}{@{}*4{c@{\ }}c}
\begin{minipage}{0.28\textwidth}
\includegraphics[width=\textwidth]{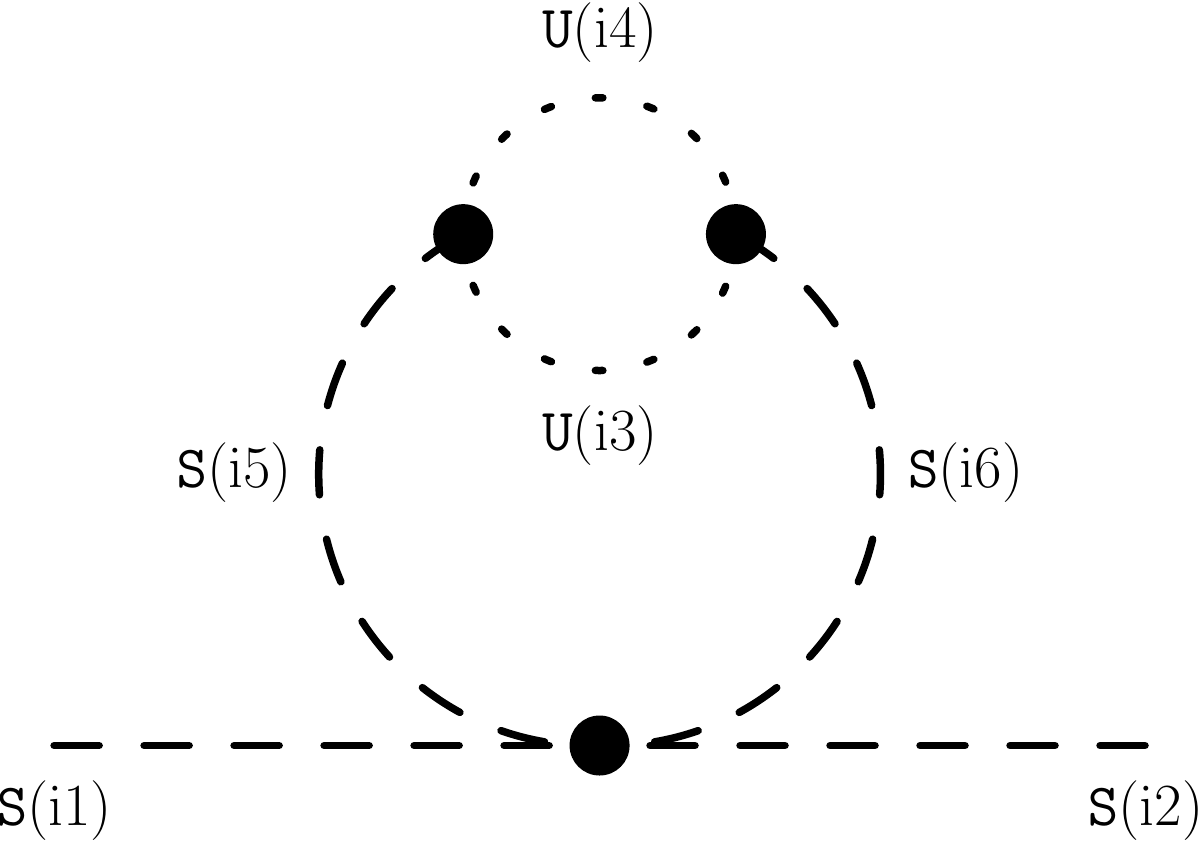}
\end{minipage}
&\begin{minipage}{0.03\textwidth}
$\longrightarrow$
\end{minipage}
&\begin{minipage}{0.28\textwidth}
\includegraphics[width=\textwidth]{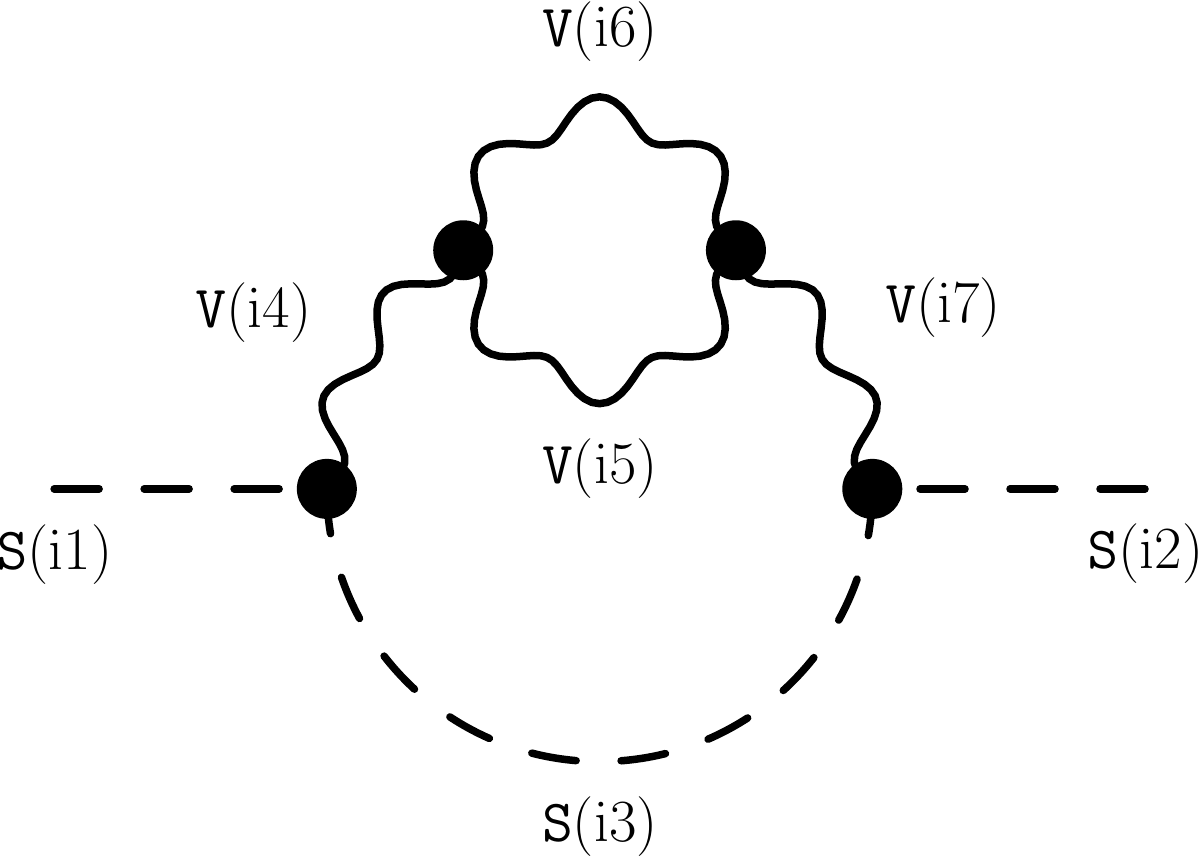}
\end{minipage}
&\begin{minipage}{0.02\textwidth}
$+$
\end{minipage}
& \begin{minipage}{0.28\textwidth}
\includegraphics[width=\textwidth]{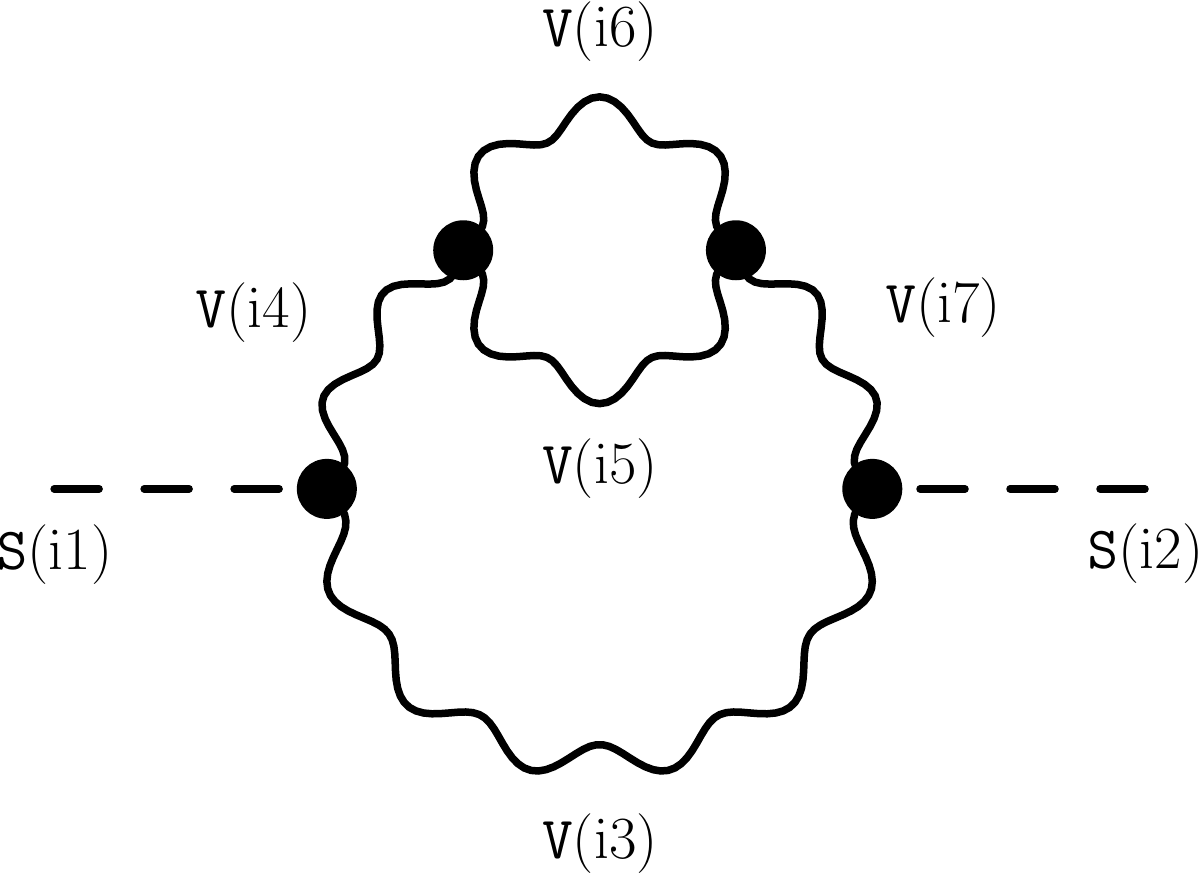}
\end{minipage} \\\quad\\
& \begin{minipage}{0.03\textwidth}
$+$\end{minipage}
&\begin{minipage}{0.28\textwidth}
\includegraphics[width=\textwidth]{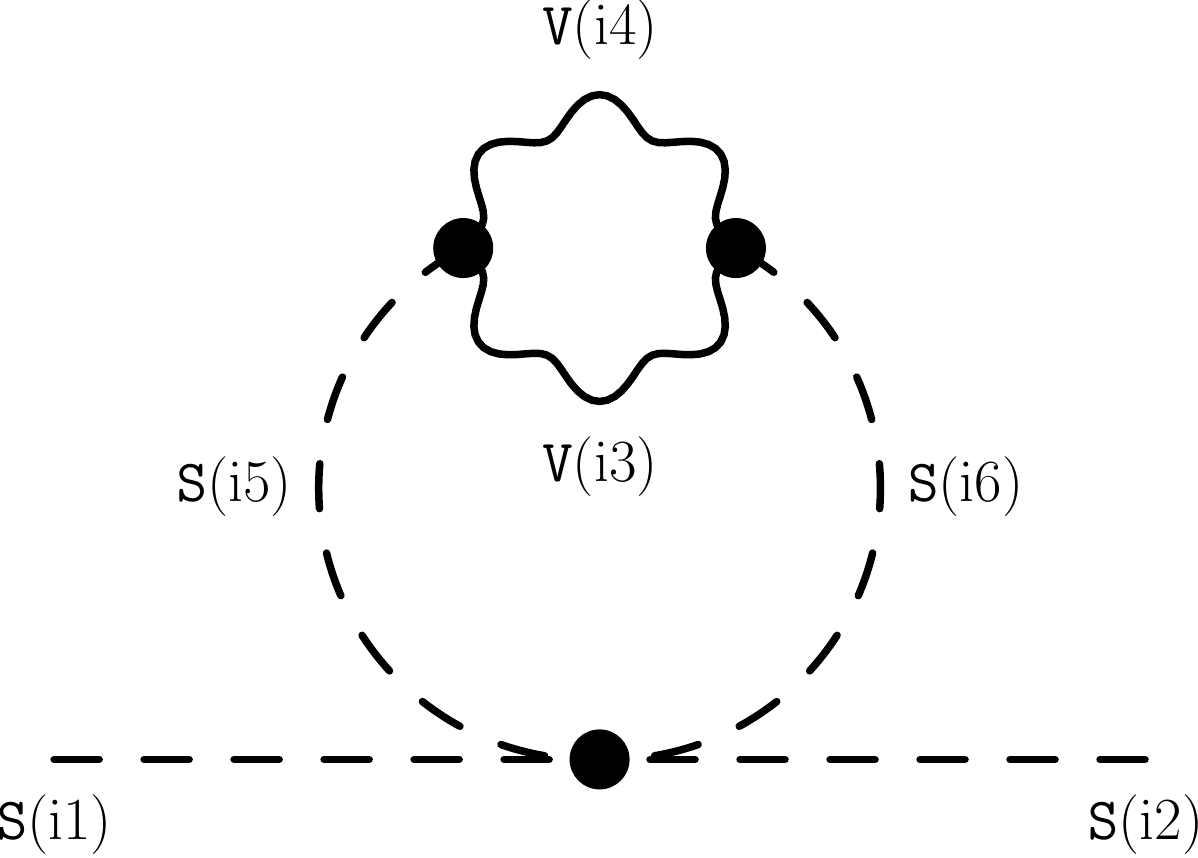}
\end{minipage}
&\begin{minipage}{0.02\textwidth}
$+$
\end{minipage}
& \begin{minipage}{0.18\textwidth}
\includegraphics[width=\textwidth]{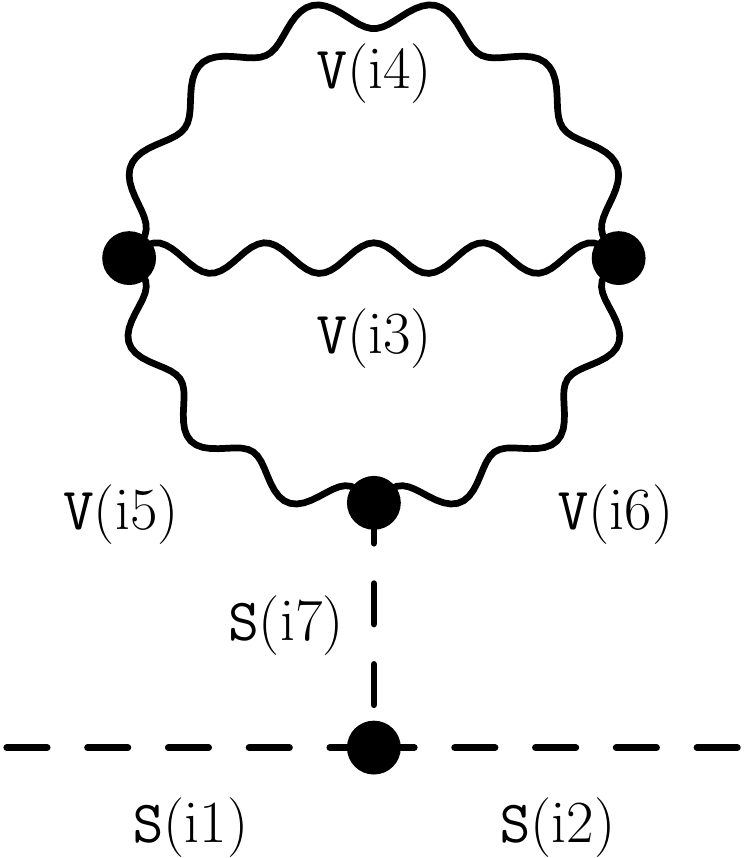}
\end{minipage}
\end{tabular}
\end{center}

There is therefore some ambiguity in how to treat these two special
classes. In the first case, it is probably simpler to retain a
definition for the four-point scalar--vector interactions and work
with \lbls{107} rather than reducing it. For the second case, the
simpler approach depends on the treatment of tadpole diagrams: if we
use a standard approach and do not include ``internal'' propagators,
then it may be easier to retain ghost couplings/propagators for just
this one class and work with \lbls{112}. On the other hand, if we
treat tadpoles by including internal propagators, then all of the
topologies on the right hand side of the above diagram already exist
as sets of couplings that we should evaluate. In this case it must be
simpler/faster to work with the right-hand side
of \refeq{EQ:Reduce112}.


\tocsubsection[\label{SEC:LG}]{Comparison with Landau gauge}

To make the connection with the results
of \citeres{Goodsell:2015ira,Braathen:2016cqe} completely explicit, we
shall present here the expressions for the tadpoles in the Landau
gauge and make the connection with our diagrams in the appendix.

The set of diagrams that survive in the Landau gauge are
\begin{align}
  T_{i}^{(2),\xi=0} &= \sum_{n=0}^3 T^{(2,n)}_{\xi=0},
\end{align}
where the superscript $(2,n)$ indicates $\mathcal{O} (2n)$ in the
gauge couplings. Then we have
\begin{subequations}
\begin{flushright}\begin{tabular}{r@{\ }*5{c@{\ }>{$}c<{$}@{\ }}c>{\raggedleft\arraybackslash}p{.9cm}@{}}
\cmidrule{1-12}\morecmidrules\cmidrule{1-12}
$T^{(2,0)}_{\xi=0} =$& \minitab[c]{$T_{SS}$\\\lblt{01}}
&+& \minitab[c]{$T_{FFFS}$\\\lblt{05}} &+& \minitab[c]{$T_{SSFF}$\\\lblt{06}}
&+& \minitab[c]{$T_{SSSS}$\\\lblt{07}} &+& \minitab[c]{$T_{SSS}$\\\lblt{24}}\smash{\,,}
&&& \hspace{-1em}\ \refstepcounter{equation}(\theequation)\\
\cmidrule{1-12}\morecmidrules\cmidrule{1-12}
$T^{(2,1)}_{\xi=0} =$ & \minitab[c]{$\TadL{SV}$\\\Lblt{02}}
&+& \minitab[c]{$\TadL{FFFV}$\\\Lblt{10}}
&+& \minitab[c]{$\TadL{SSSV}$\\\Lblt{13}}\smash{\,,}
&&&&&&& \hspace{-1em}\ \refstepcounter{equation}(\theequation)\\
\cmidrule{1-12}\morecmidrules\cmidrule{1-12}
$T^{(2,2)}_{\xi=0} =$ & \minitab[c]{$\TadL{VS}$\\\Lblt{03}}
&+& \minitab[c]{$\TadL{VVFF}$\\\Lblt{16}}
&+& \minitab[c]{$\TadL{VVSS}$\\\Lblt{17}}
&+& \minitab[c]{$\TadL{SSVV}$\\\Lblt{19}}
&+& \minitab[c]{$\TadL{SVV}$\\\Lblt{25}}
&+& \minitab[c]{$\ov{T}_{VVVV}^{\xi=0}$\\\Lblt{23}, \Lblt{20}, \Lblt{04}}\,,
& \hspace{-1em}\ \refstepcounter{equation}(\theequation)\\ 
\cmidrule{1-12}\morecmidrules\cmidrule{1-12}
$T^{(2,3)}_{\xi=0} =$ & \minitab[c]{$\TadL{VVSV}$\\\Lblt{21}}\smash{\,,}
&&&&&&&&&&& \hspace{-1em}\ \refstepcounter{equation}(\theequation)\\
\cmidrule{1-12}\morecmidrules\cmidrule{1-12}
\end{tabular}\end{flushright}
\end{subequations}
where
\begin{align}
\ov{T}^{\xi=0}_{VVVV} &= \cpl{SVV}{\ind{1}, \ind{2}, \ind{5}, 1}\,
                      \cpl{VVV}{\ind{2}, \ind{3}, \ind{4}, 1}\,
                      \cpl{VVV}{\ind{3}, \ind{4}, \ind{5}, 1}\,\times\nn\\
                   &\quad \begin{aligned}[t] \,\Big[
                   & \tIL{23}{}{\ind{2}, \ind{3},\ind{4}, \ind{5}}
                     - 2\, \tIL{20}{}{\ind{2}, \ind{3},\ind{4}, \ind{5}}\\
                   &{-}\, \tIL{4}{1,1}{\ind{2},\ind{5},\ind{4}}
                     - \tIL{4}{1,1}{\ind{2},\ind{5},\ind{3}}
                     + \tIL{4}{1,3}{\ind{2},\ind{5},\ind{4}}
                     + \tIL{4}{1,3}{\ind{2},\ind{5},\ind{3}}
                   \Big]
                 \end{aligned}\nn\\
&\equiv \cpl{SVV}{\ind{1}, \ind{2}, \ind{5}, 1}\,
        \cpl{VVV}{\ind{2}, \ind{3}, \ind{4}, 1}\,
        \cpl{VVV}{\ind{3}, \ind{4}, \ind{5}, 1}\,
        \ov{t}_{VVVV}^{\xi=0}(\scalebox{.8}{$\ind{2},\ind{3},\ind{4},\ind{5}$})\,.
\end{align}
The topologies (and therefore labelling of the couplings) are the same
in Landau gauge as for our Feynman-gauge results in the appendix, but
we add the superscript $\xi=0$ here. We do not give the explicit
expressions for these, since several are rather long---they contain
many more integrals than the Feynman-gauge case---but they are
provided as part of \ourcode.

Ignoring for the moment cases where some masses vanish, and using the
loop functions defined in \citere{Martin:2001vx} and the notation
\begin{subequations}
\begin{align}
D_{x,y} [f(x)] &\equiv \frac{f(x) - f(y)}{x-y}\,,  \\
D_{x,x} [f(x)] &\equiv \lim_{y \rightarrow x} D_{x,y} [f(x)]\,,
\end{align}
\end{subequations}
we find the following expressions for the tadpoles:
\begin{center}
\begin{tabular}{|@{\ }>{$\vphantom{\Big(}$}c@{\,}|@{\ }l@{\,}|r@{\,}l|r@{\,}l|}
\hline
\minitab{Tadpole\\ topology} & \,Label & \multicolumn{2}{c|}{Finite Part} & \multicolumn{2}{c|}{Our function} \\
\hline\hline
\lblt{01} & $\TadL{SS}$
& $-\frac{1}{4}\, a^{(\ind{1})ij}\, \lambda^{ijkk}$ & $D_{\m_i^2, \m_j^2}{\left[f_{SS}(\m_i^2, \m_k^2)\right]} $
& $ a^{(\ind{1})ij}\, \lambda^{ijkk}$ & $\tI{1}{}{i, k, j}$\\
\arrayrulecolor{lightgray}\hline
\lblt{24} & $\TadL{SSS}$
& $-\frac{1}{6}\, \lambda^{(\ind{1})ijk}\, a^{ijk}$ & $f_{SSS}(\m_i^2,\m_j^2,\m_k^2)$
& $\lambda^{(\ind{1})ijk}\, a^{ijk}$ & $\tI{24}{}{i, j, k}$\\
\hline
\lblt{07} & $\TadL{SSSS}$
& $-\frac{1}{4}\, a^{(\ind{1})ij}\, a^{ikl}\, a^{jkl}$ & $D_{\m_i^2,\m_j^2}{\left[f_{SSS} (\m^2_i,\m^2_k,\m^2_l)\right]}$
& $a^{(\ind{1})ij}\, a^{ikl}\, a^{jkl}$ & $\tI{7}{}{i, k, l, j}$ \\
\hline
\lblt{02} & $\TadL{SV}$ & $-\frac{1}{4}\, a^{ik(\ind{1})}\, g^{aaik}$ & $D_{\m_i^2,\m_k^2}{\left[f_{VS} (\m^2_a, \m^2_i)\right]}$
& $-a^{ik(\ind{1})}\, g^{aaik}$ & $\tIL{2}{}{i, k, a}$\\
\hline
\lblt{03} & $\TadL{VS} $& $-\frac{1}{4}\, g^{abii}\, g^{ab(\ind{1})}$ & $D_{\m_a^2,\m_b^2}{\left[f_{VS}(\m^2_a, \m_i^2)\right]}$
& $g^{abii}\, g^{ab(\ind{1})}$ & $\tIL{3}{}{a, b, i} $\\
\hline
\lblt{13} & $\TadL{SSSV}$& $-\frac{1}{2}\, a^{ik(\ind{1})}\, g^{aij}\, g^{akj}$ & $D_{\m_i^2,\m_k^2}{\left[f_{SSV} (\m_i^2, \m_j^2, \m_a^2)\right]}$
& $a^{ik(\ind{1})}\, g^{aij}\, g^{akj}$ & $\tIL{13}{}{i, j, a, k} $\\
\hline
\lblt{17} & $\TadL{VVSS}$ & $-\frac{1}{4}\, g^{ab(\ind{1})}\, g^{aij}\, g^{bij}$ & $D_{\m_a^2,\m_b^2}{\left[f_{SSV} (\m_i^2, \m_j^2, \m_a^2)\right]}$
& $g^{ab(\ind{1})}\, g^{aij}\, g^{bij}$ & $\tIL{17}{}{a, i, j, b} $\\
\hline
\lblt{19} & $\TadL{SSVV}$ & $-\frac{1}{4}\, g^{abi}\, g^{abj}\, a^{ij(\ind{1})}$ & $D_{\m_i^2, \m_j^2}{\left[f_{VVS} (\m_a^2, \m_b^2,\m^2_i)\right]}$
& $ g^{abi}\, g^{abj}\, a^{ij(\ind{1})}$ & $\tIL{19}{}{i, a, b, j}$\\
\hline
\lblt{25} & $\TadL{SVV}$ & $-\frac{1}{2}\, g^{abi(\ind{1})}\, g^{abi}$ & $f_{VVS} (\m_a^2, \m_b^2, \m_i^2)$
& $g^{abi(\ind{1})}\, g^{abi}$ & $\tIL{25}{}{i, a, b}$\\
\hline
\lblt{21} & $\TadL{VVSV}$ & $-\frac{1}{2}\, g^{abi}\, g^{cbi}\, g^{ac(\ind{1})}$ & $D_{\m_a^2,\m_c^2}{\left[f_{VVS}( \m_a^2,  \m_b^2,\m_i^2)\right]} $
& $ -g^{abi}\, g^{cbi}\, g^{ac(\ind{1})}$ & $\tIL{21}{}{a, i, b, c}$\\
\hline
\lblt{23}, \lblt{20}, \lblt{04} & $\ov{T}_{VVVV}^{\xi=0} $ & $-\frac{1}{4}\, g^{abc}\, g^{dbc}\, g^{ad(\ind{1})}$ & $D_{\m_a^2,\m_d^2}{\left[f_{\rm gauge}(\m_a^2, \m_b^2,\m_c^2)\right]}$
& $ g^{abc}\, g^{dbc}\, g^{ad(\ind{1})}$ & $\ov{t}_{VVVV}^{\xi=0}(\scalebox{.8}{$a, b, c, d$}) $ \\
\arrayrulecolor{black}\hline\hline
\end{tabular}
\end{center}

The fermionic diagrams are:
{
\allowdisplaybreaks
\begin{subequations}
\begin{align}
  \fin{\TadL{FFFV}} &= 2\, g^{aJ}_{I}\, \ov{g}_{bJ}^{K}\, \Real{M_{KI'}\, y^{I'I(\ind{1})}}\, D_{\m_I^2, \m_K^2}{\left[f_{FFV} (\m^2_I, \m^2_J, \m_a^2)\right]} \nn\\
 &\quad + g^{aJ}_I\, g^{aJ'}_{I'}\, \Real{y^{II'(\ind{1})}\, M^*_{JJ'}}  \left[ f_{\ov{FF}V} (\m^2_I, \m^2_J, \m_a^2)
 + m_I^2\, D_{\m_I^2, \m_{I'}^2}{\left[f_{\ov{FF}V} (\m^2_I,  \m^2_J, \m_a^2)\right]} \right]\nn\\*
 &\quad + g^{aJ}_I\, g^{aJ'}_{I'}\, \Real{M^{IK'}\, M^{KI'}\, M^*_{JJ'}\, y_{KK'(\ind{1})}}\, D_{\m_I^2, \m_{I'}^2}{\left[f_{\ov{FF}V} (\m^2_I,  \m^2_J, \m_a^2)\right]}\,,\\
 \fin{\TadL{VVFF}} &= \frac{1}{2}\, g^{aJ}_I\, \ov{g}_{bJ}^I\, g^{ab(\ind{1})}\, D_{\m_a^2, \m_b^2}{\left[f_{FFV} (\m^2_I, \m^2_J, \m_a^2)\right]} \nn \\*
 &\quad + \frac{1}{2}\, g^{aJ}_I\, g^{bJ'}_{I'}\, g^{ab(\ind{1})}\, M^{II'}\, M^*_{JJ'}\, D_{\m_a^2, \m_b^2}{\left[f_{\ov{FF}V} (\m^2_I, \m^2_J, \m_a^2)\right]}\,. 
\end{align}
\end{subequations}
}%
In \citeres{Goodsell:2015ira,Braathen:2016cqe} the topology
$\TadL{SVV}$ was missing. Our new approach, however, pays immediate
dividends when we consider some masses to be vanishing. We find, for
example, that
\begin{subequations}
\begin{align}
  \tI{1}{}{0,k,0} &= 0\,,\\
  \fin{\tI{7}{}{0,k,l, 0}} &= -\frac{1}{4}\, R_{SS} (\m_k^2, \m_l^2)
    + \frac{1}{4}\, \binteps{\epsilon^1}{0,\m_k^2, \m_l^2}\,,  
  \label{EQ:EpsilonT7}
\end{align}
\end{subequations}
where $R_{SS} (x,y)$ is defined
in \citeres{Kumar:2016ltb,Braathen:2016cqe}; using the expressions in
the appendix we can find the finite part of all of the integrals even
in the case of vanishing masses. The second term on the right-hand
side of \refeq{EQ:EpsilonT7} clearly arises from an infra-red
divergence; such divergences are the counterparts of the expansions
found in \citere{Braathen:2016cqe}, where the equivalent
for \refeq{EQ:EpsilonT7} is
\begin{align}
  D_{\m_G^2,\m_G^2}{\left[f_{SSS} (\m^2_G,\m^2_k,\m^2_l)\right]} &=
    R_{SS}(\m_k^2, \m_l^2)
    - \binteps{\epsilon^0}{0,\m_k^2, \m_l^2}\, \log\frac{\m_G^2}{Q^2}
    + \mathcal{O}{\left( \m_G^2 \right)}
\end{align}
with the renormalisation scale $Q$. Instead of an expansion in the
mass of the Goldstone bosons, we use dimensional regularisation in
order to regularise the infra-red divergences. Then we should find
that the infra-red divergent parts in the two-loop calculation cancel
exactly against equivalent parts coming either from putting the
Goldstone-boson masses on-shell in the one-loop parts, or using
``consistent tadpole solutions'' (by using tree-level masses in the
one-loop calculation). We intend to elaborate further on this
connection elsewhere.

\tocsubsection[\label{SEC:SM}]{Standard Model calculation}

Our self-energies up to second order in the gauge coupling have been
compared with the expressions in \citere{Martin:2003it}, and, as we
have shown above, the tadpoles in the Landau gauge exactly match those
found in \citere{Goodsell:2015ira}. In order to explicitly cross-check
our Feynman-gauge result, and the combinations of classes, it would be
desirable to have models to compare to. However, as described in the
introduction, the only example where scalar electroweak corrections
have been computed is for the Higgs mass in the Standard Model. Since
the results are very long, these have been made public in the form of
computer
codes \SMH{}\,\cite{Martin:2014cxa}, \mr{}\,\cite{Kniehl:2015nwa,Kniehl:2016enc}
and \SMDR{}\,\cite{Martin:2019lqd}.

The result used in \mr has been computed including tadpoles via
internal propagators in a general gauge, and is excellent for a
numerical calculation. However, the expressions in \SMH are actually
written in a sufficiently convenient form to be translated into code
readable by \texttt{Mathematica}, and so we were able to use it for
an \emph{analytical} cross-check of our results. On the other hand,
the result there is in Landau gauge, and combines the tadpoles with
self-energy contributions via \refeq{EQ:MU2EQ}.

Therefore we have cross-checked our results against the Standard Model
by the following procedure:
\begin{itemize}
\item
We calculated the tadpoles and Higgs self-energies in the Landau gauge
using the \FeynArts model file for the~SM (with some modifications
that will be described elsewhere), and using \TwoCalc, \TARCER and our
own reduction to reduce the basis of integrals to those that can be
evaluated by \TSIL. We compared the analytic expressions with those
in \SMH \emph{and found perfect agreement in all terms}. For this
calculation we computed self-energies and counterterms separately,
rather than using the BPHZ method (in the SM all necessary
counterterms can be reduced to just evaluating two-point functions
since there are sufficiently many tree-level relationships among the
parameters).
\item
Using the same method, we performed the same calculation in Feynman
gauge. The results differ from \SMH only by the treatment of tadpole
diagrams once we take $p^2$ equal to the tree-level Higgs mass. This
leads to $899$ self-energy diagrams based on the initial $121$ generic
classes in \FeynArts, and $75$ one-loop self-energy diagrams with
one-loop counterterm insertions (plus $161$ tadpoles and $25$ one-loop
tadpoles with one-loop counterterm insertions).
\item
We performed the same calculation using only the topologies of our
$58$ combined classes with the combined loop functions as given in
this section. This yielded only $425$ self-energy diagrams (and, of
course, includes the counterterm contributions). Upon reducing to
the \TSIL basis we find \emph{exact agreement of all terms} with the
above ``brute force'' approach.
\end{itemize}

\tocsection[\label{SEC:CONCLUSIONS}]{Conclusions}

We have given expressions for scalar self-energies and tadpoles at the
complete two-loop order in Feynman gauge. With the aim of being
flexible, we have given the results for all renormalised component
diagrams, and also a much shorter version where the diagrams are
combined into just $16$ tadpole and $58$ self-energy topologies,
provided that the external scalars are not (would-be) Goldstone
bosons. The results are provided in the most compact analytical form
that we could give, but are also available electronically as a new
package \ourcode at \website (where more tools and calculations will
be added over time) so that they can be easily applied. We also
include online all of the replacement rules and degeneracies for each
diagram with routines for extracting the finite part.

This work closes a gap in the literature that has existed for at least
sixteen years. However, thanks to the technology and techniques that
we have developed, it is just the first step in a new program of
completing the electroweak corrections in generic theories. The
logical next step is an implementation in a general code such
as \SARAH so that they can be applied to any model automatically. The
following steps are:
\begin{enumerate}
\item
Gauge-boson and mixed gauge--scalar self-energies. In particular, the
latter are required for decay processes, while the former are
essential for the extraction of electroweak parameters.
\item
Threshold corrections to match general theories onto the Standard
Model via a pole-matching technique. Naively this requires matching
gauge-boson masses, but as shown in \citere{Braathen:2018htl} we only
need the derivatives of the scalar self-energies, and gauge threshold
corrections at zero Higgs expectation value.
\item
Muon decay in general theories. This is required for extraction of the
Higgs electroweak expectation value in fixed-order calculations. It is
a four-point fermion amplitude, so is combinatorially much more
complicated than two-point functions, but since it is at zero external
momentum we do not have any new problems related to $\gamma_5$, and
the loop functions all reduce to the same ones used in the tadpoles.
\item Fermion self-energies, in particular to calculate the top-quark mass. 
\end{enumerate}

We have applied our results to the Standard Model and compared to the
existing expressions in the literature, finding perfect
agreement. However, since the only analytically available calculation
was performed in Landau gauge, it was not possible to compare our
Feynman-gauge calculation with an independent check. However, by
combining the diagrams with all tadpole diagrams it should be possible
to give an explicitly gauge-independent result that could also be
compared with~\mr. Moreover, using the results
from \sect{SEC:GOLDSTONESGHOSTS} it should also be possible to
implicitly sum over all Goldstone-boson propagators in
our \emph{general} result, transforming their couplings and masses
into those of gauge bosons. We leave these developments to future
work.

\section*{\tocref{Acknowledgments}}

Both authors acknowledge support from the grant
\mbox{``HiggsAutomator''} of the Agence Nationale de la Recherche
(ANR) (ANR-15-CE31-0002), and the LABEX Institut Lagrange de Paris.  The work of SP was
supported in part by BMBF Grant No. 05H18PACC2.

\appendix
\section*{\tocref{Appendix}}

\tocsection[\label{SEC:REDUCTION}]{Integral relations}

In this appendix, we introduce the notation for the integrals that are
used in the list of results for tadpoles and self-energies in
\appx{SEC:DIAGLIST}. After mentioning some general relations among
these integrals, the reduction rules that are necessary in this
article are displayed. Finally, the UV-divergent parts of the basic
set of scalar integrals are given.

\tocsubsection[\label{SEC:INTEGRALNOTATION}]{Notation and symmetries}

We mostly follow the notation of \citere{Weiglein:1993hd},
\begin{subequations}
\begin{align}
   \tint{i_1\cdots i_n}{} &= \int \frac{\mathrm{d}^dq_1\,\mathrm{d}^dq_2}
   {\left[\imath\,\pi^2 \left(2\,\pi\,\mu\right)^{d - 4}\right]^2}\,
   \frac{1}{\left(k_{i_1}^2 - \m_{i_1}^2\right) \cdots
   \left(k_{i_n}^2 - \m_{i_n}^2\right)}\,,\\
   \yint{i_1\cdots i_n}{j_1\cdots j_o}{} &=
   \int \frac{\mathrm{d}^dq_1\,\mathrm{d}^dq_2}
   {\left[\imath\,\pi^2 \left(2\,\pi\,\mu\right)^{d - 4}\right]^2}\,
   \frac{k_{j_1}^2 \cdots k_{j_o}^2}{\left(k_{i_1}^2 - \m_{i_1}^2\right) \cdots
   \left(k_{i_n}^2 - \m_{i_n}^2\right)}\,,
\end{align}
\end{subequations}
with the indices~\mbox{$i_1,\dots, i_n, j_1,\dots, j_o \in \{1, 2, 3,
4, 5\}$}, the dimension~\mbox{$d=4-2\,\epsilon$} with the ultra-violet
regulator~$\epsilon$, the dimensional regulator~$\mu$, and the
kinematic variables
\begin{gather}
  k_1 = q_1\,,\quad k_2 = q_1 + p\,,\quad k_3 = q_2 - q_1\,,\quad
  k_4 = q_2\,,\quad k_5 = q_2 + p
\end{gather}
that depend on the loop momenta~$q_1, q_2$ and the external
momentum~\mbox{$k_6\equiv p$}. In addition, we use sub-indices~$a$
and~$b$ in order to distinguish denominators of the same kinematic
type, but with possibly different masses.

The set of different two-point integrals can be reduced by using shift
and mirror symmetries of the loop momenta
(see \citere{Weiglein:1993hd}). They can be applied by exchanging the
indices of the kinematic variables in the integrals in the following
ways,
\begin{subequations}\label{EQ:SYMMETRY}
\begin{align}
  1\leftrightarrow2 \text{ and } 4\leftrightarrow5\,, &&
  1\leftrightarrow4 \text{ and } 2\leftrightarrow5\,, &&
  1\leftrightarrow5 \text{ and } 2\leftrightarrow4\,,\\
  1\leftrightarrow3 \text{ if $2$ absent}\,, &&
  2\leftrightarrow3 \text{ if $1$ absent}\,, &&
  3\leftrightarrow4 \text{ if $5$ absent}\,, &&
  3\leftrightarrow5 \text{ if $4$ absent}\,,\\
  1\leftrightarrow2 \text{ if $3$ absent}\,, &&
  4\leftrightarrow5 \text{ if $3$ absent}\,.
\end{align}
\end{subequations}
We also introduce the following notation for one-loop integrals
appearing in counterterms:
\begin{subequations}\label{EQ:TYONELOOP}
\begin{align}
  \toneint{i_1\cdots i_n}{} &= \int \frac{\mathrm{d}^dq_1}
  {\imath\,\pi^2 \left(2\,\pi\,\mu\right)^{d - 4}}\,
  \frac{1}{\left(k_{i_1}^2 - \m_{i_1}^2\right) \cdots
  \left(k_{i_n}^2 - \m_{i_n}^2\right)}\,,\\
   \yoneint{i_1\cdots i_n}{j_1\cdots j_o}{} &=
   \int \frac{\mathrm{d}^dq_1}
   {\imath\,\pi^2 \left(2\,\pi\,\mu\right)^{d - 4}}\,
   \frac{k_{j_1}^2 \cdots k_{j_o}^2}{\left(k_{i_1}^2 - \m_{i_1}^2\right) \cdots
   \left(k_{i_n}^2 - \m_{i_n}^2\right)}\,,
\end{align}
\end{subequations}
with~$i_1,\dots, i_n, j_1,\dots, j_o \in \{1, 2\}$. For scalar
one-loop integrals with at most two propagators, we adapt the
well-known notation of Passarino and Veltman\,\cite{Passarino:1978jh}:
\begin{align}\label{EQ:PAVEONELOOP}
  \toneint{i_1}{} &= \aint{\m_{i_1}^2}\,,
  & \toneint{i_1i_1}{} &= \bint{0,\m_{i_1}^2,\m_{i_1}^2}\,,
  & \toneint{12}{} &= \bint{p^2,\m_1^2,\m_2^2}\,.
\end{align}
The reduction of one-loop integrals with more than two propagators
will be addressed together with the two-loop integral reduction
in \appx{SEC:SCALARREDUCTION}.

For convenience, we also introduce the following symbols in order to
denote a sum of one- or two-loop tensor integrals:
\begin{subequations}
\begin{align}
  \ysum{i_1\cdots i_n}{c_1[j_{11}\cdots j_{1o}]\, + \cdots
    +\, c_r[j_{r1}\cdots j_{ro}]}{} &=
    c_1\,\yint{i_1\cdots i_n}{j_{11}\cdots j_{1o}}{} + \cdots
    + c_r\,\yint{i_1\cdots i_n}{j_{r1}\cdots j_{ro}}{}\,,\\
  \yonesum{i_1\cdots i_n}{c_1[j_{11}\cdots j_{1o}]\, + \cdots
    +\, c_r[j_{r1}\cdots j_{ro}]}{} &=
    c_1\,\yoneint{i_1\cdots i_n}{j_{11}\cdots j_{1o}}{} + \cdots
    + c_r\,\yoneint{i_1\cdots i_n}{j_{r1}\cdots j_{ro}}{}
\end{align}
\end{subequations}
with integer coefficients~$c_1,\dots,c_r$ of the products of kinematic
variables in the numerators with indices~$j_{11},\dots,j_{ro}$.

\tocsubsection[\label{SEC:TENSORREDUCTION}]{Reduction of tensor integrals}

Tensor integrals that carry identical indices in the super- and
subscript can be reduced easily via
\begin{align}\label{EQ:YREDNUMDEN}
  \yint{i_1\cdots x\cdots i_n}{j_1\cdots x\cdots j_o}{} &=
    \yint{i_1\cdots i_n}{j_1\cdots j_o}{}
    + \m_x^2\,\yint{i_1\cdots x\cdots i_n}{j_1\cdots j_o}{}\,.
\end{align}
After the repeated application of \refeq{EQ:YREDNUMDEN} to a given
integral, the following cases appear: two loop integrals whose
integrand does not depend on both loop momenta are equal to zero; the
reduction of scalar integrals with a numerator equal to~$1$ is
described in \appx{SEC:SCALARREDUCTION}; the reduction of the
remaining tensor integrals is described \EG\
in \citeres{Weiglein:1993hd,Tarasov:1996br}. Reduction rules for the
integrals appearing in this article that are valid for all different
kinematical configurations (with the exception of vanishing external
momentum~$p$) are given in the following:
{
\allowdisplaybreaks
\begin{subequations}\label{EQ:YRED}
\begin{align}
  \yint{1_a1_b23}{4}{} &= \ysum{1_a1_b23}{[1]+[3]}{}\,,\\
  \yint{1_a1_b23}{5}{} &= \ysum{1_a1_b23}{[2]+[3]}{}\,,\\
  \yint{1_a1_b24}{5}{} &= \ysum{1_a1_b24}{[4]+[6]}{}\,,\\
  \yint{1_a1_b34}{2}{} &= \ysum{1_a1_b34}{[1]+[6]}{}\,,\\
  \yint{1_a1_b34}{5}{} &= \ysum{1_a1_b34}{[4]+[6]}{}\,,\\
  \label{EQ:3Y1245}
  \yint{1245}{3}{} &=
    \frac{1}{2} \left[
      \ysum{1245}{[1]+[2]+[4]+[5]-[6]}{}
      - \frac{1}{p^2}\,\ysum{1245}{([1]-[2])([4]-[5])}{}
    \right],\\
  \yint{1234}{5}{} &=
    -\frac{1}{2} \left[
      \ysum{1234}{[1]-[2]-[3]-[4]-[6]}{}
      + \ysum{11'234}{([2]-[6])([3]-[4])}{}
    \right],\\
  \yint{1_a1_b234}{5}{} &=
    -\frac{1}{2} \left[
      \ysum{1_a1_b234}{[1]-[2]-[3]-[4]-[6]}{}
      + \ysum{1_a1_b1'234}{([2]-[6])([3]-[4])}{}
    \right],\\
  \label{EQ:55Y11234}
  \begin{split}
  \yint{1_a1_b234}{55}{} &=
    \frac{1}{d - 1} \left[
      \ysum{1_a1_b234}{([2]-[3])([4]-[6])}{}
      - \ysum{1_a1_b1'234}{([2]-[3]+[4]-[6])([24]-[36])}{} \right]\\
    &\quad+ \frac{1}{4}\,\frac{d}{d - 1} \begin{aligned}[t] \,\Big[
      & \ysum{1_a1_b234}{([1]-[2]-[3]-[4]-[6])([1]-[2]-[3]-[4]-[6])}{}\\
      &{+}\, 2\,\ysum{1_a1_b1'234}{([2]-[6])([3]-[4])([1]-[2]-[3]-[4]-[6])}{}
        + \ysum{1_a1_b1'1'234}{([2]-[6])([2]-[6])([3]-[4])([3]-[4])}{}
      \Big]\,.
    \end{aligned}
  \end{split}
\end{align}
\end{subequations}
When applying \refeq{EQ:YREDNUMDEN} to \refeq{EQ:55Y11234} also the
following integrals appear after taking into account the symmetry
relations in \refeqs{EQ:SYMMETRY}:
\begin{subequations}
\begin{align}
  \yint{1_a1_b1'3}{2}{} &= \tint{1_a1_b3}{} + p^2\,\tint{1_a1_b1'3}{}\,,\\  
  \yint{1_a1_b1'3}{4}{} &= \tint{1_a1_b3}{} + \m_3^2\,\tint{1_a1_b1'3}{}\,,\\
  \yint{1_a1_b1'1'3}{4}{} &= \tint{1_a1_b1'3}{} + \m_3^2\,\tint{1_a1_b1'1'3}{}\,,\\
  \yint{1_a1_b1'1'23}{4}{} &= \tint{1_a1_b1'23}{} + \m_3^2\,\tint{1_a1_b1'1'23}{}\,,\\
  \yint{1_a1_b1'1'3}{24}{} &= \tint{1_a1_b3}{}
    + \left(\m_3^2 + p^2\right) \tint{1_a1_b1'3}{}
    + \m_3^2\,p^2\,\tint{1_a1_b1'1'3}{}\,.
\end{align}
\end{subequations}
}%

\tocsubsection[\label{SEC:SCALARREDUCTION}]{Reduction of scalar integrals}

\tocsubsubsection[\label{SEC:INTREL}]{General relations}

Our aim is to reduce all scalar two-loop integrals into the basis of
integrals that can be evaluated numerically by the
code \TSIL{}\,\cite{Martin:2005qm}. We begin by reducing to the
unrenormalised basis
\begin{gather}\label{EQ:TSILBASIS}
\begin{aligned}
  \tint{134}{} &= -\mathbf{I}{\left(\m_1^2,\m_3^2,\m_4^2\right)}\,,
  & \tint{1234}{} &= \mathbf{U}{\left(\m_2^2,\m_1^2,\m_3^2,\m_4^2\right)}\,,
  & \tint{11234}{} &= -\mathbf{V}{\left(\m_2^2,\m_1^2,\m_3^2,\m_4^2\right)}\,,\\
  \tint{234}{} &= -\mathbf{S}{\left(\m_2^2,\m_3^2,\m_4^2\right)}\,,
  & \tint{2234}{} &= \mathbf{T}{\left(\m_2^2,\m_3^2,\m_4^2\right)}\,,
  & \tint{12345}{} &= -\mathbf{M}{\left(\m_1^2,\m_5^2,\m_2^2,\m_4^2,\m_3^2\right)}\,, \\
  \toneint{i_1}{} &= - \mathbf{A} (\m_1^2)\,,
  & \toneint{12}{} &= \mathbf{B} (\m_1^2, \m_2^2)\,,
\end{aligned}
\end{gather}
where we include the one-loop integrals~$\mathbf{A}$ and~$\mathbf{B}$
that are known analytically.\footnote{The integral~$\tint{134}{}$ is
also known analytically, and is therefore available in other private
codes (such as \SARAH). Closed analytic expressions for some integrals
with special kinematic configurations can be found \EG\
in \citeres{Berends:1993ee,Scharf:1993ds}.}

While \TSIL is capable of delivering these bare integrals, it finds
them by solving differential equations for ``renormalised'' versions;
and indeed it is these renormalised integrals that we actually need
for practical applications. This finite basis
is\,\cite{Martin:2003qz}
\begin{subequations}
\begin{align}
  \begin{split}
  S(x,y,z) &= \lim_{\epsilon\rightarrow 0} \begin{aligned}[t] \bigg\{
    & \textbf{S}(x,y,z)
       - \frac{1}{\epsilon} \left[\textbf{A}(x) + \textbf{A}(y) + \textbf{A}(z)\right]
       - \frac{1}{2\,\epsilon^2} \left[x + y + z\right]\\
    &{-}\, \frac{1}{2\,\epsilon} \left[\frac{p^2}{2} - x - y - z\right]
       \!\bigg\}\,,
    \end{aligned}
  \end{split}\\  
  T(x,y,z) &\equiv - \frac{\partial}{\partial x}\, S(x,y,z)\,,\\
  I(x,y,z) &\equiv S(x,y,z)\Big|_{p^2=0}\,,\\ 
  U(x,y,z,u) &= \lim_{\epsilon\rightarrow 0} \bigg\{
    \textbf{U}(x,y,z,u)
    - \frac{1}{\epsilon}\, \textbf{B}(x,y)
    + \frac{1}{2\,\epsilon^2}
    - \frac{1}{2\,\epsilon}
    \bigg\}\,,\\
  V(x,y,z,u) &\equiv - \frac{\partial}{\partial y}\, U(x,y,z,u)\,, \\
  M(x,y,z,u,v) &= \lim_{\epsilon\rightarrow 0}\textbf{M}(x,y,z,u,v)\,.
\end{align}
\end{subequations}
while the one-loop integrals are\footnote{The symbols $A$ and $B$
should not be confused with the integrals $A_0$ and $B_0$
of \refeqs{EQ:PAVEONELOOP}.}
\begin{subequations}
\begin{align}
\label{expr_A}
  A(x) &\equiv \lim_{\epsilon\rightarrow 0}{\left[\textbf{A}(x)+\frac{x}{\epsilon}\right]} = x \left(\llog\,x - 1\right)\,,\\
  B(x,y) &\equiv \lim_{\epsilon\rightarrow 0}{\left[\textbf{B}(x,y)-\frac{1}{\epsilon}\right]} = -\llog\,p^2 - f_B(x_+) - f_B(x_-)\,,
\end{align}
\end{subequations}
where 
\begin{align}
  f_B(x) &= \log(1-x) - x\,\log{\left(1-\frac{1}{x}\right)} - 1\,,
  & x_\pm &= \frac{p^2 + x + y \pm \sqrt{\left(p^2 + x + y\right)^2 - 4\,p^2\,x}}{2\,p^2}\,,
\end{align}
and we define
\begin{gather}
  \llog\, x \equiv \log \frac{x}{Q^2}\,,
\end{gather}
where $Q$ is the renormalisation scale. 

In this appendix we shall give the expressions for diagrams in terms
of the bare integrals. As described in the text, this is because the
reduction is different for different configurations of masses
(coincident masses, vanishing masses, special kinematic
configurations, certain on-shell conditions).

After applying the symmetry relations of \refeqs{EQ:SYMMETRY}, the
following different scalar integrals remain:
\begin{subequations}\label{EQ:TLIST}\abovedisplayskip=-.5ex
\begin{alignat}{8}
  \label{EQ:TNODOUBLE}
  & &\quad&\tint{13}{}\,, &\quad&\tint{123}{}\,, &\quad&\tint{134}{}\,,
  &\quad&\tint{234}{}\,, &\quad&\tint{1234}{}\,, &\quad&\tint{1245}{}\,,
  &\quad&\tint{12345}{}\,,\\
  &\tint{1_a1_b2}{}\,, &\quad&\tint{1_a1_b3}{}\,, &\quad&\tint{1_a1_b23}{}\,,
  &\quad&\tint{1_a1_b34}{}\,, &\quad& &\quad&\tint{1_a1_b234}{}\,,\\
  & &\quad&\tint{1_a1_b1'3}{}\,, &\quad&\tint{1_a1_b1'23}{}\,,
  &\quad&\tint{1_a1_b1'34}{}\,, &\quad& &\quad&\tint{1_a1_b1'234}{}\,,\\
  & &\quad&\tint{1_a1_b1'1'3}{}\,, &\quad&\tint{1_a1_b1'1'23}{}\,,
  &\quad&\tint{1_a1_b1'1'34}{}\,, &\quad& &\quad&\tint{1_a1_b1'1'234}{}\,.
\end{alignat}
\end{subequations}
The integrals in \refeq{EQ:TNODOUBLE} differ from the others since no
index appears repeatedly. From those, the integrals that are not part
of \refeqs{EQ:TSILBASIS} can be decomposed into a product of two
one-loop integrals,
\begin{subequations}\abovedisplayskip=-.5ex
\begin{align}
  \tint{13}{} &= \aint{\m_1^2}\,\aint{\m_3^2}\,,\\
  \tint{123}{} &= \bint{p^2,\m_1^2,\m_2^2}\,\aint{\m_3^2}\,,\\
  \tint{1245}{} &= \bint{p^2,\m_1^2,\m_2^2}\,\bint{p^2,\m_4^2,\m_5^2}\,.
\end{align}
\end{subequations}
The integral~$\tint{1_a1_b2}{}$ is equal to zero in each kinematical
configuration; the remaining integrals with the repeated indices~$1_a$
and~$1_b$ can be decomposed via partial fractioning if the
corresponding masses~$\m_{1_a}$ and~$\m_{1_b}$ are not equal to each
other (the same rule applies to the one-loop integrals
of \refeq{EQ:TYONELOOP}),
\begin{align}\label{EQ:PARTFRAC}
  \tint{1_a1_bn\cdots}{} &=
  \frac{\tint{1_an\cdots}{} - \tint{1_bn\cdots}{}}{\m_{1_a}^2 - \m_{1_b}^2}\,.
\end{align}
Instead, if the masses are equal, a dedicated reduction of the
integral is required. The same rule applies to the massless
propagators with index~$1'$. In addition to these obvious
qdegeneracies, special kinematical configurations for the other masses
and/or the external momentum need to be taken into account in order to
avoid manifest poles in the reduction formulas. We do not take into
accout accidental thresholds when one mass is equal to the sum of two
others (unless one of these masses is equal to zero).

\tocsubsubsection[\label{SEC:SCALARREDUCTIONRULES}]{Reduction rules}

We have derived the reduction formulas for all integrals that appear
in 't\,Hooft--Feynman gauge with the help
of \TwoCalc, \TARCER{}\,\cite{Mertig:1998vk} and the recurrence
relations in \citere{Davydychev:1998si}. Instead of writing out the
full expression for each integral with multiple index~$1'$, we present
the respective results in terms of recurrence relations
(see \citere{Tarasov:1997kx}) in the following. In some cases closed
forms are given.\footnote{Reduction rules for integrals with higher
powers of massive propagators can be computed by evaluating the
derivative with respect to the mass at this propagator while keeping
all other masses as independent variables.} We make use of the
shorthands
\begin{subequations}
\begin{align}
  \mydelta{\m_i^2,\m_j^2,\m_k^2} &= \m_i^4 + \m_j^4 + \m_k^4
  - 2 \left(\m_i^2\,\m_j^2 + \m_j^2\,\m_k^2 + \m_k^2\,\m_i^2\right)\,,\\
  \myu{\m_i^2,\m_j^2,\m_k^2} &= \frac{1}{2}\,\frac{\partial}{\partial \m_i^2}\,
  \mydelta{\m_i^2,\m_j^2,\m_k^2} = \m_i^2 - \m_j^2 - \m_k^2\,.
\end{align}
\end{subequations}
The reduction of all scalar one-loop integrals is given
by\footnote{The extension of these rules to two-loop integrals that
factorise into a product of one-loop integrals is straightforward.}
{
\allowdisplaybreaks
\begin{subequations}\vspace{-2ex}
\begin{align}
\intertext{\textbf{\boldmath$\toneint{\mybrace{1\cdots1}{n}}{}$ and $\toneint{\mybrace{1'\cdots1'}{n}}{}$\,:}}
  \toneint{\mybrace{1\cdots1}{n}}{} &=
    \frac{d - 2\,n + 2}{2 \left(n - 1\right) \m_1^2}\,
    \toneint{\mybrace{1\cdots1}{n-1}}{}
    = \frac{\mygamma{n - \frac{d}{2}}}
      {\left(-\m_1^2\right)^{n - 1}\,\mygamma{n}\,\mygamma{1 - \frac{d}{2}}}\,
      \aint{\m_1^2}\,,\\
  \toneint{\mybrace{1'\cdots1'}{n}}{} &= 0\,,
\end{align}
\end{subequations}

\begin{subequations}\vspace{-6ex}
\begin{align}
\intertext{\textbf{\boldmath$\toneint{\mybrace{1\cdots1}{n_1}\mybrace{2\cdots2}{n_2}}{}$\,:}}
\begin{split}
  0 &= \left[\left(d - 3\,n_2\right) \myu{\m_2^2,\m_1^2,p^2}
       + 2 \left(n_1 - n_2\right) \m_1^2\right]
       \toneint{\mybrace{1\cdots1}{n_1}\mybrace{2\cdots2}{n_2}}{}
       - n_2\, \mydelta{\m_1^2,\m_2^2,p^2}\,
         \toneint{\mybrace{1\cdots1}{n_1}\mybrace{2\cdots2}{n_2 + 1}}{}\\[1ex]
    &\quad+ n_2 \left(2\,\m_2^2 - \myu{\m_1^2,\m_2^2,p^2}\right)
         \toneint{\mybrace{1\cdots1}{n_1 - 1\,}\mybrace{2\cdots2}{\,n_2 + 1}}{}
       - 2\,n_1\,\m_1^2\,
         \toneint{\mybrace{1\cdots1}{n_1 + 1\,}\mybrace{2\cdots2}{\,n_2 - 1}}{}\\[1.5ex]
    &\quad+ 2 \left(d - n_1 - n_2\right)
         \toneint{\mybrace{1\cdots1}{n_1}\mybrace{2\cdots2}{n_2 - 1}}{}
       - 2 \left(d - n_1 - n_2\right)
         \toneint{\mybrace{1\cdots1}{n_1 - 1}\mybrace{2\cdots2}{n_2}}{}\,,
\end{split}\\[2ex]
\begin{split}
  \toneint{\mybrace{1\cdots1}{n}2}{} &= \frac{d - 2\,n + 1}{n - 1}\,
    \frac{\myu{\m_1^2,\m_2^2,p^2}}{\mydelta{\m_1^2,\m_2^2,p^2}}\,
    \toneint{\mybrace{1\cdots1}{n-1}2}{}
    + \frac{d - n}{n - 1}\,\frac{1}{\mydelta{\m_1^2,\m_2^2,p^2}}\,
      \toneint{\mybrace{1\cdots1}{n-2}2}{}\\[1ex]
    &\quad+ \frac{\myu{p^2,\m_1^2,\m_2^2}}{\mydelta{\m_1^2,\m_2^2,p^2}}\,
      \toneint{\mybrace{1\cdots1}{n}}{}
    - \frac{n - 2}{n - 1}\,\frac{1}{\mydelta{\m_1^2,\m_2^2,p^2}}
      \toneint{\mybrace{1\cdots1}{n-1}}{}\,,
\end{split}\\[1ex]
  \toneint{\mybrace{1\cdots1}{n}2'}{p^2=\m_1^2} &=
    \frac{\left(d - 2\,n\right) \left(d - n - 1\right)}
    {2 \left(d - n - 2\right) \left(n - 1\right) \m_1^2}\,
    \toneint{\mybrace{1\cdots1}{n-1}2'}{p^2=\m_1^2}\,,\\[1ex]
  \toneint{\mybrace{1'\cdots1'}{n}2}{p^2=\m_2^2} &=
    \frac{d - n - 1}{2 \left(d - 2\,n - 1\right) \m_2^2}\,
    \toneint{\mybrace{1'\cdots1'}{n-1}2}{p^2=\m_2^2}\,,\\[1ex]
  \toneint{\mybrace{1'\cdots1'}{n}2'}{} &=
    -\frac{d - n - 1}{\left(n - 1\right) p^2}\,
    \toneint{\mybrace{1'\cdots1'}{n-1}2'}{}\,.
\end{align}
\end{subequations}

\newpage

The required reduction rules for the scalar two-loop integrals follow
from
\begin{subequations}\vspace{-2ex}
\label{EQ:T113}
\begin{align}
\intertext{\textbf{\boldmath$\tint{\mybrace{1\cdots1}{n}3}{}$ and $\tint{\mybrace{1'\cdots1'}{n}3}{}$\,:}}
  \tint{\mybrace{1\cdots1}{n}3}{} &=
  \frac{\mygamma{n - \frac{d}{2}}}
  {\left(-\m_1^2\right)^{n - 1}\,\mygamma{n}\,\mygamma{1 - \frac{d}{2}}}
  \,\tint{13}{}\,,\\
  \tint{\mybrace{1'\cdots1'}{n}3}{} &= 0\,,
\end{align}
\end{subequations}

\begin{subequations}\vspace{-6ex}
\begin{align}
\intertext{\textbf{\boldmath$\tint{1123}{}$ and $\tint{\mybrace{1'\cdots1'}{n}23}{}$\,:}}
  \tint{1123}{} &=
  \frac{d - 2}{\mydelta{\m_1^2,\m_2^2,p^2}} \left(
  \tint{23}{} + \frac{\myu{p^2,\m_1^2,\m_2^2}}{2\,\m_1^2}\,\tint{13}{}\right)
  + \frac{\myu{\m_1^2,\m_2^2,p^2}}{\mydelta{\m_1^2,\m_2^2,p^2}} \left(d - 3\right)
  \tint{123}{}\,,\\
  \tint{112'3}{p^2=\m_1^2} &= \frac{d - 2}{4\,\m_1^4}\,\tint{13}{}\,,\\
  \tint{\mybrace{1'\cdots1'}{n}23}{} &=
  \frac{1}{\left(\m_2^2 - p^2\right)^2} \left[
  \frac{d - n}{n - 1}\,\tint{\mybrace{1'\cdots1'}{n-2}23}{}
  - \frac{d - 2\,n + 1}{n - 1} \left(\m_2^2 + p^2\right)
  \tint{\mybrace{1'\cdots1'}{n-1}23}{}\right],\\
  \tint{\mybrace{1'\cdots1'}{n}23}{p^2=\m_2^2} &=
  \frac{\sqrt{\pi}\ \mygamma{n + 2 - d}}
  {2^{2\,n + 1 - d}\,\m_2^{2\,n}\,\mygamma{n + \frac{3}{2} - \frac{d}{2}}\,
  \mygamma{1 - \frac{d}{2}}}\,\tint{23}{p^2=\m_2^2}\,,
\end{align}
\end{subequations}

\begin{subequations}\vspace{-6ex}
\begin{align}
\intertext{\textbf{\boldmath$\tint{\mybrace{1\cdots1}{n}34}{}$ and $\tint{\mybrace{1'\cdots1'}{n}34}{}$\,:}}
  \begin{split}
  \tint{\mybrace{1\cdots1}{n}34}{} &= \frac{1}{\mydelta{\m_1^2,\m_3^2,\m_4^2}}
  \begin{aligned}[t] \,\bigg\{
  & \frac{d - 2\,n + 1}{n - 1}\,\myu{\m_1^2,\m_3^2,\m_4^2}\,
    \tint{\mybrace{1\cdots1}{n - 1}34}{}
    + \frac{d - n}{n - 1}\,\tint{\mybrace{1\cdots1}{n - 2}34}{}\\[.5ex]
  &{+}\, \myu{\m_3^2,\m_1^2,\m_4^2}\,\tint{\mybrace{1\cdots1}{n}3}{}
    - \frac{n - 2}{n - 1}\,\tint{\mybrace{1\cdots1}{n - 1}3}{}
    + \myu{\m_4^2,\m_1^2,\m_3^2}\,\tint{\mybrace{1\cdots1}{n}4}{}
    - \frac{n - 2}{n - 1}\,\tint{\mybrace{1\cdots1}{n - 1}4}{}
    \bigg\}\,,
    \end{aligned}
  \end{split}\\[1ex]
  \tint{\mybrace{1\cdots1}{n}34'}{\m_3^2=\m_1^2} &=
  \frac{\mygamma{n + 1 - \frac{d}{2}}}{\left(-\m_1^2\right)^n
  \left(d - n - 1\right)\mygamma{n}\mygamma{1 - \frac{d}{2}}}
  \,\tint{13}{\m_3^2=\m_1^2}\,,\\
  \tint{\mybrace{1'\cdots1'}{n}34}{} &=
  \frac{1}{\left(\m_3^2 - \m_4^2\right)^2} \left[
  \frac{d - n}{n - 1}\,\tint{\mybrace{1'\cdots1'}{n-2}34}{}
  - \frac{d - 2\,n + 1}{n - 1} \left(\m_3^2 + \m_4^2\right)
  \tint{\mybrace{1'\cdots1'}{n-1}34}{}\right],\\[.5ex]
  \tint{\mybrace{1'\cdots1'}{n}34}{\m_4^2=\m_3^2} &=
  \frac{\sqrt{\pi}\ \mygamma{n + 2 - d}}
  {2^{2\,n + 1 - d}\,\m_3^{2\,n}\,\mygamma{n + \frac{3}{2} - \frac{d}{2}}\,
  \mygamma{1 - \frac{d}{2}}}\,\tint{34}{\m_4^2=\m_3^2}\,,\\
  \tint{\mybrace{1'\cdots1'}{n}3'4'}{} &= 0\,,
\end{align}
\end{subequations}

\begin{subequations}\vspace{-6ex}
\begin{align}
\intertext{\textbf{\boldmath$\tint{11234}{}$\,:}}
  \begin{split}
  \tint{11234}{} &= \left[-\frac{d - 2}{2\,\m_1^2} + \left(d - 3\right) \left(
  \frac{\myu{\m_1^2,\m_2^2,p^2}}{\mydelta{\m_1^2,\m_2^2,p^2}}
  + \frac{\myu{\m_1^2,\m_3^2,\m_4^2}}{\mydelta{\m_1^2,\m_3^2,\m_4^2}}\right)\right]
  \tint{1234}{}\\
  &\quad + \frac{\m_3^2}{\m_1^2} \left(
  \frac{\myu{p^2,\m_1^2,\m_2^2}}{\mydelta{\m_1^2,\m_2^2,p^2}}
  - \frac{\myu{\m_3^2,\m_1^2,\m_4^2}}{\mydelta{\m_1^2,\m_3^2,\m_4^2}}\right)
  \tint{2334}{}
  + \frac{\m_4^2}{\m_1^2} \left(
  \frac{\myu{p^2,\m_1^2,\m_2^2}}{\mydelta{\m_1^2,\m_2^2,p^2}}
  - \frac{\myu{\m_4^2,\m_1^2,\m_3^2}}{\mydelta{\m_1^2,\m_3^2,\m_4^2}}\right)
  \tint{2344}{}\\
  &\quad+ \begin{aligned}[t]
  \Bigg\{
    & \frac{\myu{p^2,\m_1^2,\m_2^2}}{2\,\m_1^2\,\mydelta{\m_1^2,\m_2^2,p^2}} \left[
    \left(d - 2\right) \tint{134}{} - \left(3\,d - 8\right) \tint{234}{}
    \right]
    - \frac{2\,\m_2^2 \left(\m_2^2 - p^2\right)}{\m_1^2\,\mydelta{\m_1^2,\m_2^2,p^2}}\,
    \tint{2234}{}\\
    &{+}\, \frac{d - 2}{2\,\m_1^2} \left(
    \frac{\myu{\m_3^2,\m_1^2,\m_4^2}}{\mydelta{\m_1^2,\m_3^2,\m_4^2}}\,\tint{123}{}
    + \frac{\myu{\m_4^2,\m_1^2,\m_3^2}}{\mydelta{\m_1^2,\m_3^2,\m_4^2}}\,\tint{124}{}
    \right)
    \!\!\Bigg\}\,,
    \end{aligned}
  \end{split}\\[1ex]
  \begin{split}
  \tint{112'34}{p^2=\m_1^2} &= \begin{aligned}[t]
  \frac{1}{4\,\m_1^4}\,\frac{1}{d - 3} \Bigg[
  & \!\left(d - 2\right)^2 \left(
  \frac{\myu{\m_3^2,\m_1^2,\m_4^2}}{\mydelta{\m_1^2,\m_3^2,\m_4^2}}\,\tint{13}{}
  + \frac{\myu{\m_4^2,\m_1^2,\m_3^2}}{\mydelta{\m_1^2,\m_3^2,\m_4^2}}\,\tint{14}{}
  \right)
  - \left(d - 2\right) \tint{134}{}\\
  &{+} \left(3\,d - 8\right) \tint{2'34}{p^2=\m_1^2}
  - 2\,\m_3^2\,\tint{2'334}{p^2=\m_1^2} - 2\,\m_4^2\,\tint{2'344}{p^2=\m_1^2}
  \Bigg]
  \end{aligned}\\
  &\quad- \frac{1}{2\,\m_1^2}\,\tint{2'2'34}{p^2=\m_1^2}
  + \frac{2\,\m_3^2}{\m_1^2}\,\frac{\m_1^2 - \m_3^2}{\mydelta{\m_1^2,\m_3^2,\m_4^2}}\,
  \tint{2'334}{p^2=\m_1^2}
  + \frac{2\,\m_4^2}{\m_1^2}\,\frac{\m_1^2 - \m_4^2}{\mydelta{\m_1^2,\m_3^2,\m_4^2}}\,
  \tint{2'344}{p^2=\m_1^2}\\
  &\quad+ \frac{1}{2\,\m_1^2}\,
  \frac{\myu{\m_1^2,\m_3^2,\m_4^2}}{\mydelta{\m_1^2,\m_3^2,\m_4^2}}
  \left[\left(d - 2\right) \tint{134}{}
  - \left(3\,d - 8\right) \tint{2'34}{p^2=\m_1^2}\right],
  \end{split}\\[1ex]
  \begin{split}
  \tint{11234'}{\m_3^2=\m_1^2} &= -\frac{1}{2\,\m_1^2}\,\frac{1}{d - 3}
  \left[\frac{d - 2}{2\,\m_1^2}\,\tint{123}{\m_3^2=\m_1^2}
  + \tint{234'4'}{\m_3^2=\m_1^2} - \tint{2334'}{\m_3^2=\m_1^2}\right]\\
  &\quad+ \begin{aligned}[t]
  \frac{d - 2}{2\,\m_1^2\,\mydelta{\m_1^2,\m_2^2,p^2}} \bigg[
  & \frac{d - 2}{d - 3}
  \left(\frac{\myu{p^2,\m_1^2,\m_2^2}}{2\,\m_1^2}\,\tint{13}{\m_3^2=\m_1^2}
  + \tint{23}{\m_3^2=\m_1^2}\right)
  + \myu{\m_1^2,\m_2^2,p^2}\,\tint{123}{\m_3^2=\m_1^2}
  \bigg]\,,
  \end{aligned}
  \end{split}\\[1ex]
  \begin{split}
  \tint{112'34'}{p^2=\m_1^2,\m_3^2=\m_1^2} &= \frac{1}{8\,\m_1^4}\,
  \frac{\left(3\,d - 10\right) \left(3\,d - 8\right)}
  {\left(2\,d - 7\right)\left(d - 3\right)}\,\tint{2'34'}{p^2=\m_1^2,\m_3^2=\m_1^2}
  + \frac{1}{8\,\m_1^6}\,
  \frac{\left(d - 4\right) \left(d - 2\right)^2}
  {\left(d - 3\right)^2}\,\tint{13}{\m_3^2=\m_1^2}\,,
  \end{split}
\end{align}
\end{subequations}

\begin{subequations}\vspace{-6ex}
\begin{align}
\intertext{\textbf{\boldmath$\tint{\mybrace{1'\cdots1'}{n}234}{}$\,:}}
  \begin{split}
  \tint{\mybrace{1'\cdots1'}{n}234}{} &= \begin{aligned}[t]
  \frac{1}{\m_2^2 - p^2}\,\Bigg[
  & \frac{3\,d - 2\,n - 6}{d + 2\,n - 4} \left(
  \tint{\mybrace{1'\cdots1'}{n-1}234}{}
  - \tint{\mybrace{1'\cdots1'}{n}34}{}\right)
  - \frac{4\,\m_2^2}{d + 2\,n - 4}\,
  \tint{\mybrace{1'\cdots1'}{n-1}2234}{}\\
  &{-}\, \frac{2\,\m_3^2}{d + 2\,n - 4} \left(
  \tint{\mybrace{1'\cdots1'}{n-1}2334}{}
  - \tint{\mybrace{1'\cdots1'}{n}334}{}\right)\\
  &{-}\, \frac{2\,\m_4^2}{d + 2\,n - 4} \left(
  \tint{\mybrace{1'\cdots1'}{n-1}2344}{}
  - \tint{\mybrace{1'\cdots1'}{n}344}{}\right)
  \!\Bigg]
  \end{aligned}\\
  &\quad+ \begin{aligned}[t]
  \frac{1}{\m_3^2 - \m_4^2}\,\Bigg[
  & \frac{d - 2}{d + 2\,n - 4} \left(
  \tint{\mybrace{1'\cdots1'}{n}23}{}
  - \tint{\mybrace{1'\cdots1'}{n}24}{}\right)\\
  &{-}\, \frac{2\,\m_3^2}{d + 2\,n - 4}\,
  \tint{\mybrace{1'\cdots1'}{n-1}2334}{}
  + \frac{2\,\m_4^2}{d + 2\,n - 4}\,
  \tint{\mybrace{1'\cdots1'}{n-1}2344}{}
  \Bigg]\,,
  \end{aligned}
  \end{split}\\[1ex]
  \begin{split}
  \tint{\mybrace{1'\cdots1'}{n}234}{p^2=\m_2^2} &=
  -\frac{1}{2\,n - 1}\,\tint{\mybrace{1'\cdots1'}{n-1}2234}{p^2=\m_2^2}
  + \begin{aligned}[t]
  \frac{1}{\m_3^2 - \m_4^2}\,\Bigg[
  & \frac{d - 2}{2\,n - 1} \left(
  \tint{\mybrace{1'\cdots1'}{n}23}{p^2=\m_2^2}
  -\tint{\mybrace{1'\cdots1'}{n}24}{p^2=\m_2^2}\right)\\
  &{-}\, \frac{2\,\m_3^2}{2\,n - 1}\,
  \tint{\mybrace{1'\cdots1'}{n-1}2334}{p^2=\m_2^2}
  + \frac{2\,\m_4^2}{2\,n - 1}\,
  \tint{\mybrace{1'\cdots1'}{n-1}2344}{p^2=\m_2^2}
  \Bigg]\,,
  \end{aligned}
  \end{split}\\[1ex]
  \begin{split}
  \tint{\mybrace{1'\cdots1'}{n}234}{\m_4^2=\m_3^2} &=
  \frac{1}{2\,\m_3^2}\,\frac{d - 2}{2\,n - 1}\,
  \tint{\mybrace{1'\cdots1'}{n}23}{}
  - \frac{1}{2\,n - 1}\,
  \tint{\mybrace{1'\cdots1'}{n-1}2334}{\m_4^2=\m_3^2}\\[.5ex]
  &\quad+ \begin{aligned}[t]
  \frac{1}{\m_2^2 - p^2}\,\Bigg[
  & \frac{3\,d - 2\,n - 6}{2\,n - 1} \left(
  \tint{\mybrace{1'\cdots1'}{n-1}234}{\m_4^2=\m_3^2}
  - \tint{\mybrace{1'\cdots1'}{n}34}{\m_4^2=\m_3^2}\right)
  - \frac{4\,\m_2^2}{2\,n - 1}\,
  \tint{\mybrace{1'\cdots1'}{n-1}2234}{\m_4^2=\m_3^2}\\[.5ex]
  &{-}\, \frac{4\,\m_3^2}{2\,n - 1} \left(
  \tint{\mybrace{1'\cdots1'}{n-1}2334}{\m_4^2=\m_3^2}
  - \tint{\mybrace{1'\cdots1'}{n}334}{\m_4^2=\m_3^2}\right)
  \!\Bigg]\,,
  \end{aligned}
  \end{split}\\[1ex]
  \begin{split}
  \tint{\mybrace{1'\cdots1'}{n}234}{p^2=\m_2^2,\m_4^2=\m_3^2} &=
  \frac{2\,d - 2\,n - 3}{d - 2\,n - 2}\,\frac{\m_2^2 + \m_3^2}{4\,\m_2^2\,\m_3^2}\,
  \tint{\mybrace{1'\cdots1'}{n-1}234}{p^2=\m_2^2,\m_4^2=\m_3^2}
  - \frac{3\,d - 2\,n - 4}{16\,\m_2^2\,\m_3^2 \left(d - 2\,n - 2\right)}\,
  \tint{\mybrace{1'\cdots1'}{n-2}234}{p^2=\m_2^2,\m_4^2=\m_3^2}\\[1ex]
  &\quad- \frac{\left(d - 2\right) \left(d - 1\right)}
  {8 \left(d - 2\,n - 2\right) \left(d - n - 1\right)} \left(
  \frac{2}{\m_3^2}\,\tint{\mybrace{1'\cdots1'}{n}23}{p^2=\m_2^2}
  + \frac{1}{\m_2^2}\,
  \tint{\mybrace{1'\cdots1'}{n}34}{\m_4^2=\m_3^2}\right),
  \end{split}\\[1ex]
  \tint{\mybrace{1'\cdots1'}{n}23'4'}{} &=
  \frac{2 \left(2\,d - 2\,n - 3\right)}{d - 2\,n - 2}\,
  \frac{\m_2^2 + p^2}{\left(\m_2^2 - p^2\right)^2}\,
  \tint{\mybrace{1'\cdots1'}{n-1}23'4'}{}
  - \frac{3\,d - 2\,n - 4}{d - 2\,n - 2}\,\frac{1}{\left(\m_2^2 - p^2\right)^2}\,
  \tint{\mybrace{1'\cdots1'}{n-2}23'4'}{}\,,\\[1ex]
  \tint{\mybrace{1'\cdots1'}{n}23'4'}{p^2=\m_2^2} &=
  \frac{3\,d - 2\,n - 6}{2\,d - 2\,n - 5}\,\frac{1}{4\,\m_2^2}\,
  \tint{\mybrace{1'\cdots1'}{n-1}23'4'}{p^2=\m_2^2}\,.
\end{align}
\end{subequations}
Additional integrals are introduced by the previous reductions. They
can be further reduced by the following relations:
\begin{subequations}\vspace{-2ex}
\label{EQ:TREDADD_FIRST}
\begin{align}
\intertext{\textbf{\boldmath$\tint{\mybrace{1'\cdots1'}{n}2234}{}$\,:}}
  \begin{split}
  \tint{\mybrace{1'\cdots1'}{n}2234}{} &= \begin{aligned}[t]
  \frac{1}{\m_2^2 - p^2}\,\Bigg[
  & \!\left(2\,d - 2\,n - 5\right)
  \tint{\mybrace{1'\cdots1'}{n}234}{}
  - \tint{\mybrace{1'\cdots1'}{n-1}2234}{}\\
  &{-}\, 2\,\m_3^2\,\tint{\mybrace{1'\cdots1'}{n}2334}{}
  - 2\,\m_4^2\,\tint{\mybrace{1'\cdots1'}{n}2344}{}
  \Bigg]\,,
  \end{aligned}
  \end{split}\\
  \begin{split}
  \tint{\mybrace{1'\cdots1'}{n}2234}{p^2=\m_2^2} &= \begin{aligned}[t]
  \frac{1}{4\,\m_2^2}\,\Bigg[
  & \!\left(3\,d - 2\,n - 8\right) \left(
    \tint{\mybrace{1'\cdots1'}{n}234}{p^2=\m_2^2}
    - \tint{\mybrace{1'\cdots1'}{n+1}34}{}\right)\\
  &{-}\, 2\,\m_3^2 \left(
  \tint{\mybrace{1'\cdots1'}{n}2334}{p^2=\m_2^2}
  - \tint{\mybrace{1'\cdots1'}{n+1}334}{}\right)
  - 2\,\m_4^2 \left(
  \tint{\mybrace{1'\cdots1'}{n}2344}{p^2=\m_2^2}
  - \tint{\mybrace{1'\cdots1'}{n+1}344}{}\right)
  \!\Bigg]\,,
  \end{aligned}
  \end{split}
\end{align}
\end{subequations}

\begin{subequations}\vspace{-6ex}
\begin{align}
\intertext{\textbf{\boldmath$\tint{\mybrace{1'\cdots1'}{n}2334}{}$\,:}}
  \begin{split}
  \tint{\mybrace{1'\cdots1'}{n}2334}{} &=
  \frac{d - 3}{\m_3^2 - \m_4^2}\,
  \tint{\mybrace{1'\cdots1'}{n}234}{}
  + \frac{d - 2}{\left(\m_3^2 - \m_4^2\right)^2}\,
  \tint{\mybrace{1'\cdots1'}{n}24}{}
  - \frac{2\,\m_4^2}{\left(\m_3^2 - \m_4^2\right)^2}\,
  \tint{\mybrace{1'\cdots1'}{n-1}2344}{}\\[.5ex]
  &\quad+ \frac{\m_3^2 + \m_4^2}{\left(\m_3^2 - \m_4^2\right)^2} \left(
  \tint{\mybrace{1'\cdots1'}{n-1}2334}{}
  - \frac{d - 2}{2\,\m_3^2}\,\tint{\mybrace{1'\cdots1'}{n}23}{}
  \right),
  \end{split}\\[1ex]
  \begin{split}
  \tint{\mybrace{1'\cdots1'}{n}2334}{\m_4^2=\m_3^2} &=
  \begin{aligned}[t]
  \frac{d - 2\,n - 2}{\left(2\,n - 1\right) \left(\m_2^2 - p^2\right)}\,\Bigg[
  & \frac{d - 2}{4\,\m_3^2}\,
  \tint{\mybrace{1'\cdots1'}{n}34}{\m_4^2=\m_3^2}
  - \frac{3\,d - 2\,n - 6}{4\,\m_3^2}\,
  \tint{\mybrace{1'\cdots1'}{n-1}234}{\m_4^2=\m_3^2}\\
  &{+}\, \frac{\m_2^2}{\m_3^2}\,
  \tint{\mybrace{1'\cdots1'}{n-1}2234}{\m_4^2=\m_3^2}
  + \tint{\mybrace{1'\cdots1'}{n-1}2334}{\m_4^2=\m_3^2}
  \!\Bigg]
  \end{aligned}\\
  &\quad+ \begin{aligned}[t]
  \frac{1}{4\,\m_3^2}\,\Bigg[
  & \frac{d - 3}{2\,n - 1} \left(
  \tint{\mybrace{1'\cdots1'}{n-1}2334}{\m_4^2=\m_3^2}
  - \frac{d - 2}{2\,\m_3^2}\,
  \tint{\mybrace{1'\cdots1'}{n}23}{}\right)
  + \left(2\,d - 2\,n - 5\right)
  \tint{\mybrace{1'\cdots1'}{n}234}{\m_4^2=\m_3^2}
  \Bigg]\,,\hspace{-1em}
  \end{aligned}
  \end{split}\\[1ex]
  \tint{\mybrace{1'\cdots1'}{n}2334}{p^2=\m_2^2,\m_4^2=\m_3^2} &=
  \frac{1}{4\,\m_3^2} \left[\left(2\,d - 2\,n - 5\right)
  \tint{\mybrace{1'\cdots1'}{n}234}{p^2=\m_2^2,\m_4^2=\m_3^2}
  - \tint{\mybrace{1'\cdots1'}{n-1}2234}{p^2=\m_2^2,\m_4^2=\m_3^2}\right],
\end{align}
\end{subequations}

\begin{subequations}\vspace{-6ex}
\label{EQ:TREDADD_LAST}
\begin{align}
\intertext{\textbf{\boldmath$\tint{\mybrace{1'\cdots1'}{n}334}{}$\,:}}
  \tint{\mybrace{1'\cdots1'}{n}334}{} &=
  \frac{d - 2\,n - 1}{\m_3^2 - \m_4^2}\,
  \tint{\mybrace{1'\cdots1'}{n}34}{}
  - \frac{1}{\m_3^2 - \m_4^2}\,
  \tint{\mybrace{1'\cdots1'}{n-1}334}{}\,,\\[1ex]
  \tint{\mybrace{1'\cdots1'}{n}334}{\m_4^2=\m_3^2} &=
  \frac{d - n - 2}{2\,\m_3^2}\,
  \tint{\mybrace{1'\cdots1'}{n}34}{\m_4^2=\m_3^2}\,,
\end{align}
\end{subequations}
}%
The integrals with multiple index~$4$ can be reduced after using the
symmetry relation~\mbox{$3\leftrightarrow4$}. All remaining integrals
belong to the set of \refeq{EQ:TSILBASIS} after applying the symmetry
relations of \refeqs{EQ:SYMMETRY}.

\tocsubsection[\label{SEC:SCALARREDUCTIONP2ZERO}]{Vanishing external momentum}

In the limit of vanishing external momentum,~$p\to 0$, many of the
reduction rules cannot be applied directly due to manifest poles
in~$p^2$. However, if this condition is imposed onto the integrals in
the first place, their reduction becomes much easier. Symbolically,
the implementation of this limit amounts to performing the index
substitutions~$2\to1$ and~$5\to4$ (the masses are relabeled
accordingly, but in order to distinguish the possibly different masses
of repeated indices, new subscripts are introduced).

For the tensor integrals, only the first reduction
in \refeq{EQ:3Y1245} requires some special care, since it contains an
explicit pole in~$p^2$. The momentum-free equation reads
\begin{align}\label{EQ:YP2ZERO}
  \yint{1245}{3}{p^2=0} \to \yint{1_a1_b4_a4_b}{3}{} &= \ysum{1_a1_b4_a4_b}{[1]+[4]}{}
\end{align}
which corresponds to the first term on the right-hand side
of \refeq{EQ:3Y1245} with~$p\to0$. For all other tensor reductions
in \refeqs{EQ:YRED} the limit~$p\to0$ can be taken explicitly.

In addition to the types of integrals in \refeqs{EQ:TLIST}, the
following scalar integrals appear after taking into account the
symmetry relations of \refeq{EQ:SYMMETRY} and partial fractioning
via \refeq{EQ:PARTFRAC}:
\begin{gather}\label{EQ:TP2ZERO}
  \tint{1113}{}\,, \quad \tint{1133}{}\,, \quad \tint{11134}{}\,, \quad
  \tint{11334}{}\,.
\end{gather}
Further integrals with multiple massless propagators with index~$1'$
can be solved with the recurrence relations given above. In general,
the solutions for the integrals of \refeq{EQ:TP2ZERO} with repeated
index~$i$ can be derived by computing the derivative with respect to
the mass~$\m_i$ of the scalar integral with fewer repetitions
of~$i$. The results that are not contained
in \refeqs{EQ:T113}--\eqref{EQ:TREDADD_LAST} read
{
\allowdisplaybreaks
\begin{subequations}\vspace{-2ex}
\begin{align}
\intertext{\textbf{\boldmath$\tint{\mybrace{1\cdots1}{n_1}\mybrace{3\cdots3}{n_3}}{}$ and $\tint{\mybrace{1'\cdots1'}{n_1}\mybrace{3\cdots3}{n_3}}{}$\,:}}
  \tint{\mybrace{1\cdots1}{n_1}\mybrace{3\cdots3}{n_3}}{} &=
  \frac{\mygamma{n_1 - \frac{d}{2}}}
  {\left(-\m_1^2\right)^{n_1 - 1}\,\mygamma{n_1}\,\mygamma{1 - \frac{d}{2}}}\,
  \frac{\mygamma{n_3 - \frac{d}{2}}}
  {\left(-\m_3^2\right)^{n_3 - 1}\,\mygamma{n_3}\,\mygamma{1 - \frac{d}{2}}}
  \,\tint{13}{}\,,\\
  \tint{\mybrace{1'\cdots1'}{n_1}\mybrace{3\cdots3}{n_3}}{} &= 0\,,
\end{align}
\end{subequations}


\begin{subequations}\vspace{-6ex}
\begin{align}
\intertext{\textbf{\boldmath$\tint{\mybrace{1\cdots1}{n_1}\mybrace{3\cdots3}{n_3}4}{}$ and $\tint{\mybrace{1'\cdots1'}{n_1}\mybrace{3\cdots3}{n_3}4}{}$\,:}}
  \begin{split}
  0 &=
  \left(n_3 - 1\right) \myu{\m_4^2,\m_1^2,\m_3^2}\,
    \tint{\mybrace{1\cdots1}{n_1}\mybrace{3\cdots3}{n_3}4}{}
  - \left(n_3 - 1\right)\,
    \tint{\mybrace{1\cdots1}{n_1-1}\mybrace{3\cdots3}{n_3}4}{}
  + \left(n_3 - 1\right)\,
    \tint{\mybrace{1\cdots1}{n_1}\mybrace{3\cdots3}{n_3}}{}\\[2ex]
  &\quad+ \left(d - 2\,n_1 - n_3 + 1\right)
  \tint{\mybrace{1\cdots1}{n_1}\mybrace{3\cdots3}{n_3-1}4}{}
  - 2\,n_1\,\m_1^2\,
    \tint{\mybrace{1\cdots1}{n_1+1\,}\mybrace{3\cdots3}{\,n_3-1}4}{}\,,
  \end{split}\\[2ex]
  \tint{\mybrace{1\cdots1}{n_1}\mybrace{3\cdots3}{n_3}4'}{\m_3^2=\m_1^2} &=
  \frac{1}{2\,\m_1^2} \left[\frac{d - 2\,n_1}{d - n_1 - n_3 - 1}\,
  \tint{\mybrace{1\cdots1}{n_1}\mybrace{3\cdots3}{n_3}\vphantom{4'}}{\m_3^2=\m_1^2}
  - 2\,\frac{d - n_1 - n_3}{d - n_1 - n_3 - 1}\,
  \tint{\mybrace{1\cdots1}{n_1}\mybrace{3\cdots3}{\,n_3 - 1}4'}{\m_3^2=\m_1^2}
  \right],\\[1ex]
  \tint{\mybrace{1'\cdots1'}{n_1}\mybrace{3\cdots3}{n_3}4}{\m_4^2=\m_3^2} &=
  \frac{d - 2\,n_1 - n_3 - 1}{n_3 - 1}\,
  \tint{\mybrace{1'\cdots1'}{n_1 + 1}\,\mybrace{3\cdots3}{\,n_3 - 1}4}{\m_4^2=\m_3^2}\,.
\end{align}
\end{subequations}
}%

\tocsubsection[\label{SEC:UVDIVERGENCES}]{UV divergences}

The remaining scalar integrals are UV~divergent in general. In
dimensional regularisation, these UV~divergences can be parametrised
by the regulator~$\epsilon$. For the two-loop two-point integrals
appearing in this article, the terms that diverge in the
limit~$\epsilon\to 0$ are known analytically; the finite terms are
evaluated numerically with the help of \TSIL. In the following we list
these UV~divergences explicitly with the considered order
in~$\epsilon$ indicated in the superscript of the integrals:
{
\allowdisplaybreaks
\begin{subequations}\vspace{-2ex}
\begin{align}
\intertext{\textbf{\boldmath$\tint{134}{}$\,:}}
  \tint{134}{\epsilon^{-2}} &= \frac{1}{2} \left(\m_1^2 + \m_3^2 + \m_4^2\right),\\
  \tint{134}{\epsilon^{-1}} &= \frac{1}{2} \left(\m_1^2 + \m_3^2 + \m_4^2\right)
    + \ainteps{\epsilon^0}{\m_1^2}
    + \ainteps{\epsilon^0}{\m_3^2}
    + \ainteps{\epsilon^0}{\m_4^2}\,,
\end{align}
\end{subequations}

\begin{subequations}\vspace{-6ex}
\begin{align}
\intertext{\textbf{\boldmath$\tint{234}{}$\,:}}
  \tint{234}{\epsilon^{-2}} &= \frac{1}{2} \left(\m_2^2 + \m_3^2 + \m_4^2\right),\\
  \tint{234}{\epsilon^{-1}} &= \frac{1}{2} \left(\m_2^2 + \m_3^2 + \m_4^2\right)
    + \ainteps{\epsilon^0}{\m_2^2}
    + \ainteps{\epsilon^0}{\m_3^2}
    + \ainteps{\epsilon^0}{\m_4^2} - \frac{1}{4}\,p^2\,,
\end{align}
\end{subequations}

\begin{subequations}\vspace{-6ex}
\begin{align}
\intertext{\textbf{\boldmath$\tint{1234}{}$\,:}}
  \tint{1234}{\epsilon^{-2}} &= \frac{1}{2}\,,\\
  \tint{1234}{\epsilon^{-1}} &= \frac{1}{2}
  + \binteps{\epsilon^0}{p^2,\m_1^2,\m_2^2}\,,
\end{align}
\end{subequations}

\begin{subequations}\vspace{-6ex}
\begin{align}
\intertext{\textbf{\boldmath$\tint{2234}{}$ and $\tint{2'2'34}{}$\,:}}
  \tint{2234}{\epsilon^{-2}} &= \frac{1}{2}\,,\\
  \tint{2234}{\epsilon^{-1}} &= -\frac{1}{2}
  + \frac{1}{\m_2^2}\,\ainteps{\epsilon^0}{\m_2^2}\,,\\
  \tint{2'2'34}{\epsilon^{-2}} &= -\frac{1}{2}\,,\\
  \tint{2'2'34}{\epsilon^{-1}} &= \frac{1}{2}
  - \binteps{\epsilon^0}{p^2,\m_3^2,\m_4^2}\,,
\end{align}
\end{subequations}

\begin{subequations}\vspace{-6ex}
\begin{align}
\intertext{\textbf{\boldmath$\tint{11234}{}$\,:}}
  \tint{11234}{\epsilon^{-2}} &= 0\,,\\
  \begin{split}
  \tint{11234}{\epsilon^{-1}} &= \frac{1}{\mydelta{\m_1^2,\m_2^2,p^2}}
  \begin{aligned}[t] \bigg[
    & \frac{\myu{p^2,\m_1^2,\m_2^2}}{\m_1^2}\,\ainteps{\epsilon^0}{\m_1^2}
      + 2\,\ainteps{\epsilon^0}{\m_2^2}\\
    &{+}\, \myu{\m_1^2,\m_2^2,p^2} \left(
    \binteps{\epsilon^0}{p^2,\m_1^2,\m_2^2} - 1\right)
    \!\bigg]\,,
    \end{aligned}
  \end{split}\\
  \tint{112'34}{p^2=\m_1^2,\,\epsilon^{-2}} &= \frac{1}{2\,\m_1^2}\,,\\
  \begin{split}
  \tint{112'34}{p^2=\m_1^2,\,\epsilon^{-1}} &= \frac{1}{\m_1^2}
  \begin{aligned}[t] \,\bigg\{
    & \frac{1}{\mydelta{\m_1^2,\m_3^2,\m_4^2}} \begin{aligned}[t] \,\bigg[
      & \frac{3}{8}\, \myu{\m_1^2,\m_3^2,\m_4^2} \left(
        \m_1^2 + 4\,\ainteps{\epsilon^0}{\m_3^2} + 4\,\ainteps{\epsilon^0}{\m_4^2}
        \right)\\
      &{-}\, 4\,\m_3^2\,\m_4^2
        - \ainteps{\epsilon^0}{\m_3^2}\,\ainteps{\epsilon^0}{\m_4^2}
        \bigg]
      \end{aligned}\\
    &{+}\, \frac{1}{2\,\m_1^2}\,\ainteps{\epsilon^0}{\m_1^2} - \frac{3}{2}
    \bigg\}\,,
    \end{aligned}
  \end{split}\\
  \tint{112'34'}{p^2=\m_1^2,\m_3^2=\m_1^2,\,\epsilon^{-1}} &=
    \frac{1}{\m_1^4}\,\ainteps{\epsilon^0}{\m_1^2}\,,
\end{align}
\end{subequations}

\begin{subequations}\vspace{-6ex}
\begin{align}
\intertext{\textbf{\boldmath$\tint{1'1'234}{}$\,:}}
  \tint{1'1'234}{\epsilon^{-2}} &= \frac{1}{\m_2^2 - p^2}\,,\\
  \begin{split}
  \tint{1'1'234}{\epsilon^{-1}} &=
    \frac{1}{\m_2^2 - p^2}\,\binteps{\epsilon^0}{0,\m_3^2,\m_4^2}\\
    &\quad+ \frac{1}{\left(\m_2^2 - p^2\right)^2} \left[2\,p^2
    + 2\,\ainteps{\epsilon^0}{\m_2^2}
    - \left(\m_2^2 + p^2\right) \binteps{\epsilon^0}{p^2,0,\m_2^2}\right],
  \end{split}\\
  \tint{1'1'234}{p^2=\m_2^2,\,\epsilon^{-2}} &= -\frac{1}{2\,\m_2^2}\,,\\
  \tint{1'1'234}{p^2=\m_2^2,\,\epsilon^{-1}} &= -\frac{1}{2\,\m_2^2} \left[
    \frac{1}{\m_2^2}\,\ainteps{\epsilon^0}{\m_2^2}
    + \binteps{\epsilon^0}{0,\m_3^2,\m_4^2} - 3\right],\\
  \tint{1'1'23'4'}{\epsilon^{-2}} &= \frac{1}{2\left(\m_2^2 - p^2\right)}\,,\\
  \tint{1'1'23'4'}{\epsilon^{-1}} &= \frac{1}{2\left(\m_2^2 - p^2\right)}
  + \frac{1}{\left(\m_2^2 - p^2\right)^2} \left[
    2\,p^2 + 2\,\ainteps{\epsilon^0}{\m_2^2}
    - \left(\m_2^2 + p^2\right) \binteps{\epsilon^0}{p^2,0,\m_2^2}\right],\\
  \tint{1'1'23'4'}{p^2=\m_2^2,\,\epsilon^{-2}} &= -\frac{1}{4\,\m_2^2}\,,\\
  \tint{1'1'23'4'}{p^2=\m_2^2,\,\epsilon^{-1}} &= -\frac{1}{2\,\m_2^2} \left[
    \frac{1}{\m_2^2}\,\ainteps{\epsilon^0}{\m_2^2} - 2\right],    
\end{align}
\end{subequations}

\begin{subequations}\vspace{-6ex}
\begin{align}
\intertext{\textbf{\boldmath$\tint{12345}{}$\,:}}
  \tint{12345}{\epsilon^{-2}} &= 0\,,\\
  \tint{12345}{\epsilon^{-1}} &= 0\,.
\end{align}
\end{subequations}
}%

\addtocontents{toc}{\protect\newpage}

\tocsection[\label{SEC:DIAGLIST}]{List of renormalised Feynman diagrams}

The complete list of renormalised two-loop diagrams (with polynomial
remainder in the UV~regulator~$1/\epsilon$) for tadpoles and
self-energies in 't\,Hooft--Feynman gauge is shown in the
following. Contributions that only exist in the \MS scheme are marked
by~$\deltaMS$. At first, we repeat the previously known results of
\citeres{Martin:2003it,Goodsell:2015ira} in our nomenclature, and then
we list all new results.

\tocsubsection[\label{SEC:TADLIST}]{Tadpole diagrams}

\tocsubsubsection[\label{SEC:TADLISTMARK}]{Known results}

\vspace{-2ex}
\begin{longtable}{@{}ccl@{}}
  \tblt{01}{.865}\\
  \tblt{07}{.863}\\
  \tblt{24}{.76}\\
  \tblt{05}{.867}\\
  \tblt{06}{.867}\\\hline
\end{longtable}

\Needspace{5\baselineskip}
\tocsubsubsection[\label{SEC:TADLISTSCALARVECTOR}]{New results with vectors}

\vspace{-2ex}
\begin{longtable}{@{}ccl@{}}
  \tblt{02}{.885}\\
  \tblt{03}{.865}\\
  \tblt{04}{.883}\\
  \tblt{12}{.862}\\
  \tblt{13}{.865}\\
  \tblt{17}{.86}\\
  \tblt{18}{.865}\\
  \tblt{19}{.885}\\
  \tblt{21}{.865}\\
  \tblt{22}{.885}\\
  \tblt{23}{.885}\\
  \tblt{25}{.76}\\\hline
\end{longtable}

\Needspace{14\baselineskip}
\tocsubsubsection[\label{SEC:TADLISTFERMION}]{New results with fermions and vectors}

\vspace{-2ex}
\begin{longtable}{@{}ccl@{}}
  \tblt{10}{.868}\\
  \tblt{16}{.868}\\
  \tblt{11}{.868}\\\hline
\end{longtable}

\Needspace{5\baselineskip}
\tocsubsubsection[\label{SEC:TADLISTGHOST}]{New results with ghosts}

\vspace{-2ex}
\begin{longtable}{@{}ccl@{}}
  \tblt{08}{.865}\\
  \tblt{09}{.86}\\
  \tblt{14}{.865}\\
  \tblt{15}{.865}\\
  \tblt{20}{.865}\\\hline
\end{longtable}

\Needspace{5\baselineskip}
\tocsubsection[\label{SEC:SELIST}]{Self-energy diagrams}

\tocsubsubsection[\label{SEC:SELISTMARTINSCALAR}]{Known results with only scalars}

\vspace{-2ex}
\begin{longtable}{@{}ccl@{}}
  \tbls{001}\\
  \tbls{009}\\
  \hline\lbls{015} & \igself[015] & \lbls{009}$\Big|_{\ind{1}\leftrightarrow\ind{2}}$\\
  \tbls{024}\\
  \tbls{059}\\
  \tbls{101}\\
  \tbls{105}\\
  \tbls{111}\\
  \tbls{120}\\\hline
\end{longtable}

\Needspace{5\baselineskip}
\tocsubsubsection[\label{SEC:SELISTMARTINSCALARFERMION}]{Known results with scalars and fermions}

\vspace{-2ex}
\begin{longtable}{@{}ccl@{}}
  \tbls{021}\\
  \tbls{022}\\
  \hline\lbls{023} & \igself[023] & \lbls{022}$\Big|_{\ind{1}\leftrightarrow\ind{2},\,\ind{3}\leftrightarrow\ind{4},\,\ind{6}\leftrightarrow\ind{7}}$\\
  \tbls{057}\\
  \tbls{058}\\
  \tbls{110}\\\hline
\end{longtable}

\Needspace{12\baselineskip}
\tocsubsubsection[\label{SEC:SELISTMARTINSCALARVECTOR}]{Known results with scalars and one vector}

\vspace{-2ex}
\begin{longtable}{@{}ccl@{}}
  \tbls{002}\\
  \tbls{003}\\
  \tbls{010}\\
  \hline\lbls{016} & \igself[016] & \lbls{010}$\Big|_{\ind{1}\leftrightarrow\ind{2}}$\\
  \tbls{031}\\
  \hline\lbls{032} & \igself[032] & \lbls{031}$\Big|_{\ind{1}\leftrightarrow\ind{2},\,\ind{3}\leftrightarrow\ind{4},\,\ind{6}\leftrightarrow\ind{7}}$\\
  \tbls{033}\\
  \tbls{066}\\
  \tbls{067}\\
  \hline\lbls{068} & \igself[068] & \lbls{067}$\Big|_{\ind{1}\leftrightarrow\ind{2},\,\ind{4}\leftrightarrow\ind{7}}$\\
  \tbls{069}\\
  \tbls{102}\\
  \tbls{113}\\\hline
\end{longtable}

\Needspace{5\baselineskip}
\tocsubsubsection[\label{SEC:SELISTMARTINFERMIONVECTOR}]{Known results with fermions and one vector}

\vspace{-2ex}
\begin{longtable}{@{}ccl@{}}
  \tbls{028}\\
  \tbls{062}\\\hline
\end{longtable}

\Needspace{5\baselineskip}
\tocsubsubsection[\label{SEC:SELISTMARTINSCALARFERMIONVECTOR}]{Known results with scalars, fermions and one vector}

\vspace{-2ex}
\begin{longtable}{@{}ccl@{}}
  \tbls{029}\\
  \hline\lbls{030} & \igself[030] & \lbls{029}$\Big|_{\ind{1}\leftrightarrow\ind{2},\,\ind{3}\leftrightarrow\ind{4},\,\ind{6}\leftrightarrow\ind{7}}$\\
  \tbls{063}\\
  \tbls{064}\\
  \hline\lbls{065} & \igself[065] & \lbls{064}$\Big|_{\ind{1}\leftrightarrow\ind{2},\,\ind{4}\leftrightarrow\ind{7}}$\\\hline
\end{longtable}

\Needspace{5\baselineskip}
\tocsubsubsection[\label{SEC:SELISTSCALARVECTOR}]{New results with vectors}

\vspace{-2ex}
\begin{longtable}{@{}ccl@{}}
  \tbls{004}\\
  \tbls{005}\\
  \tbls{006}\\
  \tbls{007}\\
  \tbls{008}\\
  \tbls{011}\\
  \hline\lbls{017} & \igself[017] & \lbls{011}$\Big|_{\ind{1}\leftrightarrow\ind{2}}$\\
  \tbls{012}\\
  \hline\lbls{018} & \igself[018] & \lbls{012}$\Big|_{\ind{1}\leftrightarrow\ind{2}}$\\
  \tbls{013}\\
  \hline\lbls{019} & \igself[019] & \lbls{013}$\Big|_{\ind{1}\leftrightarrow\ind{2}}$\\
  \tbls{014}\\
  \hline\lbls{020} & \igself[020] & \lbls{014}$\Big|_{\ind{1}\leftrightarrow\ind{2}}$\\
  \tbls{039}\\
  \tbls{040}\\
  \hline\lbls{042} & \igself[042] & \lbls{040}$\Big|_{\ind{1}\leftrightarrow\ind{2},\,\ind{3}\leftrightarrow\ind{4},\,\ind{6}\leftrightarrow\ind{7}}$\\
  \tbls{041}\\
  \tbls{043}\\
  \hline\lbls{044} & \igself[044] & \lbls{043}$\Big|_{\ind{1}\leftrightarrow\ind{2},\,\ind{3}\leftrightarrow\ind{4},\,\ind{6}\leftrightarrow\ind{7}}$\\
  \tbls{047}\\
  \hline\lbls{048} & \igself[048] & \lbls{047}$\Big|_{\ind{1}\leftrightarrow\ind{2},\,\ind{3}\leftrightarrow\ind{4},\,\ind{6}\leftrightarrow\ind{7}}$\\
  \tbls{049}\\
  \tbls{050}\\
  \hline\lbls{052} & \igself[052] & \lbls{050}$\Big|_{\ind{1}\leftrightarrow\ind{2},\,\ind{3}\leftrightarrow\ind{4},\,\ind{6}\leftrightarrow\ind{7}}$\\
  \tbls{051}\\
  \tbls{053}\\
  \tbls{054}\\
  \hline\lbls{055} & \igself[055] & \lbls{054}$\Big|_{\ind{1}\leftrightarrow\ind{2},\,\ind{3}\leftrightarrow\ind{4},\,\ind{6}\leftrightarrow\ind{7}}$\\
  \tbls{056}\\
  \tbls{077}\\
  \hline\lbls{078} & \igself[078] & \lbls{077}$\Big|_{\ind{1}\leftrightarrow\ind{2},\,\ind{4}\leftrightarrow\ind{7}}$\\
  \tbls{079}\\
  \tbls{080}\\
  \tbls{081}\\
  \hline\lbls{082} & \igself[082] & \lbls{081}$\Big|_{\ind{1}\leftrightarrow\ind{2},\,\ind{4}\leftrightarrow\ind{7}}$\\
  \tbls{083}\\
  \tbls{088}\\
  \tbls{089}\\
  \hline\lbls{090} & \igself[090] & \lbls{089}$\Big|_{\ind{1}\leftrightarrow\ind{2},\,\ind{4}\leftrightarrow\ind{7}}$\\
  \tbls{091}\\
  \tbls{092}\\
  \tbls{093}\\
  \hline\lbls{094} & \igself[094] & \lbls{093}$\Big|_{\ind{1}\leftrightarrow\ind{2},\,\ind{4}\leftrightarrow\ind{7}}$\\
  \tbls{096}\\
  \tbls{097}\\
  \hline\lbls{098} & \igself[098] & \lbls{097}$\Big|_{\ind{1}\leftrightarrow\ind{2},\,\ind{4}\leftrightarrow\ind{7}}$\\
  \tbls{099}\\
  \tbls{100}\\
  \tbls{103}\\
  \tbls{104}\\
  \tbls{106}\\
  \hline\lbls{108} & \igself[108] & \lbls{106}$\Big|_{\ind{1}\leftrightarrow\ind{2},\,\ind{3}\leftrightarrow\ind{5},\,\ind{4}\leftrightarrow\ind{6}}$\\
  \tbls{107}\\
  \tbls{109}\\
  \tbls{115}\\
  \tbls{116}\\
  \tbls{118}\\
  \tbls{119}\\
  \tbls{121}\\\hline
\end{longtable}

\Needspace{5\baselineskip}
\tocsubsubsection[\label{SEC:SELISTFERMION}]{New results with fermions and vectors}

\vspace{-2ex}
\begin{longtable}{@{}ccl@{}}
  \tbls{037}\\
  \hline\lbls{038} & \igself[038] & \lbls{037}$\Big|_{\ind{1}\leftrightarrow\ind{2},\,\ind{3}\leftrightarrow\ind{4},\,\ind{6}\leftrightarrow\ind{7}}$\\
  \tbls{074}\\
  \hline\lbls{075} & \igself[075] & \lbls{074}$\Big|_{\ind{1}\leftrightarrow\ind{2},\,\ind{4}\leftrightarrow\ind{7}}$\\
  \tbls{076}\\
  \tbls{087}\\
  \tbls{114}\\\hline
\end{longtable}

\Needspace{7\baselineskip}
\tocsubsubsection[\label{SEC:SELISTGHOST}]{New results with ghosts}

\vspace{-2ex}
\begin{longtable}{@{}ccl@{}}
  \tbls{025}\\
  \hline\lbls{026} & \igself[026] & \lbls{025}$\Big|_{\ind{1}\leftrightarrow\ind{2},\,\ind{3}\leftrightarrow\ind{4},\,\ind{6}\leftrightarrow\ind{7}}$\\
  \tbls{027}\\
  \tbls{034}\\
  \hline\lbls{035} & \igself[035] & \lbls{034}$\Big|_{\ind{1}\leftrightarrow\ind{2},\,\ind{3}\leftrightarrow\ind{4},\,\ind{6}\leftrightarrow\ind{7}}$\\
  \tbls{036}\\
  \tbls{045}\\
  \hline\lbls{046} & \igself[046] & \lbls{045}$\Big|_{\ind{1}\leftrightarrow\ind{2},\,\ind{3}\leftrightarrow\ind{4},\,\ind{6}\leftrightarrow\ind{7}}$\\
  \tbls{060}\\
  \tbls{061}\\
  \tbls{070}\\
  \tbls{071}\\
  \hline\lbls{072} & \igself[072] & \lbls{071}$\Big|_{\ind{1}\leftrightarrow\ind{2},\,\ind{4}\leftrightarrow\ind{7}}$\\
  \tbls{073}\\
  \tbls{084}\\[-.2ex]
  \hline\lbls{085} & \igself[085] & \lbls{084}$\Big|_{\ind{1}\leftrightarrow\ind{2},\,\ind{4}\leftrightarrow\ind{7}}$\\[-.2ex]
  \tbls{086}\\[-.2ex]
  \tbls{095}\\[-.2ex]
  \tbls{112}\\[-.2ex]
  \tbls{117}\\[-.2ex]\hline
\end{longtable}

\begingroup
\let\secfnt\undefined
\newfont{\secfnt}{ptmb8t at 10pt}
\setstretch{.5}
\bibliographystyle{h-physrev}
\bibliography{literature}
\endgroup

\end{document}